\documentclass[12pt]{article}
\usepackage{cite}
\usepackage{multirow}
\usepackage{amssymb, amsmath, amsopn, amsthm}
\usepackage{bm,amsfonts,array,calc,rotating}
\usepackage{epsfig}
\usepackage{graphicx}
\usepackage{color}

\definecolor{orange}{cmyk}{0,0.5,1,0}

\newcommand{\p}{\partial}

\newcommand{\ra}{\rightarrow}
\usepackage[T1]{fontenc}

\numberwithin{equation}{section}

\long\def\@makecaption#1#2{%
  \vskip\abovecaptionskip
  \sbox\@tempboxa{{\bf #1:} #2}%
  \ifdim \wd\@tempboxa >\hsize
    {\small\bf #1:} {\small #2}\par
  \else
    \global \@minipagefalse
    \hb@xt@\hsize{\hfil\box\@tempboxa\hfil}%
  \fi
  \vskip\belowcaptionskip}

\font\cmss=cmss10 \font\cmsss=cmss10 at 7pt
\def\IZ{\relax\ifmmode\mathchoice
{\hbox{\cmss Z\kern-.4em Z}}{\hbox{\cmss Z\kern-.4em Z}}
{\lower.9pt\hbox{\cmsss Z\kern-.4em Z}} {\lower1.2pt\hbox{\cmsss
Z\kern-.4em Z}}\else{\cmss Z\kern-.4em Z}\fi}

\def\sqr#1#2{{\vcenter{\vbox{\hrule height.#2pt
 \hbox{\vrule width.#2pt height#1pt \kern#1pt
 \vrule width.#2pt}\hrule height.#2pt}}}}

\begin{document}

\begin{flushright}
\baselineskip=12pt \normalsize
{MIFP-09-43}\\
\smallskip
\end{flushright}

\begin{center}
\Large {\textbf{Abelian Gauge Fluxes and Local Models in F-Theory}} \\[2cm]
\normalsize  Yu-Chieh Chung
\\[.25in]
\textit{Department of Physics, Texas A$\&$M University, \\College Station, TX 77843, USA} \\[.5cm]
\tt \footnotesize ycchung@physics.tamu.edu\\[2.5cm]
\end{center}

\begin{abstract}
We analyze the Abelian gauge fluxes in local F-theory models with
$G_S=SU(6)$ and $SO(10)$. For the case of $G_S=SO(10)$, there is a
no-go theorem which states that for an exotic-free spectrum, there
are no solutions for $U(1)^2$ gauge fluxes. We explicitly
construct the $U(1)^2$ gauge fluxes with an exotic-free bulk
spectrum for the case of $G_S=SU(6)$. We also analyze the
conditions for the curves supporting the given field content and
discuss non-minimal spectra of the MSSM with doublet-triplet
splitting.

\end{abstract}
\vspace{4cm}


\newpage
\setcounter{page}{1}

\pagenumbering{arabic}

\pagestyle{plain}

\section{Introduction}

String theory is so far the most promising candidate for a unified
theory. Building realistic models of particle physics to answer
fundamental questions is one of the challenges in string theory.
One of the main issues to be addressed from particle physics is
the unification of gauge couplings. The natural solution to this
question is the framework of grand unified theory (GUT). One task
for string theory is whether it can accommodate GUT models. String
theory makes contact with four-dimensional physics through various
compactifications. There are two procedures to realize GUTs in
string theory compactifications. The first is the top-down
procedure in which the full compactification is consistent with
the global geometry of extra dimensions and then the spectrum is
close to GUT after breaking some symmetries \cite{GSW}. In the
bottom-up procedure, the gauge breaking can be understood in the
decoupling limit of gravity \cite{Aldazabal:2000sa,
Verlinde:2005jr}, particularly in the framework that D-branes are
introduced on the local regions within the extra dimensions in
type IIB compactification \cite{Aldazabal:2000sa, Verlinde:2005jr,
Blumenhagen:2005mu}. In this case we can neglect the effects from
the global geometry for the time being, which makes the procedure
more flexible and efficient. In addition, the construction of the
local models can reveal the requirements for the global geometry.
Eventually the local models need to be embedded into some compact
geometry for UV completion.

In $SU(5)$ GUTs, there are two important Yukawa couplings, $\bf
10\bf 10{\bf 5}_{H}$ and $\bf 10{\bf\bar 5}_{M}{\bf\bar 5}_{H}$.
It is well-known that $\bf 10\bf 10{\bf 5}_{H}$ is forbidden in
perturbative type IIB theory. However, it was shown in
\cite{BlumenhagenIbanezKachru,Blumenhagen:2008zz} that the Yukawa
coupling $\bf 10\bf 10{\bf 5}_{H}$ can be achieved by introducing
non-perturbative corrections. From this perspective, the
non-perturbative property is intrinsic for GUT model building in
type IIB theory. F-theory is a non-perturbative 12-d theory built
on the type IIB framework with an auxiliary two-torus
(\cite{Vafa:1996xn}, see \cite{Denef:2008wq} for review). The
ordinary string extra dimensions are regarded as a base manifold
and the two-torus is equivalent to an elliptic curve as a fiber on
this base manifold. The modulus of the elliptic curve is
identified as axion-dilaton in type IIB theory. Due to the SL(2,Z)
monodromy of the modulus, F-theory is essentially non-perturbative
in type IIB language. There are elegant correspondences between
physical objects in type IIB and geometry in F-theory. The modular
parameter of the elliptical fiber, identified with the
axion-dilaton in type IIB, varies over the base. Singularities
develop when the fibers degenerate. The loci of the singular
fibers indicate the locations of the seven-branes in type IIB and
the type of the singularity determines the gauge group of the
world-volume theory of the seven-branes \cite{Bershadsky:1996nh}.
According to the classification of the singular fibration, there
are singularities of types $A$, $D$, and $E$. The first two types
have perturbative descriptions in Type IIB. More precisely,
$A$-type and $D$-type singularities correspond to configurations
of the $D7$-branes and $D7$-branes along $O$-planes, respectively
\cite{Sen:1996vd}. For the singularity of type $E$, there is no
perturbative description in type IIB, which means that F-theory
captures a non-perturbative part of the type IIB theory. Under
geometric assumptions, the full F-theory can decouple from gravity
\cite{BHV:2008I,BHV:2008II,Donagi:local}. In this way, one can
focus on the gauge theory descending from world-volume theory of
the seven-branes supported by the local geometry of the
discriminant loci in the base manifold of a elliptically fibered
Calabi-Yau fourfold. Recently some local supersymmetric GUT models
have been built in this F-theory context \cite{BHV:2008I,
BHV:2008II,Donagi:local,Heckman:local,Heckman:localFlavor01,
Heckman:localFlavor02,Heckman:localFlavor03,Ibanez:locaFlavorl01,
Ibanez:localFlavor02,Randall:localFlavor,Nanopoulos:local,Blumenhagen:local,Bourjaily:local,
Chen:2009me,Conlon:local01,Conlon:local02,Vafa:NoncommutativeandYukawa},
and some progress has been made in constructing global models
\cite{Caltech:global,Donagi:global,Cordova:Decouplinggeom,Other:global,
Tartar:globalFlavor01,Tartar:globalFlavor02}. Supersymmetry
breaking has been discussed in
\cite{Buchbinder:SUSYbreaking,Caltech:SUSYbreaking,Blumenhagen:SUSYbreaking},
and the application to cosmology has been studied in
\cite{Heckman:cosmology}. It has become more clear that F-theory
provides a very promising framework for model building of
supersymmetric GUTs. To build local $SU(5)$ GUTs in F-theory, one
can start with engineering a Calabi-Yau fold with an $A_4$
singularity. To decouple from gravity, it is required that the
volume of $S$, which is a component of the discriminant locus and
is wrapped by seven-branes is contractible to zero
size.\footnote{There are two ways in which we could take $V_S\ra
0$. The first way is by requiring $S$ to contract to a point, and
the second is by requiring $S$ to contract to a curve of
singularities. See \cite{Donagi:global,Cordova:Decouplinggeom} for
the details.} We assume that $S$ can contract to a point and thus
possesses an ample canonical bundle $K^{-1}_S$
\cite{BHV:2008I,BHV:2008II,Donagi:local}. In particular, we focus
on the case that $S$ is a del Pezzo surface wrapped by
seven-branes, which engineers an eight-dimensional supersymmetric
gauge theory with gauge group $G_S=SU(5)$ in
$\mathbb{R}^{3,1}\times S$. Other components $S'_i$ of the
discriminant locus intersect $S$ along the curves $\Sigma_i$. Due
to the collision of the singularities, the gauge group $G_S$ will
be enhanced to $G_{\Sigma_i}$ on $\Sigma_i$ and the matter in the
bi-fundamental representations will be localized on the curves
\cite{Katz:1996xe}. It was shown in
\cite{BHV:2008I,BHV:2008II,Donagi:local} that the spectrum is
given by the bundle-valued cohomology groups. In
\cite{BHV:2008I,BHV:2008II,Donagi:local}, the minimal $SU(5)$ GUT
has been studied. In that case, with non-trivial $U(1)_Y$ gauge
flux, the GUT group is broken into $G_{\rm std}\equiv SU(3)\times
SU(2)\times U(1)_Y$. Furthermore, one can obtain an exotic-free
spectrum of the minimal supersymmetric Standard Model (MSSM) from
those curves with doublet-triplet splitting but no rapid proton
decay. The success of the minimal $SU(5)$ GUT model motivates us
to pursue other local GUT models from higher rank gauge groups.
The next simplest one is gauge group of rank five, namely $SO(10)$
and $SU(6)$. These two non-minimal $SU(5)$ GUTs have been studied
in \cite{Chen:2009me}. For the latter, one can get an exotic-free
spectrum, but due to the lack of an extra $U(1)$ flux, the GUT
group cannot be broken into $G_{\rm std}$. To avoid this
difficulty, it is natural to study local F-theory models of
$G_S=SU(6)$ and $G_S=SO(10)$ with supersymmetric $U(1)^2$ gauge
fluxes, which consist of two supersymmetric $U(1)$ gauge fluxes
and are associated with rank two polystable bundles over $S$. The
aim of the present paper is to construct explicitly the
supersymmetric $U(1)^2$ gauge fluxes in local F-theory models of
$G_S=SU(6)$ and $SO(10)$ and study the matter spectrum of the
MSSM.

For the case of $G_S=SO(10)$, there is a no-go theorem
\cite{BHV:2008II} which states that for an exotic-free spectrum,
there are no solutions for $U(1)^2$ gauge fluxes. For the case of
$G_S=SU(6)$, we can explicitly construct supersymmetric $U(1)^2$
gauge fluxes. It turns out that each flux configuration contains
two fractional lines bundles. One of the gauge fluxes is universal
and has the same $U(1)_Y$ hypercharge flux as the minimal $SU(5)$
GUT \cite{BHV:2008I,BHV:2008II,Donagi:local}. The second one
varies along with the configurations of the bulk zero modes. With
suitable supersymmetric $U(1)^2$ gauge fluxes, the bulk spectrum
will be exotic-free and the chiral matter will come from the
curves. The restriction of these $U(1)^2$ fluxes to the curves
induce $U(1)$ fluxes over the curves, which breaks the enhanced
gauge group $G_{\Sigma}$ into $G_{\rm std}\times U(1)$. In this
case, the Higgs fields can be localized on the curves
$\Sigma_{SU(7)}$ and $\Sigma_{SO(12)}$. On the $\Sigma_{SU(7)}$,
non-trivial induced fluxes break $SU(7)$ into $G_{\rm std}\times
U(1)$. With suitable fluxes, doublet-triplet splitting can be
achieved. However, the situations become more complicated on the
curves with $G_{\Sigma}=SO(12)$. Since the dimension of the
adjoint representation of $SO(12)$ is higher than $SU(7)$, one
gets more constraints to solve for given field configurations,
which results in difficulties for doublet-triplet splitting. By
explicitly solving the allowed field configurations, one can find
that there are still a few solutions with doublet-triplet
splitting. From the analysis, it is clear that if one engineers
the Higgs fields on the curve $\Sigma_{SU(7)}$ instead of
$\Sigma_{SO(12)}$, this is the case. To obtain a complete matter
spectrum of the MSSM, we analyze the case of $\Sigma_{E_6}$ in
addition to $\Sigma_{SU(7)}$ and $\Sigma_{SO(12)}$. It is
extremely difficult to obtain the minimal spectrum of the MSSM
without exotic fields. However, we found that in some cases, the
exotic fields can form trilinear couplings with the doublets or
triplets on the curves with $G_{\Sigma}=SU(7)$. When these fields
get vacuum expectation values (vevs), the exotic fields will be
decoupled from the low-energy spectrum. A way to do this is that
we introduce extra curves supporting the doublets or triplets,
which intersect the curves hosting the exotic fields to form the
couplings. With the help of these doublets or triplets, it turns
out that the non-minimal spectrum of the MSSM without
doublet-triplet splitting problem can be achieved by local
F-theory model of $G_S=SU(6)$ with supersymmetric $U(1)^2$ gauge
fluxes.

The organization of the rest of the paper is as follows: in
section $2$, we briefly review the construction of local F-theory
model and local geometry, in particular the geometry of the del
Pezzo surfaces. In section $3$, we include a brief review of the
$SU(5)$ GUTs with $G_S=SU(5)$, $SO(10)$, and $SU(6)$. We also
introduce the notion of stability of the vector bundle, in
particular, that of the polystable bundle of rank two in section
$4$. In section $5$, we review a no-go theorem for the case of
$G_S=SO(10)$ and construct explicitly supersymmetric $U(1)^2$
gauge fluxes for the case of $G_S=SU(6)$. We also give examples
for non-minimal spectra of the MSSM with doublet-triplet
splitting. We conclude in section $6$.

\section{F-theory and Local Geometry}
In this section we shall review some important ingredients of the
local F-theory models and local geometry, and in particular the
geometry of the del Pezzo surfaces.

\subsection{F-theory}
Consider F-theory compactified on an elliptically fibred
Calabi-Yau fourfold, $T^2\ra X\ra B$ with sections, which can be
realized in the Weierstrass form,
\begin{eqnarray}
y^{2}=x^{3}+fx+g,
\end{eqnarray}
where $x$ and $y$ are the complex coordinates on the fiber, $f$
and $g$ are sections of the suitable line bundles over the base
manifold $B$. The degrees of $f$ and $g$ are determined by the
Calabi-Yau condition, $c_{1}(X)=0$. The degenerate locus of fibers
is given by the discriminant $\Delta=4f^{3}+27g^{2}=0$, which is
in general a codimension one reducible subvariety in the base $B$.
For local models, we focus on one component $S$ of the
discriminant locus $\Delta=0$, which will be wrapped by a stack of
the seven-branes and supports the GUT model. In order to decouple
from the gravitational sector, the anti-canonical bundle
$K^{-1}_{S}$ of the surface $S$ is assumed to be ample. According
to the classification theorem of algebraic surfaces, the surface
$S$ is a del Pezzo surface and birational to the complex
projective plane $\mathbb{P}^{2}$ . There are ten del Pezzo
surfaces: $\mathbb{P}^1\times \mathbb{P}^1$, $\mathbb{P}^2$, and
$dP_k,\;k=1,2,...,8$, which are blow-ups of $k$ generic points on
$\mathbb{P}^{2}$. In this paper we shall focus on the case of
$S=dP_k,\; 2\leqslant k\leqslant 8$ with $(-2)$
2-cycles\footnote{A $(-2)$ 2-cycle is a $2$-cycle with
self-intersection number $-2$.}. In the vicinity of $S$, the
geometry of $X$ may be regarded approximately an ALE fibration
over $S$. The singularity of the ALE fiberation determines the
gauge group $G_S$ of 8d $\mathcal{N}=1$ super-Yang-Mills theory.
After compacifying on $S$ and partially twisting, the resulting
effective theory is 4d $\mathcal{N}=1$ super-Yang-Mills theory
whose gauge group is the commutant of structure group of the
vector bundle over $S$ in $G_S$
\cite{BHV:2008I,BHV:2008II,Donagi:local}. Let $V$ be a holomorphic
vector bundle over $S$. The unbroken gauge group in 4d is the
commutant $\Gamma_S$ of $H_S$ in $G_{S}$, where $H_S$ is the
structure group of the bundle $V$. In order to preserve
supersymmetry, the bundle $V$ has to admit a hermitian connection
$A$ satisfying the Donaldson-Uhlenbeck-Yau (DUY) equation
\cite{GSW}
\begin{eqnarray}
F_{mn}=F_{\bar m\bar n}=0,\;\;g^{m\bar n}F_{m\bar n}=0,\label{DUY}
\end{eqnarray}
where $g_{m\bar n}$ is a K\"ahler metric on $S$, and $F$ is the
curvature of the connection $A$. It was shown in
\cite{Donalson,UY} that a bundle admitting a hermitian connection
solving Eq. (\ref{DUY}) is equivalent to a (semi) stable bundle,
which is guaranteed by the Donaldson-Uhlebecker-Yau theorem. We
shall in the next section define the stability of vector bundles
and briefly review some facts about the equivalence. The spectrum
from the bulk is given by the bundle-valued cohomology groups
$H_{\bar\p}^{i}(S,R_{k})$ and their duals, where
$R_{k}=V,\;\wedge^{k}V$, or ${\rm EndV}$. The spectrum of the bulk
transforms in the adjoint representation of $G_{S}$. The
decomposition of ${\rm ad} G_{S}$ into representations of
$\Gamma_{S}\times H_{S}$ is
\begin{eqnarray}
{\rm ad}G_{S}=\bigoplus_{k}\rho_{k}\otimes \mathcal{R}_{k},
\end{eqnarray}
where $\rho_{k}$ and $\mathcal{R}_{k}$ are representations of
$\Gamma_S$ and $H_{S}$, respectively. The matter fields are
determined by the zero modes of the Dirac operator on $S$. It was
shown in \cite{BHV:2008II,Donagi:local} that the chiral and
anti-chiral spectrum is determined by the bundle-valued cohomology
groups
\begin{equation}
H^{0}_{\bar\p}(S,R_{k}^{\vee})^{\vee}\oplus
H_{\bar\p}^{1}(S,R_{k})\oplus
H_{\bar\p}^{2}(S,R_{k}^{\vee})^{\vee}
\end{equation}
and
\begin{equation}
H_{\bar\p}^{0}(S,R_{k})\oplus
H_{\bar\p}^{1}(S,R_{k}^{\vee})^{\vee}\oplus
H_{\bar\p}^{2}(S,R_{k})
\end{equation}
respectively, where $\vee$ stands for the dual bundle and $R_{k}$
is the vector bundle on $S$ whose sections transform in the
representation $\mathcal{R}_{k}$ of the structure group $H_{S}$.
By the vanishing theorem of del Pezzo surfaces \cite{BHV:2008II},
the number of chiral fields $\rho_{k}$ and anti-chiral fields
${\rho^{\ast}_{k}}$ can be calculated by
\begin{equation}
N_{\rho_{k}}=-\chi(S,R_{k})\label{numberdiff}
\end{equation}
and
\begin{equation}
N_{\rho^{\ast}_{k}}=-\chi(S,R_{k}^{\vee}),
\end{equation}
respectively. In particular, when $V=L_1\oplus L_2$ with structure
group $U(1)\times U(1)$, according to Eq. (\ref{numberdiff}), the
chiral spectrum of $\rho_{r,s}$ is determined by
\begin{equation}
N_{\rho_{r,s}}=-\chi(S,{L_1}^{r}\otimes L_2^{s}),\label{EulerChar}
\end{equation}
where $r$ and $s$ correspond respectively to the $U(1)_1$ and
$U(1)_2$ charges of the representations in the group theory
decomposition. In order to preserve supersymmetry, the gauge
bundle $V$ has to obey the DUY equation (\ref{DUY}), which is
equivalent to the polystability conditions, namely
\begin{equation}
J_{S}\wedge c_{1}(L_1)=J_{S}\wedge c_{1}(L_2)=0,\label{BPS}
\end{equation}
where $J_{S}$ is the K\"ahler form on $S$. We will discuss the
polystability conditions in more detail in section $4$.

Another way to obtain chiral matter is from intersecting
seven-branes along a curve, which is a Riemann surface. Let $S$
and $S'$ be two components of the discriminant locus $\Delta$ with
gauge groups $G_{S}$ and $G_{S'}$, respectively. The gauge group
on the curve $\Sigma$ will be enhanced to $G_{\Sigma}$, where
$G_{\Sigma}\supset G_{S}\times G_{S'}$. Therefore, chiral matter
appears as the bi-fundamental representations in the decomposition
of ${\rm ad}G_{\Sigma}$
\begin{equation}
{\rm ad}G_{\Sigma}={\rm ad}G_{S}\oplus {\rm ad}G_{S'}\oplus_{k}
({\cal U}_{k}\otimes {\cal U'}_{k}).\label{bifundamental}
\end{equation}
As mentioned above, the presence of $H_{S}$ and $H_{S'}$ will
break $G_{S}\times G_{S'}$ to the commutant subgroup when
non-trivial gauge bundles on $S$ and $S'$ with structure groups
$H_{S}$ and $H_{S'}$ are turned on. Let
$\Gamma=\Gamma_{S}\times\Gamma_{S'}$ and $H=H_{S}\times H_{S'}$,
the decomposition of ${\cal U}\otimes {\cal U'}$ into irreducible
representation is
\begin{equation}
{\cal U}\otimes {\cal U'}={\bigoplus}_{k}(v_{k}, {\cal V}_{k}),
\end{equation}
where $v_{k}$ and ${\cal V}_{k}$ are representations of $\Gamma$
and $H$, respectively. The light chiral fermions in the
representation $v_{k}$ are determined by the zero modes of the
Dirac operator on $\Sigma$. It is shown in
\cite{BHV:2008II,Donagi:local} that the net number of chiral
fields $v_{k}$ and anti-chiral fields $v^{\ast}_{k}$ is given by
\begin{eqnarray}
N_{v_{k}}-N_{v^{\ast}_{k}}=\chi(\Sigma,K^{1/2}_{\Sigma}\otimes
V_{k}),
\end{eqnarray}
where $V_{k}$ is the vector bundle whose sections transform in the
representation ${\cal V}_{k}$ of the structure group $H$. In
particular, if $H_S$ and $H_{S'}$ are $U(1)\times U(1)$ and
$U(1)$, respectively, $G_{\Sigma}$ can be broken into $G_M\times
U(1)\times U(1)\times U(1)\subset G_S\times U(1)$. In this case,
the bi-fundamental representations in Eq. (\ref{bifundamental})
will be decomposed into
\begin{equation}
\bigoplus_{j}(\sigma_{j})_{r_{j},s_j,r'_{j}},
\end{equation}
where $r_{j}$, $s_j$ and $r'_{j}$ correspond to the $U(1)$ charges
of the representations in the group theory decomposition and
$\sigma_{j}$ are representations in $G_M$. The representations
$(\sigma_{j})_{r_{j},s_j,r'_{j}}$ are localized on $\Sigma$
\cite{Katz:1996xe,BHV:2008II,Donagi:local} and as shown in
\cite{BHV:2008II,Donagi:local}, the generation number of the
representations $(\sigma_{j})_{r_{j},s_j,r'_{j}}$ and
$(\bar\sigma_{j})_{-r_{j},-s_j,-r'_{j}}$ can be calculated by
\begin{equation}
N_{(\sigma_{j})_{r_{j},s_j,r'_{j}}}=h^{0}(\Sigma,K^{1/2}_{\Sigma}
\otimes {L}_{1\Sigma}^{r_{j}}\otimes {L}_{2\Sigma}^{s_{j}}\otimes
{L'}_{\Sigma}^{r'_{j}})
\end{equation}
and
\begin{equation}
N_{(\bar\sigma_{j})_{-r_{j},-s_j,-r'_{j}}}=h^{0}(\Sigma,K^{1/2}_{\Sigma}
\otimes {L}_{1\Sigma}^{-r_{j}}\otimes
{L}_{2\Sigma}^{-s_{j}}\otimes {L'}_{\Sigma}^{-r'_{j}}),
\end{equation}
where ${L}_{1\Sigma}\equiv L_1|_{\Sigma}$, ${L}_{2\Sigma}\equiv
L_2|_{\Sigma}$, and ${L'}_{\Sigma}\equiv L'|_{\Sigma}$ are the
restrictions of the line bundles $L_1$, $L_2$ and $L'$ to the
curve $\Sigma$, respectively. Note that from Eq. (\ref{genus
zero}) below, if $c_1({L}_{1\Sigma}^{r_{j}}\otimes
{L}_{2\Sigma}^{s_{j}}\otimes {L'}_{\Sigma}^{r'_{j}})=0$, then
$N_{(\sigma_{j})_{r_{j},s_j,r'_{j}}}=N_{(\bar\sigma_{j})_{-r_{j},-s_j,-r'_{j}}}=0$.
If $c_1({L}_{1\Sigma}^{r_{j}}\otimes {L}_{2\Sigma}^{s_{j}}\otimes
{L'}_{\Sigma}^{r'_{j}})\neq 0$, then only one of them is
non-vanishing. Using these properties, we can solve the
doublet-triplet splitting problem with suitable line bundles. In
addition to the analysis of the spectrum, the pattern of Yukawa
couplings also has been studied
\cite{BHV:2008I,BHV:2008II,Donagi:local, Tartar:globalFlavor01}.
By the vanishing theorem of del Pezzo surfaces
\cite{BHV:2008II,Donagi:local}, Yukawa couplings can form in two
different ways. In the first way, the coupling comes from the
interaction between two fields on the curves and one field on the
bulk $S$. In the second way, all three fields are localized on the
curves which intersect at a point where the gauge group $G_p$ is
further enhanced by two ranks. Recently, flavor physics in
F-theory models has been studied in
\cite{Heckman:localFlavor01,Heckman:localFlavor02,Heckman:localFlavor03,
Ibanez:locaFlavorl01,Ibanez:localFlavor02,Randall:localFlavor,Tartar:globalFlavor01,
Tartar:globalFlavor02,Conlon:local02,Vafa:NoncommutativeandYukawa}.
When one turns on bulk three-form fluxes, the structure of the
Yukawa couplings will be distorted and non-commutative geometry
will emerge \cite{Vafa:NoncommutativeandYukawa}.  The case of
${\rm rk(V)}=1$ and minimal $SU(5)$ GUT model has been studied in
\cite{BHV:2008I,BHV:2008II,Donagi:local}. In this article, we
shall focus on the case that $V$ is a polystable bundle of rank
two. We will study non-minimal cases, namely $G_S=SU(6)$ and
$SO(10)$ with these rank two polystable bundles and the spectrum
of the MSSM.

\subsection{Local Geometry}

To make the present paper self-contained, in this section we
include a brief review of the geometry of the del Pezzo surfaces,
curves on the surfaces and some useful formulae.
\subsubsection{Del Pezzo Surfaces}

As mentioned in the previous section, in local models we require
that the anti-canonical bundle $K^{-1}_S$ of the surface $S$
wrapped by the seven-branes be ample. An algebraic surface with
ample anti-canonical bundle is called a del Pezzo surface. It was
shown that there are ten families of del Pezzo surfaces:
$\mathbb{P}^{1}\times \mathbb{P}^1$, $\mathbb{P}^2$ and the
blow-ups of $\mathbb{P}^2$ at $k$ generical points, where
$1\leqslant k\leqslant 8$.\cite{delPezzo:01,delPezzo:02}. In what
follows, we shall briefly review the geometry of the del Pezzo
surfaces.

The del Pezzo surface $S$ is an algebraic surface with ample
anti-canonical bundle, namely $K_S^{-1}>0$. It follows that
$h^{1}(S,\mathcal{O}_{S})=h^{2}(S,\mathcal{O}_{S})=0$\footnote{It
can be easily seen by the Kodaira vanishing theorem which states
that for any ample line bundle $\mathcal{L}$, $h^{i}(S,K_S\otimes
\mathcal{L})=0,\;\forall i>0$.} and that
$\chi(S,\mathcal{O}_{S})=\sum_{i=0}^{2}(-1)^{i}h^{i}(S,\mathcal{O}_{S})=1$.
According to the classification theorem of algebraic surfaces,
these surfaces are birational to the complex projective plane
$\mathbb{P}^2$. It was shown in
\cite{BHV:2008I,BHV:2008II,Donagi:local} that to obtain an
exotic-free bulk spectrum, the gauge fluxes have to correspond to
the dual of $(-2)$ 2-cycle in $S$. The Picard group of
$\mathbb{P}^2$ is generated by hyperplane divisor $H$ with
intersection number $H\cdot H=1$. Thus there is no $(-2)$ 2-cycle
in $\mathbb{P}^2$. Let us turn to the case of $dP_k$. The Picard
group of $dP_k$ is generated by the hyperplane divisor $H$, which
is inherited from $\mathbb{P}^2$ and the exceptional divisors
$E_i,\;i=1,2,..,k$ from the blow-ups with intersection numbers
$H\cdot H=1$, $H\cdot E_i=0$, and $E_i\cdot
E_j=-\delta_{ij},\;\forall\;i,j$. It is easy to see that $dP_1$
contains no $(-2)$ 2-cycles. It follows that the candidates of the
del Pezzo surfaces containing $(-2)$ 2-cycles are $dP_k$ with
$2\leqslant k\leqslant 8$. In what follows, I shall focus on the
del Pezzo surfaces $dP_k$ with $2\leqslant k\leqslant 8$. The
canonical divisor of $dP_k$ is $K_S=-3H+E_1+,...,+E_k$. The first
term comes from $K_{\mathbb{P}^2}=-3H$ and the rest comes from the
blow-ups, which lead to the exceptional divisors
$E_1,\;E_2,...,\;E_k$. For local models in F-theory, the curves
supporting matter fields are required to be effective. Next we
shall define effective curves and the Mori cone. Consider a
complex surface $Y$ and its homology group $H_2(Y,\mathbb{Z})$.
Let $C$ be a holomorphic curve in $Y$. Then $[C]\in
H_2(Y,\mathbb{Z})$ is called an effective class if $[C]$ is
equivalent to $C$. The Mori cone $\overline{{\rm NE}}(Y)$ is
spanned by a countable number of generators of the effective
classes \cite{Hartshone:01,Griffith:01}. The Mori cones
$\overline{{\rm NE}}(dP_k)$ of the del Pezzo surfaces $dP_k$ are
all finitely generated \cite{delPezzo:01}. To be concrete, we list
the generators of the Mori cones of $dP_k,\; 2\leqslant k\leqslant
8$ in Table \ref{MoriCone}.

\begin{table}[h]
\begin{center}
\renewcommand{\arraystretch}{1.25}
\begin{tabular}{|c|c|c|c|c|c|c|c|} \hline

Mori Cone & Generators & Number \\
\hline\hline

$\overline{{\rm NE}}(dP_2)$ & $E_i,\;H-E_1-E_2$& 3\\
\hline
$\overline{{\rm NE}}(dP_3)$ & $E_i,\;H-\sum_{m=1}^{2}E_{i_{m}}$& 6\\
\hline
$\overline{{\rm NE}}(dP_4)$ & $E_i,\;H-\sum_{m=1}^{2}E_{i_{m}}$ & 10\\
\hline
$\overline{{\rm NE}}(dP_5)$ & $E_i,\;H-\sum_{m=1}^{2}E_{i_{m}},\;2H-\sum_{n=1}^{5}E_{i_{n}}$ & 16\\
\hline
$\overline{{\rm NE}}(dP_6)$ & $E_i,\;H-\sum_{m=1}^{2}E_{i_{m}},\;2H-\sum_{n=1}^{5}E_{i_{n}}$ & 27\\
\hline
$\overline{{\rm NE}}(dP_7)$ & $E_i,\;H-\sum_{m=1}^{2}E_{i_{m}},\;2H-\sum_{n=1}^{5}E_{i_{n}},\;3H-2E_i-\sum_{p=1}^{6}E_{i_{p}}$ & 56\\
\hline
 & $E_i,\;H-\sum_{m=1}^{2}E_{i_{m}},\;2H-\sum_{n=1}^{5}E_{i_{n}},\;3H-2E_i-\sum_{p=1}^{6}E_{i_{p}},$ & 240\\
$\overline{{\rm NE}}(dP_8)$ &
$4H-2\sum_{q=1}^{3}E_{i_{q}}-\sum_{r=1}^{5}E_{i_{r}},\;5H-2\sum_{l=1}^{6}E_{i_{l}}-E_r-E_s,$ &\\
& $6H-3E_i-2\sum_{m=1}^{7}E_{i_{m}}$ & \\
\hline
\end{tabular}
\caption{The generators of the Mori cone $\overline{{\rm
NE}}(dP_k)$ for $k=2,...8$, where all indices are distinct.}
\label{MoriCone}
\end{center}
\end{table}
With the Mori cone, one can easily check that the anti canonical
divisor $-K_S$ is ample.\footnote{Here we can apply the
Nakai-Moishezon criterion which states that for any divisor $D$,
$D$ is ample if and only if $D\cdot D>0$ and $D\cdot
C_{\alpha}>0$, where $C_{\alpha}$ are generators of the Mori
cone.} The dual of the Mori cone is the ample cone, denoted by
${\rm Amp}(dP_k)$, which is defined by ${\rm
Amp}(dP_k)=\{\omega\in
 H_2(dP_k,\mathbb{R})|\;\omega\cdot\zeta>0,\;\forall\zeta\in
\overline{{\rm NE}}(dP_k)\}$. Each ample divisor $\omega$ in the
ample cone is associated with a K\"ahler class $J_S$. In this
article we choose ``large volume polarization'', namely
$\omega=AH-\sum_{i=1}^{k}a_kE_k$ with $A\gg a_k>0$
\cite{BHV:2008I,BHV:2008II}. It is easy to check that $\omega$ is
ample. For the del Pezzo surfaces $S$ and a line bundle
$\mathcal{L}$ over $S$, there are two useful theorems. One is the
Riemann-Roch theorem \cite{Hartshone:01,Griffith:01}, which says
that
\begin{equation}
\chi(S,\mathcal{L})=1+\frac{1}{2}c_1({\mathcal{L}})^2-\frac{1}{2}c_1({\mathcal{L}})\cdot
K_S.\label{R-R Surface}
\end{equation}
Another one is the vanishing theorem (\cite{BHV:2008I}, also see
\cite{YM:01}), which states that for a non-trivial holomorphic
vector bundle $\mathcal{V}$ over $S$ satisfying the DUY equation
(\ref{DUY}),
\begin{equation}
H_{\bar\p}^{0}(S,\mathcal{V})=H_{\bar\p}^{2}(S,\mathcal{V})=0.
\label{Vanishing theorem}
\end{equation}
These two theorems simplify the calculation of the spectrum. Note
that the vanishing theorem (\ref{Vanishing theorem}) holds when
$\mathcal{V}$ is a line bundle. It follows from Eq. (\ref{R-R
Surface}) and Eq. (\ref{Vanishing theorem}) that
$h^{1}(S,\mathcal{L})=-\chi(S,\mathcal{L})=-(1-\frac{1}{2}c_1({\mathcal{L}})\cdot
K_S+\frac{1}{2}c_1({\mathcal{L}})^2)$. The number of zero modes
will be determined by the intersection numbers
$c_1({\mathcal{L}})\cdot K_S$ and $c_1({\mathcal{L}})^2$.

For local models, we require that all curves be effective. That
is, the homological classes of the curves in $H_2(S,\mathbb{Z})$
can be written as non-negative integral combinations of the
generators of the Mori cone, namely
$\Sigma=\sum_{\beta}n_{\beta}\mathcal{C}_{\beta}$ with
$n_{\beta}\in \mathbb{Z}_{\geqslant 0}$\footnote{By abuse of
notation, we use $\Sigma$ to denote the homological class of the
curve $\Sigma$.}. To calculate the genus of the curve, we can
apply the adjunction formula, which says that for a smooth,
irreducible curve of genus $g$, the following equation holds
\begin{equation}
\Sigma\cdot (\Sigma+K_S)=2g-2.
\end{equation}
In the present paper, we shall choose genus zero curves to support
the matter in the GUTs or MSSM, which means that all matter curves
satisfy the equation $\Sigma\cdot (\Sigma+K_S)=-2$. To calculate
the spectrum from the curves, we also need the Rieman-Roch theorem
\cite{Hartshone:01,Griffith:01} for the algebraic curves. For the
case of the algebraic curve $\Sigma$, the Rieman-Roch theorem
states that for a line bundle $\mathcal{L}$ over $\Sigma$,
\begin{equation}
h^{0}(\Sigma,\mathcal{L})-h^{1}(\Sigma,\mathcal{L})=1-g+c_1.(\mathcal{L}).
\end{equation}
In particular, for the case of $g=0$, we have
\begin{equation}
h^{0}(\Sigma,K^{1/2}_{\Sigma}\otimes\mathcal{L})=\left \{
\begin{array}{l}
c_1(\mathcal{L}),\;\;\;\;\;{\rm if}\;c_1(\mathcal{L})\geqslant 0\\
0,\;\;\;\;\;\;\;\;\;\;\;\;{\rm if}\;c_1(\mathcal{L})<0,
\end{array} \right.\label{genus zero}
\end{equation}
where $K^{1/2}_{\Sigma}$ is the spin bundle of $\Sigma$ and the
Serre duality \cite{Hartshone:01,Griffith:01} has been used. Eq.
(\ref{genus zero}) will be useful to calculate the spectrum from
the curves.

\section{$U(1)$ Gauge Fluxes}

In this section we briefly review some ingredients of $SU(5)$ GUT
Models with $G_S=SU(5),\;SU(10)$ and $SU(6)$. In these models, we
introduce a non-trivial $U(1)$ gauge flux to break gauge group
$G_S$. We are primarily interested in doublet-triple splitting and
an exotic-free spectrum of the MSSM. From now on, unless otherwise
stated, the del Pezzo surface $S$ is assumed to be $dP_8$.

\subsection{$G_S=SU(5)$}
Before discussing the case of $G_S=SO(10),\;SU(6)$, let us review
the case of $G_S=SU(5)$ \cite{BHV:2008I,BHV:2008II,Donagi:local}.
On the bulk, we consider the following breaking pattern
\cite{Slansky:1981yr}:
\begin{equation}
\begin{array}{c@{}c@{}l@{}c@{}l}
SU(5) &~\rightarrow~& SU(3)\times SU(2)\times U(1)_S\\
{\bf 24} &~\rightarrow~& {\bf (8,1)}_{0}+{\bf (1,3)}_{0}+{\bf
(3,2)}_{-5}+{\bf (\bar
3,2)}_{5}+{\bf (1,1)}_{0}.\\
\end{array}
\end{equation}
The bulk zero modes are given by
\begin{equation}
{\bf(3,2)}_{-5}\in H_{\bar\p}^{0}(S,L^{5})^{\vee}\oplus
H_{\bar\p}^{1}(S,L^{-5})\oplus
H_{\bar\p}^{2}(S,L^{5})^{\vee}\label{SU(5)bulk01}
\end{equation}
\begin{equation}
{\bf(\bar 3,2)}_{5}\in H_{\bar\p}^{0}(S,L^{-5})^{\vee}\oplus
H_{\bar\p}^{1}(S,L^{5})\oplus
H_{\bar\p}^{2}(S,L^{-5})^{\vee},\label{SU(5)bulk02}
\end{equation}
where $\vee$ stands for the dual and $L$ is the supersymmetric
line bundle associated with $U(1)_S$. Let $N_{{\bf (A,B)}_{c}}$ be
the number of the fields in the representation ${\bf (A,B)}_{c}$
under $SU(3)\times SU(2)\times U(1)_S$, where $c$ is the charge of
$U(1)_S$. Note that ${\bf(3,2)}_{-5}$ and ${\bf(\bar 3,2)}_{5}$
are exotic fields in the MSSM. In order to eliminate the exotic
fields ${\bf(3,2)}_{-5}$ and ${\bf(\bar 3,2)}_{5}$, it is required
that $\chi(S,L^{\pm 5})=0$. It follows from the Riemann-Roch
theorem (\ref{R-R Surface}) that $c_1(L^{\pm 5})^2=-2$ and
$c_1(L^{\pm 5})$ correspond to a root of $E_8$, $E_i-E_j,\;i\neq
j$, which leads to a fractional line bundle\footnote{From now on,
all indices appearing in the divisors will be assumed to be
distinct unless otherwise stated.}
$L=\mathcal{O}_{S}(E_i-E_j)^{\pm {1/5}}$
\cite{BHV:2008I,BHV:2008II,Donagi:local}. In this case, all matter
fields must come from the curves. Now we turn to the spectrum from
the curves. In general, the gauge groups on the curves will be
enhanced at least by one rank. With $G_S=SU(5)$, the gauge groups
on the curves $G_{\Sigma}$ can be enhanced to $SU(6)$ or $SO(10)$
\cite{Katz:1996xe}. We first focus on the curves supporting the
matter fields in an $SU(5)$ GUT. To obtain complete matter
multiples of $SU(5)$ GUT, it is required that
$L_{\Sigma}=\mathcal{O}_{\Sigma}$ and
$L'_{\Sigma}\neq\mathcal{O}_{\Sigma}$, where $L'$ is a line bundle
associated with $U(1)'$. Consider the following breaking patterns:
\begin{equation}
\begin{array}{c@{}c@{}l@{}c@{}l}
SU(6) &~\rightarrow~& SU(5)\times U(1)' \\
{\bf 35} &~\rightarrow~& {\bf 24}_0+{\bf 1}_0+{\bf 5}_6+{\bf {\bar
5}}_{-6}\label{GUT SU(5)01}
\end{array}
\end{equation}
\begin{equation}
\begin{array}{c@{}c@{}l@{}c@{}l}
SO(10) &~\rightarrow~& SU(5)\times U(1)' \\
{\bf 45} &~\rightarrow~& {\bf 24}_0+{\bf 1}_0+{\bf
10}_4+{\bf{\overline{10}}}_{-4}.\label{GUT SU(5)02}
\end{array}
\end{equation}
From the patterns (\ref{GUT SU(5)01}) and (\ref{GUT SU(5)02}), it
can be seen by counting the dimension of the adjoint
representations that matter fields ${\bf 5}_6$ and ${\bf\bar
5}_{-6}$ are localized on the curves with $G_{\Sigma}={SU(6)}$
while ${\bf 10}_4$ and ${\bf \overline{10}}_{-4}$ are localized on
the curve with $G_{\Sigma}={SO(10)}$. The Higgs fields localize on
the curves with $G_{\Sigma}={SU(6)}$ as well. Since on the matter
curves $L_{\Sigma}$ is required to be trivial, the only line
bundle used to determine the spectrum is $L'_{\Sigma}$. With
non-trivial $L'_{\Sigma}$, it is not difficult to engineer three
copies of the matter fields, $3\times {\bf 5}_{6}$, $3\times
{\bf\bar 5}_{-6}$, and $3\times {\bf 10}_4$. In order to get
doublet-triplet splitting, it is required that
$L_{\Sigma}\neq\mathcal{O}_{\Sigma}$ and
$L'_{\Sigma}\neq\mathcal{O}_{\Sigma}$. With non-trivial
$L_{\Sigma}$ and $L'_{\Sigma}$, $G_{\Sigma}$ will be broken into
$G_{\rm std}\times U(1)'$. Consider the following breaking
patterns,
\begin{equation}
\begin{array}{c@{}c@{}l@{}c@{}l}
SU(6) &~\rightarrow~& SU(3)\times SU(2)\times U(1)_S\times U(1)' \\
{\bf 35} &~\rightarrow~& {\bf (8,1)}_{0,0}+{\bf (1,3)}_{0,0}+{\bf
(3,2)}_{-5,0}+{\bf (\bar 3,2)}_{5,0}+{\bf (1,1)}_{0,0}\\& &+{\bf
(1,1)}_{0,0}+{\bf (1,2)}_{3,6}+{\bf (3,1)}_{-2,6}+{\bf (1,\bar
2)}_{-3,-6}+{\bf(\bar 3,1)}_{2,-6}\label{MSSM SU(5)01}
\end{array}
\end{equation}
\begin{equation}
\begin{array}{c@{}c@{}l@{}c@{}l}
SO(10) &~\rightarrow~& SU(3)\times SU(2)\times U(1)_S\times U(1)' \\
{\bf 45} &~\rightarrow~& {\bf (8,1)}_{0,0}+{\bf (1,3)}_{0,0}+{\bf
(3,2)}_{-5,0}+{\bf (\bar 3,2)}_{5,0}+{\bf (1,1)}_{0,0}\\& &+{\bf
(1,1)}_{0,0}+[{\bf(3,2)}_{1,4}+{\bf(\bar
3,1)}_{-4,4}+{\bf(1,1)}_{6,4}+c.c].\label{MSSM SU(5)02}
\end{array}
\end{equation}
From the patterns (\ref{MSSM SU(5)01}) and (\ref{MSSM SU(5)02}),
the field content of the MSSM is identified as shown in Table
\ref{MSSM content SU(5)}.

\begin{table}[h]
\begin{center}
\renewcommand{\arraystretch}{1.25}
\begin{tabular}{|c|c|c|c|c|c|c|c|} \hline

$Q_L$ & $u_R$ & $d_R$ & $e_R$ & $L_L$ & $H_u$ & $H_d$ \\
\hline\hline

${\bf (3,2)}_{1,4}$ & ${\bf(\bar 3,1)}_{-4,4}$ & ${\bf(\bar
3,1)}_{2,-6}$ & ${\bf(1,1)}_{6,4}$ & ${\bf(1,\bar 2)}_{-3,-6}$ &
${\bf(1,2)}_{3,6}$ & ${\bf(1,\bar
2)}_{-3,-6}$ \\
\hline
\end{tabular}
\caption{Field content of the MSSM from $G_S=SU(5)$.} \label{MSSM
content SU(5)}
\end{center}
\end{table}
The superpotential is as follows:
\begin{eqnarray}
\mathcal{W}_{{\rm MSSM}}&\supset &
Q_Lu_RH_u+Q_Ld_RH_d+L_Le_RH_d+\cdots.
\end{eqnarray}
Note that the $U(1)_S$ in the patterns is consistent with $U(1)_Y$
in the MSSM and that this is the only way to consistently identify
the fields in the patterns (\ref{MSSM SU(5)01}) and (\ref{MSSM
 SU(5)02}) with
the MSSM. Now we are going to analyze the conditions for the
curves to support the field content in Table \ref{MSSM content
SU(5)}. We choose the curve $\Sigma_{SU(6)}$ to be a genus zero
curve and let $(m_1,m_2)=(N_{{\bf(\bar 3,1)}_{2,-6}},N_{{\bf
(1,\bar 2)}_{-3,-6}})$, where $N_{{\bf (A,B)}_{a,b}}$ is the
number of the fields in the representation ${\bf (A,B)}_{a,b}$
under $SU(3)\times SU(2)\times U(1)_S\times U(1)'$, and $a$, $b$
are the charges of $U(1)_S$ and $U(1)'$, respectively. Note that
${\bf(3,1)}_{-2,6}$ is exotic in the MSSM. To avoid the exotic, we
require that $m_1\in \mathbb{Z}_{\geqslant 0}$. Given $(m_1,m_2)$,
the homological class of the curve $\Sigma_{SU(6)}$ has to satisfy
the following
equation:\footnote{$L_{\Sigma_{SU(6)}}=\mathcal{O}_{\Sigma_{SU(6)}}(\frac{(m_1-m_2)}{5})$
and
$L'_{\Sigma_{SU(6)}}=\mathcal{O}_{\Sigma_{SU(6)}}(-\frac{(3m_1+2m_2)}{30})$}
\begin{equation}
(E_i-E_j)\cdot\Sigma_{SU(6)}=m_2-m_1,\label{SU(5)-MSSN SU(6) curve
cond}
\end{equation}
where $L=\mathcal{O}_{S}(E_j-E_i)^{1/5}$ has been used. By Eq.
(\ref{SU(5)-MSSN SU(6) curve cond}), we can engineer three copies
of $d_R$, three copies of $L_L$, one copy of $H_d$, and one copy
of $H_u$ on the individual curves as shown in Table \ref{MSSM
cond. from SU(5)01}.

\begin{table}[h]
\begin{center}
\renewcommand{\arraystretch}{1.10}
\begin{tabular}{|c|c|l|l|} \hline

Multiplet & $(m_1,m_2)$& ~~~~~\,Conditions & ~~~~~~~~~$\Sigma$ \\
\hline\hline

$3\times d_R$ & (3, 0) & $(E_i-E_j)\cdot\Sigma=-3$ & $5H-4E_j-E_i$\\
\hline
$3\times L_L$ & (0, 3) & $(E_i-E_j)\cdot\Sigma=$~3 & $4H+2E_j-E_i$\\
\hline
$1\times H_d$ & (0, 1) & $(E_i-E_j)\cdot\Sigma=$~1 & $H-E_i-E_l$\\
\hline
$1\times H_u$ & (0,-1) & $(E_i-E_j)\cdot\Sigma=-1$& $H-E_j-E_s$\\
\hline
\end{tabular}
\caption{Field content of the $SU(6)$ Curve from $G_S=SU(5)$.}
\label{MSSM cond. from SU(5)01}
\end{center}
\end{table}
Note that all field configurations in Table \ref{MSSM cond. from
SU(5)01} obey the conditions, $L_{\Sigma}\neq
\mathcal{O}_{\Sigma}$ and $L'_{\Sigma}\neq \mathcal{O}_{\Sigma}$.
In local models, the curves are required to be effective. With
Table \ref{MoriCone}, it is not difficult to check that all curves
in Table \ref{MSSM cond. from SU(5)01} are effective. The results
in Table \ref{MSSM cond. from SU(5)01} show that the triplet and
double states in ${\bf 5}_6$ or ${\bf\bar 5}_{-6}$ of $SU(5)$ can
be separated by the restrictions of the supersymmetric line
bundles to the curves. Next let us turn to the curve with
$G_\Sigma=SO(10)$. Set
$(l_1,l_2,l_3)=(N_{{\bf(3,2)}_{1,4}},N_{{\bf(\bar
3,1)}_{-4,4}},N_{{\bf(1,1)}_{6,4}})$. To avoid exotics in the
MSSM, it is required that $l_k\in \mathbb{Z}_{\geqslant
0},\;k=1,2,3$. Given $(l_1,l_2,l_3)$, the curve $\Sigma_{SO(10)}$
has to satisfy the following
equations:\footnote{$L_{\Sigma_{SO(10)}}=\mathcal{O}_{\Sigma_{SO(10)}}(\frac{(l_1-l_2)}{5})$
and
$L'_{\Sigma_{SO(10)}}=\mathcal{O}_{\Sigma_{SO(10)}}(\frac{(4l_1+l_2)}{20})$}
\begin{equation}
\left\{\begin{array}{l} (E_i-E_j)\cdot\Sigma_{SO(10)}=l_2-l_1\\
l_3=2l_1-l_2.\label {SU(5)-MSSN SO(10) curve cond}
\end{array}   \right.
\end{equation}
To obtain the minimal spectrum of the MSSM, we require that
$l_1,l_2\leqslant 3$. Taking the conditions, $L_{\Sigma}\neq
\mathcal{O}_{\Sigma}$ and $L'_{\Sigma}\neq \mathcal{O}_{\Sigma}$
into account, we have the following configurations:
\begin{equation}
(l_1,l_2,l_3)=\left\{\begin{array}{l}
(1,2,0),(1,0,2),(2,1,3),(2,3,1) \label {Field content SO(10)}
\end{array}   \right\}.
\end{equation}
From the configurations in (\ref{Field content SO(10)}), it is
clear that unlike with $G_{\Sigma}=SU(6)$, it is impossible to
engineer the matter fields $3\times Q_L$, $3\times u_R$, and
$3\times e_R$ on the individual curves with $G_{\Sigma}=SO(10)$,
which correspond to $(l_1,l_2,l_3)=(3,0,0)$, $(0,3,0)$, and
$(0,0,3)$, respectively, without extra matter fields. Fortunately,
in this case all Higgs fields come from $\Sigma_{SU(6)}$ instead
of $\Sigma_{SO(10)}$. Although the field content on
$\Sigma_{SO(10)}$ is more complicated than that on
$\Sigma_{SU(6)}$, we can engineer the spectrum of the MSSM as
shown in Table \ref{MSSM cond. from SU(5)02}.

\begin{table}[h]
\begin{center}
\renewcommand{\arraystretch}{1.25}
\begin{tabular}{|c|c|l|c|l|l|} \hline

Multiplet & Curve & ~~~~~~~~~$\Sigma$ & $g_{\Sigma}$ & ~~~~~~~~$L_{\Sigma}$ & ~~~~~~~~$L'_{\Sigma}$\\
\hline \hline

$1\times Q_L+2\times u_R$ & $\Sigma_{SO(10)}^{1}$ & $2H-E_2-E_3$ &
0 & $\mathcal{O}_{\Sigma_{SO(10)}^{1}}(-1)^{1/5}$ &
$\mathcal{O}_{\Sigma_{SO(10)}^{1}}(1)^{3/10}$\\
\hline

$2\times Q_L+1\times u_R$ & \multirow{2}{*}{$\Sigma_{SO(10)}^{2}$}
& \multirow{2}{*}{$2H-E_1-E_4$} & \multirow{2}{*}{0} &
\multirow{2}{*}{$\mathcal{O}_{\Sigma_{SO(10)}^{2}}(1)^{1/5}$} &
\multirow{2}{*}{$\mathcal{O}_{\Sigma_{SO(10)}^{2}}(1)^{9/20}$} \\

$+3\times e_R$ &&&&& \\
\hline

$3\times d_R$ & $\Sigma_{SU(6)}^{1}$ & $5H-4E_1-E_2$ & 0 &
$\mathcal{O}_{\Sigma_{SU(6)}^{1}}(1)^{3/5}$ &
$\mathcal{O}_{\Sigma_{SU(6)}^{1}}(-1)^{3/10}$\\
\hline

$3\times L_L$ & $\Sigma_{SU(6)}^{2}$ & $4H+2E_1-E_2$ & 0 &
$\mathcal{O}_{\Sigma_{SU(6)}^{2}}(-1)^{3/5}$ &
$\mathcal{O}_{\Sigma_{SU(6)}^{2}}(-1)^{1/5}$\\

\hline $1\times H_d$ & $\Sigma_{SU(6)}^{d}$ & $2H-E_2-E_4$ & 0 &
$\mathcal{O}_{\Sigma_{SU(6)}^{d}}(-1)^{1/5}$ &
$\mathcal{O}_{\Sigma_{SU(6)}^{d}}(-1)^{1/15}$\\
\hline

$1\times H_u$ & $\Sigma_{SU(6)}^{u}$ & $~\,H-E_1-E_3$ & 0 &
$\mathcal{O}_{\Sigma_{SU(6)}^{u}}(1)^{1/5}$ &
$\mathcal{O}_{\Sigma_{SU(6)}^{u}}(1)^{1/15}$\\
\hline

\end{tabular}
\caption{A minimal spectrum of the MSSM from $G_S=SU(5)$, where
$L=\mathcal{O}_{S}(E_{1}-E_{2})^{1/5}$.} \label{MSSM cond. from
SU(5)02}
\end{center}
\end{table}

From Table \ref{MSSM cond. from SU(5)02}, we find that for the
case of $G_S=SU(5)$, we can get an exotic-free, minimal spectrum
of the MSSM with doublet-triplet splitting. In addition, by
arranging $H_u$ and $H_d$ on different curves, rapid proton decay
can be avoided \cite{BHV:2008I,BHV:2008II,Donagi:local}.

\subsection{$G_S=SO(10)$}
For the case of $G_S=SO(10)$ \cite{Chen:2009me}, we first look at
the spectrum from the bulk. Consider the following breaking
pattern,
\begin{equation}
\begin{array}{c@{}c@{}l@{}c@{}l}
SO(10) &~\rightarrow~& SU(5)\times U(1)_S\\
{\bf 45}~&\rightarrow~& {\bf 24}_{0}+{\bf 1}_{0}+{\bf 10}_{4}+
{\bf\overline{10}}_{-4}.
\end{array}
\end{equation}
The bulk zero modes are determined by
\begin{equation}
{\bf 10}_{4}\in H_{\bar\p}^{0}(S,L^{-4})^{\vee}\oplus
H_{\bar\p}^{1}(S,L^{4})\oplus H_{\bar\p}^{2}(S,L^{-4})^{\vee}
\end{equation}
\begin{equation}
{\bf{\overline{10}}}_{-4}\in H_{\bar\p}^{0}(S,L^{4})^{\vee}\oplus
H_{\bar\p}^{1}(S,L^{-4})\oplus H_{\bar\p}^{2}(S,L^{4})^{\vee}.
\end{equation}
To eliminate ${\bf 10}_4$ and ${\bf\overline{10}}_{-4}$, it is
required that $\chi(S,L^{\pm 4})=0$, which give rise to the
fractional line bundles $L=\mathcal{O}_{S}(E_i-E_j)^{\pm {1/4}}$.
In this case, all chiral fields must come from the curves. Let us
turn to the spectrum from the curves. With $G_S=SO(10)$, the gauge
groups on the curve can be enhanced to $G_{\Sigma}=SO(12)$ or
$G_{\Sigma}=E_6$. The breaking chains and matter content from the
enhanced adjoints of the curves are
\begin{equation}
\begin{array}{c@{}c@{}l@{}c@{}l}
SO(12) &~\rightarrow~& SO(10)\times U(1)' &~\rightarrow~& SU(5)\times U(1)'\times U(1)_S\\
{\bf 66} &~\rightarrow~& {\bf 45}_0+{\bf 1}_0&~\rightarrow~&
{\bf 24}_{0,0}+{\bf 1}_{0,0}+{\bf 10}_{0,4}+{\bf\overline{10}}_{0,-4}+{\bf 1}_{0,0}\\
& &+{\bf 10}_2+{\bf\overline{10}}_{-2} & &+{\bf 5}_{2,2}+{\bf{\bar
5}}_{2,-2}+{\bf{\bar 5}}_{-2,-2}+{\bf 5}_{-2,2}
\end{array}
\end{equation}
\begin{equation}
\begin{array}{c@{}c@{}l@{}c@{}l}
E_6 &~\rightarrow~ & SO(10) \times U(1)' &~\rightarrow~& SU(5)\times U(1)'\times U(1)_S\\
{\bf 78} &~\rightarrow~ & {\bf 45}_0+{\bf 1}_0&~\rightarrow~& {\bf
24}_{0,0}+{\bf 1}_{0,0}+{\bf 10}_{0,4}+
{\bf\overline{10}}_{0,-4}+{\bf 1}_{0,0}\\
& & +{\bf 16}_{-3}+{\bf\overline{16}}_{3} & &+({\bf 10}_{-3,-1}+{\bf{\bar 5}}_{-3,3}+{\bf 1}_{-3,-5}+c.c.).\\
\end{array}
\end{equation}
Note that the $U(1)_S$ charges of the fields localized on the
curves should be conserved in each Yukawa coupling. The
superpotential is as follows:
\begin{eqnarray}
\mathcal{W}&\supset & {\bf 10}_{-3,-1} {\bf 10}_{-3,-1} {\bf
5}_{-2,2}+{\bf 10}_{-3,-1} \bar{\bf 5}_{-3,3} \bar{\bf
5}_{2,-2}+\cdots.
\end{eqnarray}
In order to get complete matter multiplets in $SU(5)$ GUT, we
require that $L_{\Sigma}$ and $L'_{\Sigma}$ are both non-trivial.
With non-trivial $L_{\Sigma}$ and $L'_{\Sigma}$, we can engineer
field content with minimal singlets as shown in Table
{\ref{SU5SO10}} \cite{Chen:2009me}.

\renewcommand{\thefootnote}{\fnsymbol{footnote}}

\begin{table}[h]
\begin{center}
\renewcommand{\arraystretch}{1.25}
\begin{tabular}{|@{}c@{}|c|l|c|l@{}|l@{}|} \hline

Multiplet & Curve & $~~~~~~~~~\Sigma$ & $g_{\Sigma}$ & ~~~~~~~$L_{\Sigma}$ & ~~~~~~~$L'_{\Sigma}$ \\
\hline \hline

$\,3\times {\bf 10}_{-3,-1}$\footnotemark[1] & $\Sigma_{E_6}^{1}$
& $4H+2E_1-E_2$ & 0 &
$\mathcal{O}_{\Sigma_{E_6}^{1}}(-1)^{3/4}$ & $\mathcal{O}_{\Sigma_{E_6}^{1}}(-1)^{3/4}$ \\
\hline

$3\times\bar {\bf 5}_{-3,3}$\footnotemark[2] & $\Sigma_{E_6}^{2}$
& $5H+3E_2-E_5$ & 0 &
$\mathcal{O}_{\Sigma_{E_6}^{2}}(1)^{3/4}$ & $\mathcal{O}_{\Sigma_{E_6}^{2}}(-1)^{1/4}$ \\
\hline

$1\times {\bf 5}_{-2,2}$ & $\Sigma_{SO(12)}^{1}$ & $3H+E_3-E_1$ &
0 &
$\mathcal{O}_{\Sigma_{SO(12)}^{1}}(1)^{1/4}$ & $\mathcal{O}_{\Sigma_{SO(12)}^{1}}(-1)^{1/4}$ \\
\hline

$1\times\bar{\bf 5}_{2,-2}$ & $\Sigma_{SO(12)}^{2}$ &
$~\,H-E_2-E_3$
& 0 & $\mathcal{O}_{\Sigma_{SO(12)}^{2}}(-1)^{1/4}$ & $\mathcal{O}_{\Sigma_{SO(12)}^{2}}(1)^{1/4}$ \\
\hline
\end{tabular}
\caption{An $SU(5)$ GUT model from $G_S=SO(10)$, where
$L=\mathcal{O}_{S}(E_{1}-E_{2})^{1/4}$.}\label{SU5SO10}
\end{center}
\end{table}
\footnotetext[1]{With six additional singlets}
\footnotetext[2]{With three additional singlets}

\renewcommand{\thefootnote}{\arabic{footnote}}

However, because of the lack of extra $U(1)$ gauge fluxes or
Wilson lines, the doublet-triplet splitting is not achievable in
the present case. This motivates us to consider supersymmetric
$U(1)^2$ fluxes.

\subsection{$G_S=SU(6)$}

To look at the spectrum from the bulk , we consider the following
breaking pattern,
\begin{equation}
\begin{array}{c@{}c@{}l@{}c@{}l}
SU(6) &~\rightarrow~& SU(5)\times U(1)_S\\
{\bf 45}~&\rightarrow~& {\bf 24}_{0}+{\bf 1}_{0}+{\bf 5}_{6}+
{\bf{\bar 5}}_{-6}.
\end{array}
\end{equation}
The bulk zero modes are given by
\begin{equation}
{\bf 5}_{6}\in H_{\bar\p}^{0}(S,L^{-6})^{\vee}\oplus
H_{\bar\p}^{1}(S,L^{6})\oplus H_{\bar\p}^{2}(S,L^{-6})^{\vee}
\end{equation}
\begin{equation}
{\bf{\bar {5}}}_{-6}\in H_{\bar\p}^{0}(S,L^{6})^{\vee}\oplus
H_{\bar\p}^{1}(S,L^{-6})\oplus H_{\bar\p}^{2}(S,L^{6})^{\vee}.
\end{equation}
To eliminate ${\bf 5}_6$ and ${\bf\overline{{5}}}_{-6}$, it is
required that $\chi(S,L^{\pm 6})=0$, which gives rise to the
fractional line bundles $L=\mathcal{O}_{S}(E_i-E_j)^{\pm {1/6}}$
\cite{Chen:2009me}. In this case, all chiral fields must come from
the curves. Let us turn to the spectrum from the curves. With
$G_S=SU(6)$, the gauge groups on the curve can be enhanced to
$G_{\Sigma}=SU(7)$, $G_{\Sigma}=SO(12)$ or $G_{\Sigma}=E_6$. The
breaking chains and matter content from the enhanced adjoints of
the curves are
\begin{equation}
\begin{array}{c@{}c@{}l@{}c@{}l}
SU(7) &~\rightarrow~& SU(6)\times U(1)' &~\rightarrow~& SU(5)\times U(1)'\times U(1)_S\\
{\bf 48} &~\rightarrow~& {\bf 35}_0+{\bf 1}_0+{\bf
6}_{-7}+{\bf\bar 6}_{7}&~\rightarrow~& {\bf 24}_{0,0}+{\bf
1}_{0,0}+{\bf 5}_{0,6}+{\bf\bar 5}_{0,-6}+{\bf 1}_{0,0} \\ & & &
&+{\bf 5}_{-7,1}+{\bf 1}_{-7,-5}+{\bf\bar 5}_{7,-1}+{\bf
1}_{7,5}\label{SU(6)sigmaSU(7)}
\end{array}
\end{equation}
\begin{equation}
\begin{array}{c@{}c@{}l@{}c@{}l}
SO(12) &~\rightarrow~& SU(6)\times U(1)' &~\rightarrow~&
SU(5)\times U(1)'\times U(1)_S\\ {\bf 66} &~\rightarrow~& {\bf
35}_0+{\bf 1}_0+{\bf 15}_2+{\bf\overline{15}}_{-2}&~\rightarrow~&
{\bf 24}_{0,0}+{\bf 1}_{0,0}+{\bf 5}_{0,6}+{\bf\bar 5}_{0,-6}+{\bf 1}_{0,0}\\
& & & &+{\bf 10}_{2,2}+{\bf
5}_{2,-4}+{\bf\overline{10}}_{-2,-2}+{\bf\bar
5}_{-2,4}\label{SU(6)sigmaSO(12)}
\end{array}
\end{equation}
\begin{equation}
\begin{array}{c@{}c@{}l@{}c@{}l}
E_{6} &~\rightarrow~& SU(6)\times U(1)' &~\rightarrow~& SU(5)\times U(1)'\times U(1)_S\\
{\bf 78} &~\rightarrow~& {\bf 35}_0+{\bf 1}_0+{\bf
1}_{\pm2}&~\rightarrow~&
{\bf 24}_{0,0}+2\times {\bf 1}_{0,0}+{\bf 5}_{0,6}+{\bf \bar{5}}_{0,-6} +{\bf 1}_{\pm2,0}\\
& & +{\bf 20}_1+{\bf {20}}_{-1} & &+{\bf
10}_{1,-3}+{\bf\overline{10}}_{1,3}+{\bf 10}_{-1,-3}+{\bf
\overline{10}}_{-1,3}\label{SU(6)sigmaE_6}.
\end{array}
\end{equation}
In this case, the $U(1)_S$ charges of the fields localized on the
curves should be conserved in each Yukawa coupling. The
superpotential is:
\begin{eqnarray}
\mathcal{W}&\supset & {\bf 10}_{2,2} {\bf 10}_{2,2} {\bf 5}_{2,-4}
+{\bf 10}_{2,2} \bar{\bf 5}_{7,-1} \bar{\bf 5}_{7,-1}+\cdots.
\end{eqnarray}
With non-trivial $L_{\Sigma}$ and $L'_{\Sigma}$, we can engineer
configurations of the curves with desired field content but
without any exotic fields as shown in Table {\ref{SU(5)SU(6)S}}
\cite{Chen:2009 me}.
\begin{table}[h]
\begin{center}
\renewcommand{\arraystretch}{1.15}
\begin{tabular}{|c|c|c|c|c|c|c|c|} \hline

Multiplet & Curve & $\Sigma$ & $g_{\Sigma}$ & $L_{\Sigma}$ & $L'_{\Sigma}$ \\
\hline \hline

$3\times {\bf 10}_{2,2}$ & $\Sigma_{SO(12)}^{1}$ & $4H+2E_2-E_1$ &
0 &
$\mathcal{O}_{\Sigma_{SO(12)}^{1}}(1)^{1/2}~\;\,$ & $\mathcal{O}_{\Sigma_{SO(12)}^{1}}(1)~~~$ \\
\hline

$3\times\bar{\bf 5}_{7,-1}$ & $\Sigma_{SU(7)}^{1}$ & $5H+3E_1-E_6$
& 0 &
$\mathcal{O}_{\Sigma_{SU(7)}^{1}}(-1)^{1/2}\;$ & $\mathcal{O}_{\Sigma_{SU(7)}^{1}}(1)^{5/14}$ \\
\hline

$1\times {\bf 5}_{2,-4}$ & $\Sigma_{SO(12)}^{2}$ & $3H+E_1-E_3~$ &
0 &
$\mathcal{O}_{\Sigma_{SO(12)}^{2}}(-1)^{1/6}$ & $\mathcal{O}_{\Sigma_{SO(12)}^{2}}(1)^{1/6}$ \\
\hline

$1\times\bar{\bf 5}_{7,-1}$ & $\Sigma_{SU(7)}^{2}$ & $H-E_2-E_3$ &
0 &
$\mathcal{O}_{\Sigma_{SU(7)}^{2}}(-1)^{1/6}\;$ & $\mathcal{O}_{\Sigma_{SU(7)}^{2}}(1)^{5/42}$ \\
\hline
\end{tabular}
\caption{An $SU(5)$ GUT model from $G_S=SU(6)$, where
$L=\mathcal{O}_{S}(E_{1}-E_{2})^{1/6}$.} \label{SU(5)SU(6)S}
\end{center}
\end{table}

Although in this case one can obtain an exotic-free spectrum in an
$SU(5)$ GUT, the doublet-triplet splitting can not be achieved,
similar to the case of $G_S=SO(10)$. Again this motivates us to
consider supersymmetric $U(1)^2$ gauge fluxes. On the other hand,
to get the spectrum of the MSSM, we also need some mechanisms to
break $SU(5)\subset G_{\Sigma}$ into $SU(3)\times SU(2)\times
U(1)_Y$. One possible way is to consider supersymmetric $U(1)^2$
gauge fluxes instead of U(1) fluxes. These supersymmetric $U(1)^2$
gauge fluxes correspond to polystable bundles of rank two with
structure group $U(1)^2$. In the next section we shall discuss
polystable bundles of rank two.

\section{Gauge Bundles}
In this section we shall briefly review the notion of stability of
the vector bundle and the relation between (semi) stable bundles
and the DUY equation. In addition, we also discuss the semi-stable
bundles of rank two, in particular, polystable bundles over $S$.

\subsection{Stability}
Let $E$ be a holomorphic vector bundle over a projective surface
$S$ and $J_S$ be a K\"ahler form on $S$. The slope $\mu (E)$ is
defined by
\begin{eqnarray}
\mu (E)=\frac{\int_{S}c_{1}(E)\wedge J_S}{{\rm
rk}(E)}{\label{Slope}}.
\end{eqnarray}
The vector bundle $E$ is (semi)stable if for every subbundle or
subsheaf $\mathcal{E}$ with ${\rm rk}(\mathcal{E})<{\rm rk}(E)$,
the following inequality holds
\begin{eqnarray}
\mu (\mathcal{E})<(\leqslant)\mu (E){\label{Stability}}.
\end{eqnarray}
Assume that $E=\oplus_i^{k} \mathcal{E}_i$, then $E$ is polystable
if each $\mathcal{E}_i$ is a stable bundle with
$\mu(\mathcal{E}_1)=\mu(\mathcal{E}_2)=...=\mu(\mathcal{E}_k)$
\cite{Donalson,UY}. It is clear that every line bundle is stable
and polystable bundle is a type of semistable bundle. The
Donaldson-Uhlenbeck-Yau theorem \cite{Donalson,UY} states that a
(split) irreducible holomorphic bundle $E$ admits a hermitian
connection satisfying Eq. (\ref{DUY}) if and only if $E$ is
(poly)stable. As mentioned in section $2.1$, to preserve
supersymmetry, the connection of the bundle has to obey the $\rm
DUY$ equation (\ref{DUY}), which is equivalent to the (poly)
stable bundle. In particular, when the bundle is split,
supersymmetry requires that the bundle is polystable. In the next
section we primarily focus on polystable bundles of rank two over
$S$.

\subsection{Rank Two Polystable Bundle}
Here we are interested in the case $S=dP_{k}$. Consider the case
of $V=L_1\oplus L_2$, where $L_1$ and $L_2$ are line bundles over
$S$ and set $L_{i}=\mathcal{O}_{S}(D_{i}),\;i=1,2$, where $D_{i}$
are divisors in $S$. Before writing down a more explicit
expression for the bundle $V$, we first consider the stability
condition of the polystable bundle. Recall that the bundle $V$ is
polystable if $\mu (L_1)=\mu (L_2)$ where $\mu $ is slope defined
by Eq. (\ref{Slope}). To solve the DUY equation Eq. (\ref{DUY}),
it is required that $\mu (L_1)=\mu (L_2)=0$. It follows that
$c_1({L_1})\wedge J_S=c_1(L_2)\wedge J_S=0$ or equivalently,
\begin{eqnarray}
D_1\cdot \omega=D_2\cdot\omega=0 \label{Polarization},
\end{eqnarray}
where $\omega$ is the dual ample divisor of K\"ahler form $J_S$ in
the K\"ahler cone. In particular, in this case we choose ``large
volume polarization'', namely
$\omega=AH-\sum_{i=1}^{k}a_{i}E_{i},\;A\gg a_i>0$
\cite{BHV:2008I,BHV:2008II}. Note that Eq. (\ref{Polarization}) is
exactly the BPS equations, $c_{1}(L_{i})\wedge J_S=0,\;i=1,2$ for
supersymmetric line bundles. So the polystable bundle $V$ is a
direct sum of the supersymmetric line bundles $L_1$ and $L_2$. In
section $5.2$ we shall apply physical constraints to the
polystable bundle that satisfies the Eq. (\ref{Polarization}) and
derive the explicit expression of the $U(1)^2$ gauge fluxes $L_1$
and $L_2$.

\subsection{Supersymmetric $U(1)^2$ Gauge Fluxes}

Each supersymmetric $U(1)^2$ gauge flux configuration contains two
fractional line bundles, which may not be well-defined themselves.
It is natural to ask whether it makes sense for these
configurations to be polystable vector bundles of rank two. In
what follows, we shall show that supersymmetric $U(1)^2$ gauge
fluxes can be associated with polystable vector bundles of rank
two. Let us consider the case of $G_S=SU(6)$ and the breaking
pattern through $SU(6)\ra SU(5)\times U(1)\ra SU(3)\times
SU(2)\times U(1)_1\times U(1)_2$. Let $L_1$ and $L_2$ be two
supersymmetric line bundles, which associate to $U(1)_1$ and
$U(1)_2$, respectively. Write $L_i=\mathcal{O}_{S}(D_i),\;i=1,2$,
where $D_i$ are in general ``$\mathbb{Q}$-divisors'' which means
that $D_i$ are the linear combinations of the divisors in $S$ with
rational coefficients. Now we consider the rotation of the $U(1)$
charges, $U(1)_1$ and $U(1)_2$, given by
\begin{equation}
\widetilde{\mathbb{U}}={\rm M}\mathbb{U}\label{U(1)rotation}
\end{equation}
with $\mathbb{U}=(U(1)_1,U(1)_2)^{t}$,
$\widetilde{\mathbb{U}}=(\widetilde{U(1)}_1,\widetilde{U(1)}_2)^{t}$,
and ${\rm M}\in GL(2,\mathbb{Q})$, where $t$ represents the
transpose. We define $\widetilde{L}_1$ and $\widetilde{L}_2$ to be
two line bundles which associate to $\widetilde{U(1)}_1$ and
$\widetilde{U(1)}_2$, respectively and write
$\widetilde{L}_i=\mathcal{O}_{S}(\widetilde{D}_i),\;i=1,2$. Let
${\bf(A,B)}_{c,d}$ and ${\bf(A,B)}_{\widetilde{c},\widetilde{d}}$
be representations in the breaking patten $SU(6)\ra SU(3)\times
SU(2)\times U(1)_1\times U(1)_2$ and $SU(6)\ra SU(3)\times
SU(2)\times \widetilde{U(1)}_1\times \widetilde{U(1)}_2$,
respectively. Up to a linear combination of $U(1)$ charges, we
have
$N_{{\bf(A,B)}_{c,d}}=N_{{\bf(A,B)}_{\widetilde{c},\widetilde{d}}}$,
which requires that the corresponding divisors be transferred as
follows:
\begin{equation}
\widetilde{\mathbb{D}}=({\rm M}^{-1})^{t}\mathbb{D},\label{Divisor
Tansformation}
\end{equation}
where $\mathbb{D}=(D_1,D_2)^{t}$,
$\widetilde{\mathbb{D}}=(\widetilde{D}_1,\widetilde{D}_2)^{t}$. In
general, $\widetilde{D}_i$ are $\mathbb{Q}$-divisors via the
rotation (\ref{Divisor Tansformation}). However, it is possible to
get integral divisors $\widetilde{D}_i$ by a suitable choice of
the matrix ${\rm M}={\rm M}_{\ast}$. Once this is done, we obtain
two corresponding line bundles, $\widetilde{L}_1$ and
$\widetilde{L}_2$ since $\widetilde{D}_i\in
H_{2}(S,\mathbb{Z}),\;i=1,2$. Moreover, if $\mu
(\widetilde{L}_1)=\mu (\widetilde{L}_2)=0$, we can construct the
polystable bundle $V=\widetilde{L}_1\oplus\widetilde{L}_2$. Note
that when $L_i$ are supersymmetric, which means that they satisfy
the BPS condition (\ref{Polarization}), by the transformation
(\ref{Divisor Tansformation}) we have $\mu (\widetilde{L}_1)=\mu
(\widetilde{L}_2)=0$. As a result, each supersymmetric $U(1)^2$
gauge fluxes is associated with a polystable vector bundle of rank
two if the suitable matrix ${\rm M}_{\ast}$ exists. To be
concrete, let us consider the case of $G_S=SU(6)$. The breaking
pattern via $G_{\rm std}\times U(1)$ is as follows:
\begin{equation}
\begin{array}{c@{}c@{}l@{}c@{}l}
SU(6) &~\rightarrow~& SU(3)\times SU(2)\times U(1)_1\times U(1)_2 \\
{\bf 35} &~\rightarrow~& {\bf (8,1)}_{0,0}+{\bf (1,3)}_{0,0}+{\bf
(3,2)}_{-5,0}+{\bf (\bar 3,2)}_{5,0}+{\bf (1,1)}_{0,0}\\& &+{\bf
(1,1)}_{0,0}+{\bf (1,2)}_{3,6}+{\bf (3,1)}_{-2,6}+{\bf (1,\bar
2)}_{-3,-6}+{\bf (\bar 3,1)}_{2,-6}.\label{breaking SU(6)01}
\end{array}
\end{equation}
Let $L_1$ and $L_2$ be the supersymmetric line bundles associated
to $U(1)_1$ and $U(1)_2$, respectively. Note that $U(1)_1$ can be
identified as $U(1)_Y$ in the MSSM. The exotic-free spectrum from
the bulk requires that $L_1$ and $L_2$ are fractional line
bundles. The details could be found in section $5.2$. Now consider
the rotation

\begin{eqnarray}
{\rm M}=\left(\begin{array}{cc} -\frac{1}{5} & \frac{1}{10}
\\0 &
\frac{1}{6}
\end{array}\right).\label{rotation01}
\end{eqnarray}
Then we obtain
\begin{equation}
\begin{array}{c@{}c@{}l@{}c@{}l}
SU(6) &~\rightarrow~& SU(3)\times SU(2)\times U(1)_1\times U(1)_2 \\
{\bf 35} &~\rightarrow~& {\bf (8,1)}_{0,0}+{\bf (1,3)}_{0,0}+{\bf
(3,2)}_{1,0}+{\bf (\bar 3,2)}_{-1,0}+{\bf (1,1)}_{0,0}\\& &+{\bf
(1,1)}_{0,0}+{\bf (1,2)}_{0,1}+{\bf (3,1)}_{1,1}+{\bf (1,\bar
2)}_{0,-1}+{\bf (\bar 3,1)}_{-1,-1}\label{breaking SU(6)02}
\end{array}
\end{equation}
with $\widetilde{L}_1=L_1^{-5}$ and $\widetilde{L}_2=L_1^3\otimes
L_2^6$. It is clear that
$N_{{\bf(A,B)}_{c,d}}=N_{{\bf(A,B)}_{\widetilde{c},\widetilde{d}}}$
with respect to (\ref{breaking SU(6)01}) and (\ref{breaking
SU(6)02}). It turns out that $\widetilde{L}_1$ and
$\widetilde{L}_2$ are truly line bundles. Furthermore, one can
show that BPS condition (\ref{Polarization}) for $(L_1,L_2)$ is
equivalent to the stability conditions of the polystable bundle
$V=\widetilde{L}_1\oplus\widetilde{L}_2$ by the transformation
(\ref{Divisor Tansformation}). In this case, we know that
supersymmetric $U(1)^2$ gauge fluxes are associated with
polystable bundles of rank two with the same number of zero modes
charged under $U(1)^2$. With this correspondence, we can avoid
talking about the gauge bundle defined by the direct sum of two
fractional line bundles. In other words, a supersymmetric $U(1)^2$
gauge flux $(L_1,L_2)$ is well-defined in the sense that it can be
associated with a well-defined polystable bundle of rank two. Form
now on, we shall simply use the phrase $U(1)^2$ gauge fluxes in
stead of polystable bundle in the following sections.

\section{$U(1)^{2}$ Gauge Fluxes}

In this section we consider $U(1)^2$ gauge fluxes in local
F-theory models, in particular we focus on the case of
$G_S=SO(10)$ and $SU(6)$. With the gauge fluxes, $G_S$ can be
broken into $G_{\rm std}\times U(1)$. For the case of
$G_S=SO(10)$, there is a no-go theorem which states that there do
not exist $U(1)^2$ gauge fluxes such that the spectrum is
exotic-free. This result was first shown in \cite{BHV:2008II}. We
review the case in section $5.1$ for completeness. For the case of
$G_S=SU(6)$, with appropriate physical conditions, we shall show
that there are finitely many supersymmetric $U(1)^2$ gauge fluxes
with an exotic-free bulk spectrum and we obtain the explicit
expression of these gauge fluxes as well. With these explicit flux
configurations, we study doublet-triplet splitting and the
spectrum of the MSSM. The details can be found in section $5.2$
and $5.3$.

\subsection{$G_S=SO(10)$}

\subsubsection{$U(1)^2$ Gauge Flux Configurations}

The maximal subgroups of $SO(10)$ which contain $G_{\rm std}$ and
the consistent MSSM spectrum are as follows \cite{BHV:2008II}:
\begin{equation}
SO(10)\supset SU(5)\times U(1)\supset G_{\rm std}\times
U(1)\label{maxgroupSO(10)01}
\end{equation}
\begin{equation}
SO(10)\supset SU(2)\times SU(2)\times SU(4)\supset G_{\rm
std}\times U(1)\label{maxgroupSO(10)02}
\end{equation}
For the latter, one of $SU(2)$ groups needs to be broken into
$U(1)\times U(1)$ to get the consistent $U(1)_Y$ charge in the
MSSM. It follows from the patterns (\ref{maxgroupSO(10)01}) and
(\ref{maxgroupSO(10)02}) that up to linear combinations of the
$U(1)$ charges in the breaking patterns, it is enough to analyze
the case of $U(1)^2$ gauge fluxes which breaks $SO(10)$ via the
sequence $SO(10)\ra SU(5)\times U(1)\ra G_{\rm std}\times U(1)$.
The breaking pattern is as follows:
\begin{equation}
\begin{array}{c@{}c@{}l@{}c@{}l}
SO(10) &~\rightarrow~& SU(3)\times SU(2)\times U(1)_1\times U(1)_{2} \\
{\bf 45} &~\rightarrow~& {\bf (8,1)}_{0,0}+{\bf (1,3)}_{0,0}+{\bf
(3,2)}_{-5,0}+{\bf (\bar 3,2)}_{5,0}+{\bf (1,1)}_{0,0}\\& &+{\bf
(1,1)}_{0,0}+{\bf (1,1)}_{6,4}+{\bf (\bar 3,1)}_{-4,4}+{\bf
(3,2)}_{1,4}+{\bf (1,1)}_{-6,-4}\\& &+{\bf (3,1)}_{4,-4}+{\bf
(\bar 3, 2)}_{-1,-4}.\label{breaking SO(10)}
\end{array}
\end{equation}
Note that $U(1)_1$ can be identified with $U(1)_Y$ in the MSSM.
Let ${\widetilde{ L}}_3$ and $\widetilde{L}_4$ be non-trivial
supersymmetric line bundles associated with $U(1)_1$ and $U(1)_2$,
respectively, in the breaking pattern (\ref{breaking SO(10)}). The
bulk zero modes are given by
\begin{equation}
{\bf (3,2)}_{-5,0}\in
H_{\bar\p}^{0}(S,\widetilde{L}_3^{5})^{\vee}\oplus
H_{\bar\p}^{1}(S,\widetilde{L}_3^{-5})\oplus
H_{\bar\p}^{2}(S,\widetilde{L}_3^{5})^{\vee}
\end{equation}
\begin{equation}
{\bf (\bar 3,2)}_{5,0}\in
H_{\bar\p}^{0}(S,\widetilde{L}_3^{-5})^{\vee}\oplus
H_{\bar\p}^{1}(S,\widetilde{L}_3^{5})\oplus
H_{\bar\p}^{2}(S,\widetilde{L}_3^{-5})^{\vee}
\end{equation}
\begin{equation}
{\bf (3,2)}_{1,4}\in H_{\bar\p}^{0}(S,\widetilde{L}_3^{-1}\otimes
\widetilde{L}_4^{-4} )^{\vee}\oplus
H_{\bar\p}^{1}(S,\widetilde{L}_3^{1}\otimes
\widetilde{L}_4^{4})\oplus
H_{\bar\p}^{2}(S,\widetilde{L}_3^{-1}\otimes
\widetilde{L}_4^{-4})^{\vee}
\end{equation}
\begin{equation}
{\bf (\bar 3,2)}_{-1,-4}\in
H_{\bar\p}^{0}(S,\widetilde{L}_3^{1}\otimes \widetilde{L}_4^{4}
)^{\vee}\oplus H_{\bar\p}^{1}(S,\widetilde{L}_3^{-1}\otimes
\widetilde{L}_4^{-4})\oplus
H_{\bar\p}^{2}(S,\widetilde{L}_3^{1}\otimes
\widetilde{L}_4^{4})^{\vee}
\end{equation}
\begin{equation}
{\bf (3,1)}_{4,-4}\in H_{\bar\p}^{0}(S,\widetilde{L}_3^{-4}\otimes
\widetilde{L}_4^{4} )^{\vee}\oplus
H_{\bar\p}^{1}(S,\widetilde{L}_3^{4}\otimes
\widetilde{L}_4^{-4})\oplus
H_{\bar\p}^{2}(S,\widetilde{L}_3^{-4}\otimes
\widetilde{L}_4^{4})^{\vee}
\end{equation}
\begin{equation}
{\bf (\bar 3,1)}_{-4,4}\in
H_{\bar\p}^{0}(S,\widetilde{L}_3^{4}\otimes \widetilde{L}_4^{-4}
)^{\vee}\oplus H_{\bar\p}^{1}(S,\widetilde{L}_3^{-4}\otimes
\widetilde{L}_4^{4})\oplus
H_{\bar\p}^{2}(S,\widetilde{L}_3^{4}\otimes
\widetilde{L}_4^{-4})^{\vee},
\end{equation}
\begin{equation}
{\bf (1,1)}_{6,4}\in H_{\bar\p}^{0}(S,\widetilde{L}_3^{-6}\otimes
\widetilde{L}_4^{-4} )^{\vee}\oplus
H_{\bar\p}^{1}(S,\widetilde{L}_3^{6}\otimes
\widetilde{L}_4^{4})\oplus
H_{\bar\p}^{2}(S,\widetilde{L}_3^{-6}\otimes
\widetilde{L}_4^{-4})^{\vee}\label{001 vectorlikeSO(10)}
\end{equation}
\begin{equation}
{\bf (1,1)}_{-6,-4}\in H_{\bar\p}^{0}(S,\widetilde{L}_3^{6}\otimes
\widetilde{L}_4^{4} )^{\vee}\oplus
H_{\bar\p}^{1}(S,\widetilde{L}_3^{-6}\otimes
\widetilde{L}_4^{-4})\oplus
H_{\bar\p}^{2}(S,\widetilde{L}_3^{6}\otimes
\widetilde{L}_4^{4})^{\vee}.\label{002 vectorlikeSO(10)}
\end{equation}
To avoid exotics, it is clear that the line bundles
$\widetilde{L}_{3}^{5}$, $\widetilde{L}_3^1\otimes
\widetilde{L}_4^{4}$, $\widetilde{L}_3^4\otimes
\widetilde{L}_4^{-4}$, and $\widetilde{L}_3^6\otimes
\widetilde{L}_4^{4}$ cannot be trivial. Let $N_{{\bf
(A,B)}_{a,b}}$ be the number of the fields in the representation
${\bf (A,B)}_{a,b}$ under $SU(3)\times SU(2)\times U(1)_1\times
U(1)_2$, where $a$ and $b$ are the charges of $U(1)_1$ and
$U(1)_2$, respectively. By the vanishing theorem (\ref{Vanishing
theorem}), the exotic-free spectrum requires that
\begin{equation}
N_{{\bf (3,2)}_{-5,0}}=-\chi(S,E)=0\label{SO(10)chi 01}
\end{equation}
\begin{equation}
N_{{\bf (\bar 3,2)}_{5,0}}=-\chi(S,E^{-1})=0\label{SO(10)chi 02}
\end{equation}
\begin{equation}
N_{{\bf (\bar 3,2)}_{-1,-4}}=-\chi(S,F^{-1})=0\label{SO(10)chi 03}
\end{equation}
\begin{equation}
N_{{\bf (3,1)}_{4,-4}}=-\chi(S,E^{-1}\otimes
F^{-1})=0\label{SO(10)chi 04}
\end{equation}
\begin{equation}
N_{{\bf (1,1)}_{-6,-4}}=-\chi(S,E^{}\otimes
F^{-1})=0\label{SO(10)chi 05}.
\end{equation}
We define
\begin{equation}
N_{{\bf (3,2)}_{1,4}}=-\chi(S,F)\equiv\beta_1,\label{SO(10)chi 06}
\end{equation}
\begin{equation}
N_{{\bf (\bar 3,1)}_{-4,4}}=-\chi(S,E\otimes
F)\equiv\beta_2\label{SO(10)chi 07}
\end{equation}
\begin{equation}
N_{{\bf (1,1)}_{6,4}}=-\chi(S,E^{-1}\otimes
F)\equiv\beta_3\label{SO(10)chi 08},
\end{equation}
where $E=\widetilde{L}_{3}^{-5}$, $F=\widetilde{L}_3^1\otimes
\widetilde{L}_4^{4}$ and $\beta_{i}\in \mathbb{Z}_{\geqslant
0},\;i=1,2,3$. By Eqs. (\ref{SO(10)chi 01})-(\ref{SO(10)chi 03}),
and Eq. (\ref{SO(10)chi 06}), we obtain the following equations
\begin{equation}
\left\{\begin{array}{l} c_{1}(E)^{2}=-2\\
c_1(F)^{2}=-\beta_1-2
\\c_1(E)\cdot K_S=0\\c_1(F)\cdot
K_S=\beta_1.\label{Exotic free conditionSO(10)}
\end{array}   \right.
\end{equation}
Then by Eq. (\ref{Exotic free conditionSO(10)}) and Eq.
(\ref{SO(10)chi 04}), we obtain
\begin{equation}
c_{1}(E)\cdot c_1(F)=1.
\end{equation}
On the other hand, using Eq. (\ref{Exotic free conditionSO(10)})
and Eq. (\ref{SO(10)chi 05}), we have
\begin{equation}
c_{1}(E)\cdot c_1(F)=-1,
\end{equation}
which leads to a contradiction. Therefore, there do not exist
solutions for given $\beta_i\in \mathbb{Z}_{\geqslant
0},\;i=1,2,3$ such that Eqs. (\ref{SO(10)chi 01})-(\ref{SO(10)chi
08}) hold. This is a no-go theorem shown in \cite{BHV:2008II}. Due
to this no-go theorem, we are not going to study this case
further. In the next section we turn to the case of $G_S=SU(6)$.

\subsection{$G_S=SU(6)$}

\subsubsection{$U(1)^2$ Gauge Flux Configurations}

The maximal subgroups of $SU(6)$ which contain $G_{\rm std}$ and
the consistent MSSM spectrum are as follows \cite{BHV:2008II}:
\begin{equation}
SU(6)\supset SU(5)\times U(1)\supset G_{\rm std}\times
U(1)\label{maxgroupSU(6)01}
\end{equation}
\begin{equation}
SU(6)\supset SU(2)\times SU(4)\times U(1)\supset G_{\rm std}\times
U(1)\label{maxgroupSU(6)02}
\end{equation}
\begin{equation}
SU(6)\supset SU(3)\times SU(3)\times U(1)\supset G_{\rm std}\times
U(1).\label{maxgroupSU(6)03}
\end{equation}
It follows from Eqs.
(\ref{maxgroupSU(6)01})-(\ref{maxgroupSU(6)03}) that up to linear
combinations of the $U(1)$ charges in the breaking patterns, it is
enough to analyze the case of $U(1)^2$ gauge fluxes which break
$SU(6)$ via the sequence $SU(6)\ra SU(5)\times U(1)\ra G_{\rm
std}\times U(1)$. The breaking pattern is as follows:
\begin{equation}
\begin{array}{c@{}c@{}l@{}c@{}l}
SU(6) &~\rightarrow~& SU(3)\times SU(2)\times U(1)_1\times U(1)_2 \\
{\bf 35} &~\rightarrow~& {\bf (8,1)}_{0,0}+{\bf (1,3)}_{0,0}+{\bf
(3,2)}_{-5,0}+{\bf (\bar 3,2)}_{5,0}+{\bf (1,1)}_{0,0}\\& &+{\bf
(1,1)}_{0,0}+{\bf (1,2)}_{3,6}+{\bf (3,1)}_{-2,6}+{\bf (1,\bar
2)}_{-3,-6}+{\bf (\bar 3,1)}_{2,-6}.\label{breaking SU(6)}
\end{array}
\end{equation}
Note that $U(1)_1$ is consistent with $U(1)_Y$ in the MSSM. Let
$L_1$ and $L_2$ be non-trivial supersymmetric line bundles
associated with $U(1)_1$ and $U(1)_2$, respectively, in the
breaking pattern (\ref{breaking SU(6)}). The bulk zero modes are
given by
\begin{equation}
{\bf (3,2)}_{-5,0}\in H_{\bar\p}^{0}(S,L_1^{5})^{\vee}\oplus
H_{\bar\p}^{1}(S,L_1^{-5})\oplus H_{\bar\p}^{2}(S,L_1^{5})^{\vee}
\end{equation}
\begin{equation}
{\bf (\bar 3,2)}_{5,0}\in H_{\bar\p}^{0}(S,L_1^{-5})^{\vee}\oplus
H_{\bar\p}^{1}(S,L_1^{5})\oplus H_{\bar\p}^{2}(S,L_1^{-5})^{\vee}
\end{equation}
\begin{equation}
{\bf (1,2)}_{3,6}\in H_{\bar\p}^{0}(S,L_1^{-3}\otimes L_2^{-6}
)^{\vee}\oplus H_{\bar\p}^{1}(S,L_1^{3}\otimes L_2^{6})\oplus
H_{\bar\p}^{2}(S,L_1^{-3}\otimes L_2^{-6})^{\vee}\label{001
vectorlikeSU(7)}
\end{equation}
\begin{equation}
{\bf (1,\bar 2)}_{-3,-6}\in H_{\bar\p}^{0}(S,L_1^{3}\otimes
L_2^{6} )^{\vee}\oplus H_{\bar\p}^{1}(S,L_1^{-3}\otimes
L_2^{-6})\oplus H_{\bar\p}^{2}(S,L_1^{3}\otimes
L_2^{6})^{\vee}\label{002 vectorlikeSU(7)}
\end{equation}
\begin{equation}
{\bf (3,1)}_{-2,6}\in H_{\bar\p}^{0}(S,L_1^{2}\otimes L_2^{-6}
)^{\vee}\oplus H_{\bar\p}^{1}(S,L_1^{-2}\otimes L_2^{6})\oplus
H_{\bar\p}^{2}(S,L_1^{2}\otimes L_2^{-6})^{\vee}
\end{equation}
\begin{equation}
{\bf (\bar 3,1)}_{2,-6}\in H_{\bar\p}^{0}(S,L_1^{-2}\otimes
L_2^{6} )^{\vee}\oplus H_{\bar\p}^{1}(S,L_1^{2}\otimes
L_2^{-6})\oplus H_{\bar\p}^{2}(S,L_1^{-2}\otimes L_2^{6})^{\vee}.
\end{equation}
Note that ${\bf (3,2)}_{-5,0}$, ${\bf (\bar 3,2)}_{5,0}$, and
${\bf (3,1)}_{-2,6}$ are exotic fields in the MSSM. To avoid these
exotics, $L_1^{5}$ and $L_1^{-2}\otimes L_2^{6}$ need to be
non-trivial line bundles. If $L_1^{3}\otimes L_2^{6}$ is trivial,
it follows from Eq. (\ref{001 vectorlikeSU(7)}) and Eq. (\ref{002
vectorlikeSU(7)}) that $N_{{\bf (1,2)}_{3,6}}=N_{{\bf (1,\bar
2)}_{-3,-6}}=1$. By the vanishing theorem (\ref{Vanishing
theorem}), no exotic fields requires that
\begin{equation}
N_{{\bf (3,2)}_{-5,0}}=-\chi(S,L_1^{-5})=0\label{chi001}
\end{equation}
\begin{equation}
N_{{\bf (\bar 3,2)}_{5,0}}=-\chi(S,L_1^{5})=0\label{chi002}
\end{equation}
\begin{equation}
N_{{\bf (3,1)}_{-2,6}}=-\chi(S,L_1^{-2}\otimes
L_2^{6})=0.\label{chi003}
\end{equation}
We define
\begin{equation}
N_{{\bf (\bar 3,1)}_{2,-6}}=-\chi(S,L_1^{2}\otimes
L_2^{-6})\equiv\alpha_3\label{chi004},
\end{equation}
where $\alpha_3\in \mathbb{Z}_{\geqslant 0}$. Note that since
$L_1^{3}\otimes L_2^{6}$ is trivial, then $L_1^{2}\otimes
L_2^{-6}\cong L_1^{5}$. It follows from Eq. (\ref{chi002}) that
$\alpha_3=0$\footnote{This case will be denoted by
$(\alpha_1,\alpha_2,\alpha_3)=(1,1,0)^{\ast}$ later.}. Therefore,
the non-trivial conditions are (\ref{chi001}) and (\ref{chi002}),
namely $\chi(S,L_1^{\pm 5})=0$, which imply that $c_{1}({L_1}^{\pm
5})^2=-2$ and $c_1(L_1^{\pm 5})\cdot K_S=0$. Note that
$c_{1}({L_1}^{\pm 5})\in H_{2}(S,\mathbb{Z})={\rm
span}_{\mathbb{Z}}\{H,E_i,\; i=1,2,3,...8\}$, where $H$ and $E_i$
are the hyperplane divisor and exceptional divisors in $S=dP_8$.
Immediately we get a fractional line bundle\footnote{Note that
with $\alpha_3=0$, there is a symmetry $(L_1,L_2)\leftrightarrow
(L_1^{-1},L_2^{-1})$ in Eq. (\ref{chi001})-(\ref{chi004}). Without
loss of generality, we choose
$L_1=\mathcal{O}_{S}(E_j-E_i)^{1/5}$.}
$L_1=\mathcal{O}_{S}(E_j-E_i)^{1/5}$ and then
$L_{2}=\mathcal{O}_{S}(E_i-E_j)^{1/10}$. It is clear that $L_1$
and $L_2$ satisfy the BPS condition (\ref{Polarization}). As a
result, $(L_1,L_2)$ is a supersymmetric $U(1)^2$ gauge flux
configuration on the bulk. If $L_1^{3}\otimes L_2^{6}$ is
non-trivial, by the vanishing theorem (\ref{Vanishing theorem}),
an exotic-free bulk spectrum requires that
\begin{equation}
N_{{\bf (3,2)}_{-5,0}}=-\chi(S,L_1^{-5})=0\label{chi01}
\end{equation}
\begin{equation}
N_{{\bf (\bar 3,2)}_{5,0}}=-\chi(S,L_1^{5})=0
\end{equation}
\begin{equation}
N_{{\bf (3,1)}_{-2,6}}=-\chi(S,L_1^{-2}\otimes L_2^{6})=0
\label{chi06}.
\end{equation}
We define
\begin{equation}
N_{{\bf (1,2)}_{3,6}}=-\chi(S,L_1^{3}\otimes
L_2^{6})\equiv\alpha_1
\end{equation}
\begin{equation}
N_{{\bf (1,\bar 2)}_{-3,-6}}=-\chi(S,L_1^{-3}\otimes
L_2^{-6})\equiv\alpha_2
\end{equation}
\begin{equation}
N_{{\bf (\bar 3,1)}_{2,-6}}=-\chi(S,L_1^{2}\otimes
L_2^{-6})\equiv\alpha_3\label{chi07},
\end{equation}
where $\alpha_i\in\mathbb{Z}_{\geqslant 0},\;i=1,2,3$. To simplify
the notation, we define $C=L_{1}^{-5}$, and $D=L_1^3\otimes
L_2^{6}$. By Eqs. (\ref{chi01})-(\ref{chi07}) and the Riemann-Roch
theorem (\ref{R-R Surface}), we obtain the following equations:

\begin{equation}
\left\{\begin{array}{l} c_{1}(C)^{2}=-2\\
c_1(D)^{2}=-\alpha_1-\alpha_2-2\\c_1(C)\cdot c_1(D)=
1+\frac{1}{2}(\alpha_1+\alpha_2-\alpha_3)\\
\alpha_3=\alpha_2-\alpha_1
\\c_1(C)\cdot K_S=0\\c_1(D)\cdot
K_S=\alpha_1-\alpha_2.\label{Exotic free condition01}
\end{array}   \right.
\end{equation}
Note that $C$ and $D$ are required to be honest line bundles, in
other words, $c_1(C)$, $c_1(D)\in H_2(S,\mathbb{Z})={\rm
span}_{\mathbb{Z}}\{H,E_i,\;i=1,2,3,...8\}$. Note that ${\bf (\bar
3,1)}_{2,-6}$ is a candidate for a matter field in the MSSM.
Therefore, we shall restrict to the case of $\alpha_3\leqslant 3$.
In what follows, we shall demonstrate how to derive explicit
expressions for $U(1)^2$ gauge fluxes from Eq. (\ref{Exotic free
condition01}). For the case of $\alpha_3=0$, by the constraints in
Eq. (\ref{Exotic free condition01}), we may assume
$(\alpha_1,\alpha_2,\alpha_3)=(k,k,0)$ with $k\in
\mathbb{Z}_{\geqslant 0}$. We shall show that there is no solution
for $k\geqslant 4$. Note that in this case, Eq. (\ref{Exotic free
condition01}) reduces to
\begin{equation}
c_1(C)^{2}=-2,\;\;c_1(D)^{2}=-2k-2,\;\;c_1(C)\cdot
c_1(D)=1+k,\label{No-go}
\end{equation}
with $c_1(C)\cdot K_S=c_1(D)\cdot K_S=0$. From the conditions
$c_1(C)^{2}=-2$, $c_1(C)\cdot K_S=0$, and BPS condition
(\ref{Polarization}), it follows that
$C=\mathcal{O}_{S}(E_i-E_j)$, which is the universal line bundle
in the case of $G_S=SU(6)$ since these two conditions are
independent of $\alpha_i,\; i=1,2,3$ and always appear in Eq.
(\ref{Exotic free condition01}). Actually, the corresponding
fractional line bundle $L_1$ of $C$ is the $U(1)_Y$ hypercharge
flux in the minimal $SU(5)$ GUT
\cite{BHV:2008I,BHV:2008II,Donagi:local}. In what follows, we
shall focus on the solutions for the line bundle $D$. By Eq.
(\ref{No-go}), we can obtain the upper bound of $k$. Write
$D=\mathcal{O}_{S}(c_iE_i+c_jE_j+\tilde{D})$,\footnote{Due to the
BPS condition (\ref{Polarization}), $D$ contains no component
$H$.} where $\tilde{D}$ is a integral divisor containing no $H$,
$E_i$, and $E_j$. Note that the repeat indices are not a
summation, and $c_i$, $c_j\in \mathbb{Z}$. By Eq. (\ref{No-go}),
we get $-c_i+c_j=k+1$ and $c_1^{2}+c_2^2-{\tilde D}^2=2k+2$. Note
that ${\tilde D}^2\leqslant 0$ by the construction. Using the
inequality\footnote{In general, $(c_1(C)^2)(c_1(D)^2)\geqslant
(c_1(C)\cdot c_1(D))^2$. }
$c_1^2+c_2^2\geqslant\frac{1}{2}(c_1-c_2)^2$ and the condition
$k\in \mathbb{Z}_{\geqslant 0}$, we obtain $0\leqslant k\leqslant
3$, which implies that there is no solution $D$ for $k\geqslant
4$. Next we shall explicitly solve the configurations $(L_1,L_2)$
satisfying Eq. (\ref{Exotic free condition01}) for the case of
$(\alpha_1,\alpha_2,\alpha_3)=(k,k,0)$ with $0\leqslant k\leqslant
3$.

Let us start with the simplification of Eq. (\ref{Exotic free
condition01}). Note that in Eq. (\ref{Exotic free condition01}),
there are two conditions that are independent of $\alpha_i$,
namely,
\begin{equation}
c_1(C)^{2}=-2,\;\;c_1(C)\cdot K_S=0,\label{Universal}
\end{equation}
which gives rise to the universal line bundle,
$C=\mathcal{O}_S(E_i-E_j)$, as mentioned earlier. The remaining
conditions are
\begin{equation}
\left\{\begin{array}{l}
c_1(D)^{2}=-\alpha_1-\alpha_2-2\\c_1(C)\cdot c_1(D)=
1+\frac{1}{2}(\alpha_1+\alpha_2-\alpha_3)\\
\alpha_3=\alpha_2-\alpha_1
\\c_1(D)\cdot
K_S=\alpha_1-\alpha_2.\label{Exotic free condition01_1}
\end{array}   \right.
\end{equation}
Since $C$ is universal, all we have to do is to solve the line
bundles $D$ in Eq. (\ref{Exotic free condition01_1}) for a given
$(\alpha_1,\alpha_2,\alpha_3)$ and $C=\mathcal{O}_S(E_i-E_j)$.
When $(\alpha_1,\alpha_2,\alpha_3)=(0,0,0)$, Eq. (\ref{Exotic free
condition01_1}) reduces to
\begin{equation}
c_1(D)^{2}=-2,\;\;c_1(C)\cdot c_1(D)=1,\label{A}
\end{equation}
with $c_1(D)\cdot K_S=0$. By Eq. (\ref{A}), we have
$D=\mathcal{O}_{S}(\pm E_l-E_i)$ or $\mathcal{O}_{S}(\pm
E_l+E_j)$. The former gives rise to fractional line bundles
$L_1=\mathcal{O}_{S}(E_j-E_i)^{1/5}$ and $L_2=\mathcal{O}_{S}(\pm
5E_l-2E_i-3E_j)^{1/30}$. For the latter, we have
$L_1=\mathcal{O}_{S}(E_j-E_i)^{1/5}$ and $L_2=\mathcal{O}_{S}(\pm
5E_l+3E_i+2E_j)^{1/30}$. Recall that $K_S=-3H+\sum_{k=1}^{8}E_k$.
To solve the condition $c_1(D)\cdot K_S=0$, it is clear that $D$
has to be $\mathcal{O}_{S}(E_l-E_i)$ or
$\mathcal{O}_{S}(-E_l+E_j)$. The corresponding fractional line
bundle is $\mathcal{O}_{S}( 5E_l-2E_i-3E_j)^{1/30}$ or
$\mathcal{O}_{S}(- 5E_l+3E_i+2E_j)^{1/30}$. In addition to Eq.
(\ref{A}), these fractional line bundles need to satisfy the BPS
condition (\ref{Polarization}). More precisely, for the case of
$L_1=\mathcal{O}_{S}(E_j-E_i)^{1/5}$ and $L_2=\mathcal{O}_{S}(
5E_l-2E_i-3E_j)^{1/30}$, BPS equation (\ref{Polarization}) reduces
to
\begin{equation}
(E_{i}-E_j)\cdot\omega=0,\;\;(
5E_l-2E_i-3E_j)\cdot\omega=0.\label{ex01}
\end{equation}
It is not difficult to see that\footnote{''$...$'' in $\omega$
always stands for non-relevant terms for checking the BPS
condition Eq. (\ref{Polarization}). Of course, those terms are
relevant for the ampleness of $\omega$ and note that the choice of
the polarizations is not unique.} $\omega=AH-(E_i+E_j+E_l+...)$
solves Eq. (\ref{ex01}). Similarly, for the case of
$L_1=\mathcal{O}_{S}(E_j-E_i)^{1/5}$ and $L_2=\mathcal{O}_{S}(-
5E_l+3E_i+2E_j)^{1/30}$, $L_1$ and $L_2$ are also supersymmetric
with respect to $\omega=AH-(E_i+E_j+E_l+...)$. As a result, for
the case of $(\alpha_1,\alpha_2,\alpha_3)=(0,0,0)$, we find two
supersymmetric $U(1)^2$ gauge flux configurations $(L_1,L_2)$.

When $(\alpha_1,\alpha_2,\alpha_3)=(1,1,0)$, Eq. (\ref{Exotic free
condition01_1}) reduces to
\begin{equation}
c_1(D)^{2}=-4,\;\;c_1(C)\cdot c_1(D)=2,\label{B}
\end{equation}
with $c_1(D)\cdot K_S=0$. By Eq. (\ref{B}), $D$ can be
$\mathcal{O}_{S}(2E_j)$, $\mathcal{O}_{S}(-2E_i)$ or
$\mathcal{O}_{S}([E_l,E_m]-E_i+ E_j)$, where the bracket is
defined by $[A_1,A_2,..A_k]=\{\pm A_1\pm A_2...\pm A_k\}$. For
later use, we also define $[A_1,A_2,..A_k]'=\{\pm A_1\pm A_2...\pm
A_k\}\smallsetminus (+A_1+A_2+...+A_k)$, $[A_1,A_2,..A_k]''=\{\pm
A_1\pm A_2...\pm A_k\}\smallsetminus
\{(+A_1+A_2+...+A_k),(-A_1-A_2-...-A_k)\}$, and
$[A_1,A_2,..A_k]'''=\{(A_1+A_2...+A_{k-1}-A_k),(A_1+A_2...-A_{k-1}+A_k),...,
(-A_1+A_2...+A_{k-1}+A_k)\}$. Note that $\mathcal{O}_{S}(2E_j)$,
$\mathcal{O}_{S}(-2E_i)$, $\mathcal{O}_{S}(E_l+E_m-E_i+ E_j)$, and
$\mathcal{O}_{S}(-E_l-E_m-E_i+ E_j)$ cannot solve the equation
$c_1(D)\cdot K_S=0$. As a result,
$D=\mathcal{O}_{S}([E_l,E_m]''-E_i+ E_j)$, which correspond to the
fractional bundles $L_2=\mathcal{O}_{S}(
5[E_l,E_m]''-2E_i+2E_j)^{1/30}$. Clearly $L_1$ and $L_2$ satisfy
Eq. (\ref{Polarization}) with $\omega=AH-(E_i+E_j+E_l+E_m+...)$.

For the case of $(\alpha_1,\alpha_2,\alpha_3)=(2,2,0)$, Eq.
(\ref{Exotic free condition01_1}) becomes
\begin{equation}
c_1(D)^{2}=-6,\;\;c_1(C)\cdot c_1(D)=3,\label{C}
\end{equation}
with $c_1(D)\cdot K_S=0$. By Eq. (\ref{C}), $D$ can be
$\mathcal{O}_{S}([E_{l}]-E_i+2E_j)$ or
$\mathcal{O}_{S}([E_{l}]-2E_i+E_j)$. For the former, it is clear
that $\mathcal{O}_{S}(E_l-E_i+2E_j)$ does not satisfy the
condition $c_1(D)\cdot K_S=0$. Similarly, for the latter,
$\mathcal{O}_{S}( -E_l-2E_i+E_j)$ is not a solution as well. In
this case, the solutions are $L_2=\mathcal{O}_{S}(
-5E_l-2E_i+7E_j)^{1/30}$ or $L_2=\mathcal{O}_{S}(
5E_l-7E_i+2E_j)^{1/30}$. It is easy to see that the solutions also
satisfy the BPS condition (\ref{Polarization}). Note that for the
case of $\alpha_3=0$, taking
$\omega=AH-(\sum_{k=1}^{8}E_k)=(-K_S)+(A-3)H$, the conditions
$c_1(C)\cdot K_S=c_1(D)\cdot K_S=0$ are equivalent to Eq.
(\ref{Polarization}). Therefore, the solutions of Eq. (\ref{Exotic
free condition01}) are all supersymmetric for the case of
$\alpha_3=0$.

Next we consider the case of
$(\alpha_1,\alpha_2,\alpha_3)=(3,3,0)$. In this case, the line
bundle $D$ satisfies the following equations:
\begin{equation}
c_1(D)^{2}=-8,\;\;c_1(C)\cdot c_1(D)=4,\label{D}
\end{equation}
with $c_1(D)\cdot K_S=0$. By Eq. (\ref{D}), we obtain
$D=\mathcal{O}_{S}(2E_j-2E_i)$. The corresponding fractional line
bundle is $L_2=\mathcal{O}_{S}(E_j-E_i)^{7/30}$. Obviously, $L_2$
satisfies the condition $c_1(D)\cdot K_S=0$, and Eq.
(\ref{Polarization}) for $\omega=AH-(E_i+E_j+...)$.

Next we shall consider the case of $\alpha_3=1$. By the
constraints of Eq. ({\ref{Exotic free condition01_1}}), we may
assume that $(\alpha_1,\alpha_2,\alpha_3)=(m,m+1,1)$, where $m\in
\mathbb{Z}_{\geqslant 0}$. Then Eq. (\ref{Exotic free
condition01_1}) becomes
\begin{equation}
c_1(D)^{2}=-2m-3,\;\;c_1(C)\cdot c_1(D)=1+m,\label{No-go 3}
\end{equation}
with $c_1(D)\cdot K_S=-1$. Again the first thing we need to do is
to get the upper bound of $m$. Eq. (\ref{No-go 3}) implies that
$1-\sqrt{6}\leqslant m\leqslant 1+\sqrt{6}$. Since $m\in
\mathbb{Z}_{\geqslant 0}$, we obtain $0\leqslant m\leqslant 3$.
Therefore, the possible configurations are
$(\alpha_1,\alpha_2,\alpha_3)=(0,1,1),(1,2,1),(2,3,1)$ or
$(3,4,1)$.

Let us look at the case of $(\alpha_1,\alpha_2,\alpha_3)=(0,1,1)$.
In this case, Eq. (\ref{No-go 3}) reduces to the following
equations
\begin{equation}
c_1(D)^{2}=-3,\;\;c_1(C)\cdot c_1(D)=1.\label{ex02}
\end{equation}
It is easy to see that $D$ can be $\mathcal{O}_{S}([E_l,E_m]-E_i)$
or $\mathcal{O}_{S}([E_l,E_m]+E_j)$. Note that
$\mathcal{O}_{S}([E_l,E_m]''-E_i)$,
$\mathcal{O}_{S}(-E_l-E_m-E_i)$, $\mathcal{O}_{S}(E_l+E_m+E_j)$,
and $\mathcal{O}_{S}(-E_l-E_m+E_j)$ do not satisfy the equation
$c_1(D)\cdot K_S=-1$, so we have to eliminate these cases. It
turns out that the resulting fractional line bundles are
$\mathcal{O}_{S}(5(E_l+E_m)-2E_i-3E_j)^{1/30}$ and
$\mathcal{O}_{S}(5[E_l,E_m]''+3E_i+2E_j)^{1/30}$. In order to
preserve supersymmetry, the solutions need to solve Eq.
(\ref{Polarization}). For the case of
$L_2=\mathcal{O}_{S}(5(E_l+E_m)-2E_i-3E_j)^{1/30}$, Eq.
(\ref{Polarization}) reduces to
\begin{equation}
 (E_i-E_j)\cdot\omega=0,\;\;  [(E_l+E_m)-E_i]\cdot\omega=0.\label{001}
\end{equation}
For another fractional line bundle
$L_2=\mathcal{O}_{S}(5[E_l,E_m]''+3E_i+2E_j)^{1/30}$, Eq.
(\ref{Polarization}) becomes
\begin{equation}
(E_i-E_j)\cdot\omega=0,\;\;([E_l,E_m]''+E_i)\cdot\omega=0\label{002}
\end{equation}
It is clear that $\omega=AH-(E_l+E_m+2E_i+2E_j+...)$ solves Eq.
(\ref{001}) and $\omega=AH-(2E_l+E_m+E_i+E_j+...)$ solves Eq.
(\ref{002}) if $[E_l,E_m]''=-E_l+E_m$. For the case of
$[E_l,E_m]''=E_l-E_m$, $\omega=AH-(E_l+2E_m+E_i+E_j+...)$ is a
solution of Eq. (\ref{002}). Therefore,
$\mathcal{O}_{S}(5(E_l+E_m)-2E_i-3E_j)^{1/30}$ and
$\mathcal{O}_{S}(5[E_l,E_m]''+3E_i+2E_j)^{1/30}$ are
supersymmetric. In this case, the solutions of Eq. (\ref{ex02})
and the equations, $c_1(C)\cdot K_S=0,\;c_1(D)\cdot K_S=-1$
satisfy Eq. (\ref{Polarization}). It seems that for the case
$\alpha_3=1$, the condition $c_{1}(C)\cdot K_S=0,\;c_1(D)\cdot
K_S=-1$ is stronger than BPS condition (\ref{Polarization}). For
example, $D=\mathcal{O}_S(E_l-E_m-E_i)$ with corresponding
fractional line bundle
$L_2=\mathcal{O}_S(5E_l-5E_m-2E_i-3E_j)^{1/30}$ is supersymmetric
but does not satisfy the condition $c_1(D)\cdot K_S=-1$. Actually,
we shall see that this is not the case in the next examples.

Let us turn to the case of $(\alpha_1,\alpha_2,\alpha_3)=(3,4,1)$.
In this case, Eq. (\ref{No-go 3}) reduces to
\begin{equation}
c_1(D)^{2}=-9,\;\;c_1(C)\cdot c_1(D)=4.\label{ex03}
\end{equation}
It is not difficult to find that the solutions are
$D=\mathcal{O}_{S}([E_l]-2E_i+2E_j)$ and the corresponding
fractional line bundle are
$L_2=\mathcal{O}_{S}(5[E_l]-7E_i+7E_j)^{1/30}$. Note that only
$D=\mathcal{O}_{S}(E_l-2E_i+2E_j)$ satisfies the condition
$c_1(D)\cdot K_S=-1$. However, it is clear that it does not
satisfy the BPS condition (\ref{Polarization}), which means that
no configuration $(L_1,L_2)$ for an exotic-free spectrum exists in
this case. From this example, we know that for the case of
$\alpha_3=1$, the solutions of Eq. (\ref{Exotic free
condition01_1}) are not guaranteed to be supersymmetric and vice
versa. Therefore, in general we need to check these two conditions
for each solution in the case of $\alpha_3\in \mathbb{Z}_{>0}$.
Following a similar procedure, one can obtain all configurations
$(L_1,L_2)$ for the cases of $\alpha_3=1$. We summarize the
results of $\alpha_3=0,1$ in Table {\ref{SU(6) bulk01}} in which
all $L_1$ and $L_2$ satisfy the BPS condition (\ref{Polarization})
for suitable polarizations $\omega$ and the conditions $L_1^5\neq
\mathcal{O}_{S}$, $L_1^{-2}\otimes L_2^6\neq \mathcal{O}_{S}$ and
$L_1^3\otimes L_1^6\neq \mathcal{O}_{S}$.

\begin{table}[h]
\begin{center}
\renewcommand{\arraystretch}{1.25}
\begin{tabular}{|c|c|c|c|c|c|c|} \hline

& $(\alpha_1,\alpha_2,\alpha_3)$ & $L_2$ \\
\hline \hline

$1$ & $~(1,1,0)^{\ast}$ &
$\mathcal{O}_{S}(E_i-E_j)^{1/10}$\\
\hline

$2$ & $(0,0,0)$ &
$\mathcal{O}_{S}(5E_l-2E_i-3E_j)^{1/30}$\\
& & $\mathcal{O}_{S}(-5E_l+3E_i+2E_j)^{1/30}$\\
\hline

$3$ & $(1,1,0)$ &
$\mathcal{O}_{S}(5[E_l,E_m]''-2E_i+2E_j)^{1/30}$\\
\hline

$4$ & $(2,2,0)$ &
$\mathcal{O}_{S}(-5E_l-2E_i+7E_j)^{1/30}$\\
& & $\mathcal{O}_{S}(5E_l-7E_i+2E_j)^{1/30}$\\
\hline

$5$ & $(3,3,0)$  &
$\mathcal{O}_{S}(E_j-E_i)^{7/30}$\\
\hline

$6$ & $(0,1,1)$
& $\mathcal{O}_{S}(5[E_l,E_m]''+3E_i+2E_j)^{1/30}$\\
& & $\mathcal{O}_{S}(5(E_l+E_m)-2E_i-3E_j)^{1/30}$\\
\hline

$7$ & $(1,2,1)$ & $\mathcal{O}_{S}(-5E_l+3E_i+7E_j)^{1/30}$\\
& & $\mathcal{O}_{S}(5[E_l,E_m,E_k]'''-2E_i+2E_j)^{1/30}$\\
\hline

$8$ & $(2,3,1)$
& $\mathcal{O}_{S}(5[E_l,E_m]''-2E_i+7E_j)^{1/30}$\\
& & $\mathcal{O}_{S}(5(E_l+E_m)-7E_i+2E_j)^{1/30}$\\
\hline

$9$ & $(3,4,1)$
& No Solution\\
\hline

\end{tabular}
\caption{Flux configurations for $G_S=SU(6)$ with
$L_{1}=\mathcal{O}_{S}(E_j-E_i)^{1/5}$ and
$\alpha_3=0,1$.}\label{SU(6) bulk01}
\end{center}
\end{table}

Next we consider the case of $\alpha_3=2$. By the last constraint
of Eq. (\ref{Exotic free condition01}), we may assume
$(\alpha_1,\alpha_2,\alpha_3)=(l,l+2,2)$, where $l\in
\mathbb{Z}_{\geqslant 0}$. One can show that the necessary
condition for existence of the solutions of Eq. (\ref{Exotic free
condition01}) is $0\leqslant l\leqslant 3$. Therefore,
$(\alpha_1,\alpha_2,\alpha_3)$ can be
$(0,2,2),\;(1,3,2),\;(2,4,2)$ or $(3,5,2)$. Following the previous
procedure, one can obtain all configurations $(L_1,L_2)$ for the
case of $\alpha_3=2$.

For the case of $\alpha_3=3$, we may assume that
$(\alpha_1,\alpha_2,\alpha_3)=(n,n+3,3)$ with $n\in
\mathbb{Z}_{\geqslant 0}$. The necessary condition for existence
of the solutions of Eq. (\ref{Exotic free condition01}) is
$0\leqslant n\leqslant 4$, which implies that
$(\alpha_1,\alpha_2,\alpha_3)=(0,3,3),(1,4,3),(2,5,3)$, $(3,6,3)$,
or $(4,7,3)$. Following the previous procedure, one can obtain all
configurations $(L_1,L_2)$ for the case of $\alpha_3=3$. Let us
look at the case of $(\alpha_1,\alpha_2,\alpha_3)=(3,6,3)$. In
this case, Eq. (\ref{Exotic free condition01_1}) reduces to
\begin{equation}
c_1(D)^{2}=-11,\;\;c_1(C)\cdot c_1(D)=4,\label{ex04}
\end{equation}
with $c_1(D)\cdot K_S=-3$. It follows from Eq. (\ref{ex04}) that
$D$ can be $\mathcal{O}_{S}([E_l]-E_i+3E_j)$,
$\mathcal{O}_{S}([E_l]-3E_i+E_j)$, or
 $\mathcal{O}_{S}([E_l,E_m,E_n]-2E_i+2E_j)$. When one takes the condition $c_1(D)\cdot
 K_S=-3$ into account, there are only two solutions,
 $D=\mathcal{O}_{S}(E_l-E_i+3E_j)$ or $\mathcal{O}_{S}((E_l+E_m+E_n)-2E_i+2E_j)$,
 which corresponds to the fractional line bundles $\mathcal{O}_{S}(5E_l-2E_i+12E_j)^{1/30}$ and
 $\mathcal{O}_{S}(5(E_l+E_m+E_n)-7E_i+7E_j)^{1/30}$, respectively.
However, these two solutions cannot satisfy Eq.
(\ref{Polarization}). Therefore, in this case there do not exist
any $U(1)^2$ gauge fluxes for an exotic-free spectrum. A similar
situation occurs in the case of
$(\alpha_1,\alpha_2,\alpha_3)=(4,7,3)$. In this case, $D$ can be
$\mathcal{O}_{S}(-3E_i+2E_j)$ or $\mathcal{O}_{S}(-2E_i+3E_j)$ by
Eq. (\ref{Exotic free condition01_1}). However, they neither solve
Eq. (\ref{Polarization}) nor satisfy the condition $c_1(D)\cdot
K_S=-3$. As a result, there are no $U(1)^2$ gauge fluxes without
producing exotics in this case. We summarize the results of
$\alpha_3=2,3$ in Table {\ref{SU(6) bulk02}} in which all $L_1$
and $L_2$ satisfy the BPS condition (\ref{Polarization}) for
suitable polarizations $\omega$ and the conditions $L_1^5\neq
\mathcal{O}_{S}$, $L_1^{-2}\otimes L_2^6\neq \mathcal{O}_{S}$ and
$L_1^3\otimes L_1^6\neq \mathcal{O}_{S}$.

\begin{table}[h]
\begin{center}
\renewcommand{\arraystretch}{1.25}
\begin{tabular}{|c|c|c|c|c|c|c|} \hline

& $(\alpha_1,\alpha_2,\alpha_3)$ & $L_2$ \\
\hline \hline

$1$ & $(0,2,2)$ &
$\mathcal{O}_{S}(5(E_l+E_m+E_k)-2E_i-3E_j)^{1/30}$\\
& &  $\mathcal{O}_{S}(5[E_l,E_m,E_k]'''+3E_i+2E_j)^{1/30}$\\
\hline

$2$ & $(1,3,2)$  &  $\mathcal{O}_{S}(5[E_l,E_m]''+3E_i+7E_j)^{1/30}$\\

& &  $\mathcal{O}_{S}(5[E_l,E_m,E_n,E_k]'''-2E_i+2E_j)^{1/30}$\\
\hline

$3$ & $(2,4,2)$ &
$\mathcal{O}_{S}(5[E_l,E_m,E_k]'''-2E_i+7E_j)^{1/30}$\\
&&$\mathcal{O}_{S}(5(E_l+E_m+E_k)-7E_i+2E_j)^{1/30}$\\
\hline

$4$ & $(3,5,2)$  &  No Solution\\
\hline

$5$ & $(0,3,3)$  &  $\mathcal{O}_{S}(5(E_l+E_m+E_n+E_k)-2E_i-3E_j)^{1/30}$\\
& &  $\mathcal{O}_{S}(5[E_l,E_m,E_n,E_k]'''+3E_i+2E_j)^{1/30}$\\
\hline

$6$ & $(1,4,3)$ &  $\mathcal{O}_{S}(5[E_l,E_m,E_k]'''+3E_i+7E_j)^{1/30}$\\
& &  $\mathcal{O}_{S}(5[E_l,E_m,E_n,E_k,E_p]'''-2E_i+2E_j)^{1/30}$\\
\hline

$7$ & $(2,5,3)$  & $\mathcal{O}_{S}(5[E_l,E_m,E_n,E_k]'''-2E_i+7E_j)^{1/30}$ \\
& & $\mathcal{O}_{S}(5(E_l+E_m+E_n+E_k)-7E_i+2E_j)^{1/30}$ \\
\hline

$8$ & $(3,6,3)$ &  No Solution\\
\hline

$9$ & $(4,7,3)$ & No Solution\\
\hline

\end{tabular}
\caption{Flux configurations for $G_S=SU(6)$ with
$L_{1}=\mathcal{O}_{S}(E_j-E_i)^{1/5}$ and
$\alpha_3=2,3$.}\label{SU(6) bulk02}
\end{center}
\end{table}

\subsubsection{Spectrum from the Curves}

With $G_S=SU(6)$, to obtain matter in $SU(5)$ GUT, it is required
that $L_{\Sigma}\neq \mathcal{O}_{\Sigma}$ and
$L'_{\Sigma}\neq\mathcal{O}_{\Sigma}$. In this case, there are
three kinds of intersecting curves, $\Sigma_{SU(7)}$,
$\Sigma_{SO(12)}$ and $\Sigma_{E_{6}}$ with enhanced gauge groups
$SU(7)$, $SO(12)$, and $E_{6}$, respectively. The breaking
patterns are as shown in Eqs.
(\ref{SU(6)sigmaSU(7)})-(\ref{SU(6)sigmaE_6}). To achieve
doublet-triplet splitting and make contact with the spectrum in
the MSSM, we consider $U(1)^2$ flux configurations $(L_1,L_2)$
already solved in the previous section. In this section we shall
study the spectrum from the curves and show that the
doublet-triplet splitting and non-minimal spectrum of the MSSM can
be achieved. A detailed example can be found in section $5.2.3$.

In local F-theory models, the gauge group on the curve along which
$S$ intersects with $S'$ will be enhanced at least by one rank. In
the present case of $G_S=SU(6)$, the possible enhanced gauge
groups are $SU(7)$, $SO(12)$ and $E_6$. The matter fields
transform as fundamental representation $\bf 6$, anti-symmetric
tensor representation of rank two ${\bf 15}$, and anti-symmetric
tensor representation of rank three ${\bf 20}$ in $SU(6)$ can be
engineered to localize on the curves with gauge groups $SU(7)$,
$SO(12)$, and $E_6$, respectively. In order to split doublet and
triplet states in Higgs and obtain the spectrum of the MSSM,
$L_{1\Sigma}$, $L_{2\Sigma}$ and $L'_{\Sigma}$ have to be
non-trivial, which breaks $G_{\Sigma}$ into $G_{\rm std}\times
U(1)^2$. The breaking patterns of $SU(7)$, $SO(12)$ and $E_6$ are
as follows:

\begin{equation}
\begin{array}{c@{}c@{}l@{}c@{}l}
SU(7) &~\rightarrow~& SU(6)\times U(1)' &~\rightarrow~& SU(3)\times SU(2)\times U(1)'\times U(1)_1\times U(1)_2\\
{\bf 48} &~\rightarrow~& {\bf 35}_0+{\bf 1}_0&~\rightarrow~& {\bf
(8,1)}_{0,0,0}+{\bf (1,3)}_{0,0,0}+{\bf (3,2)}_{0,-5,0}+{\bf (\bar
3,2)}_{0,5,0}\\& & +{\bf
6}_{-7}+{\bf\bar 6}_{7} &&+{\bf (1,1)}_{0,0,0}+{\bf (1,1)}_{0,0,0}+{\bf (1,2)}_{0,3,6}+{\bf (3,1)}_{0,-2,6}\\
& & & &+{\bf (1,\bar 2)}_{0,-3,-6}+{\bf (\bar 3,1)}_{0,2,-6}+{\bf
(1,1)}_{0,0,0}+{\bf (1,2)}_{-7,3,1}\\&&&&+{\bf
(3,1)}_{-7,-2,1}+{\bf (1,1)}_{-7,0,-5}+{\bf (1,\bar
2)}_{7,-3,-1}+{\bf (\bar 3,1)}_{7,2,-1}\\&&&&+{\bf
(1,1)}_{7,0,5}\label{MSSN breaking pattern SU(7)}
\end{array}
\end{equation}
\begin{equation}
\begin{array}{c@{}c@{}l@{}c@{}l}
SO(12) &~\rightarrow~& SU(6)\times U(1)' &~\rightarrow~&
SU(3)\times SU(2)\times U(1)'\times U(1)_1\times U(1)_2\\ {\bf 66}
&~\rightarrow~& {\bf 35}_0+{\bf 1}_0 &~\rightarrow~& {\bf
(8,1)}_{0,0,0}+{\bf (1,3)}_{0,0,0}+{\bf (3,2)}_{0,-5,0}+{\bf (\bar
3,2)}_{0,5,0}\\& & +{\bf
15}_2+{\bf\overline{15}}_{-2}  &&+{\bf (1,1)}_{0,0,0}+{\bf (1,1)}_{0,0,0}+{\bf (1,2)}_{0,3,6}+{\bf (3,1)}_{0,-2,6}\\
& & & &+{\bf (1,\bar 2)}_{0,-3,-6}+{\bf (\bar 3,1)}_{0,2,-6}+{\bf
(1,1)}_{0,0,0}+{\bf (1,2)}_{2,3,-4}\\&&&&+{\bf
(3,1)}_{2,-2,-4}+{\bf (1,1)}_{2,6,2}+{\bf (\bar
3,1)}_{2,-4,2}+{\bf (3,2)}_{2,1,2}\\&&&&+{\bf (1,\bar
2)}_{-2,-3,4}+{\bf (\bar 3,1)}_{-2,2,4}+{\bf
(1,1)}_{-2,-6,-2}+{\bf (3,1)}_{-2,4,-2}\\&&&&+{\bf (\bar 3,\bar
2)}_{-2,-1,-2}\label{MSSN breaking pattern SO(12)}
\end{array}
\end{equation}
\begin{equation}
\begin{array}{c@{}c@{}l@{}c@{}l}
E_{6} &~\rightarrow~& SU(6)\times U(1)' &~\rightarrow~& SU(3)\times SU(2)\times U(1)'\times U(1)_1\times U(1)_2\\
{\bf 78} &~\rightarrow~& {\bf 35}_0+{\bf 1}_0+{\bf 1}_{\pm2}
&~\rightarrow~& {\bf (8,1)}_{0,0,0}+{\bf (1,3)}_{0,0,0}+{\bf
(3,2)}_{0,-5,0}+{\bf (\bar 3,2)}_{0,5,0}\\& &+{\bf
20}_1+{\bf {20}}_{-1} &&+{\bf (1,1)}_{0,0,0}+{\bf (1,1)}_{0,0,0}+{\bf (1,2)}_{0,3,6}+{\bf (3,1)}_{0,-2,6}\\
& & & &+{\bf (1,\bar 2)}_{0,-3,-6}+{\bf (\bar 3,1)}_{0,2,-6}+{\bf
(1,1)}_{0,0,0}+{\bf (1,1)}_{\pm2,0,0}\\&&&&+[{\bf
(1,1)}_{1,6,-3}+{\bf (\bar 3,1)}_{1,-4,-3}+{\bf
(3,2)}_{1,1,-3}+c.c]\\&&&&+[{\bf (1,1)}_{-1,6,-3}+{\bf (\bar
3,1)}_{-1,-4,-3}+{\bf (3,2)}_{-1,1,-3}+c.c].\label{MSSN breaking
pattern E_6}
\end{array}
\end{equation}
Due to non-trivial $U(1)^2$ flux configurations on the bulk $S$,
the last two $U(1)$ charges of the fields on the curves should be
conserved in each Yukawa coupling. From the breaking patterns, we
list possible Yukawa couplings of type $\Sigma\Sigma S$ and
$\Sigma\Sigma\Sigma$ in Table \ref{MSSMYukawaSU(6)}. According to
Table \ref{MSSMYukawaSU(6)}, the possible field content is shown
in Table \ref{MSSM Field content SU(6)}. In what follows, we shall
focus on the case of $\Sigma\Sigma\Sigma$-type couplings and find
all possible field configurations supported by the curves
$\Sigma_{SU(7)}$, $\Sigma_{SO(12)}$, and $\Sigma_{E_6}$ with given
$U(1)^2$ flux configuration $(L_1,L_2)$.

\begin{table}[h]
\begin{center}
\renewcommand{\arraystretch}{1.10}
\begin{tabular}{|c|c|c|c|c|c|c|c|} \hline

Coupling & Representation & Configuration\\
\hline \hline

& ${\bf (3,2)}_{2,1,2}{\bf (\bar
3,1)}_{1,-4,-3}{\bf (1,2)}_{-7,3,1}$ & $\Sigma_{SO(12)}\Sigma_{E_6}\Sigma_{SU(7)}$\\
& ${\bf (3,2)}_{2,1,2}{\bf (\bar
3,1)}_{2,-4,2}{\bf (1,2)}_{2,3,-4}$ & $\Sigma_{SO(12)}\Sigma_{SO(12)}\Sigma_{SO(12)}$\\
$Q_Lu_RH_u$ & ${\bf (3,2)}_{1,1,-3}{\bf (\bar 3,1)}_{2,-4,2}{\bf
(1,2)}_{-7,3,1}$ &
$\Sigma_{E_6}\Sigma_{SO(12)}\Sigma_{SU(7)}$ \\
& ${\bf (3,2)}_{-1,1,-3}{\bf (\bar
3,1)}_{2,-4,2}{\bf (1,2)}_{-7,3,1}$ & $\Sigma_{E_6}\Sigma_{SO(12)}\Sigma_{SU(7)}$\\
& ${\bf (3,2)}_{1,1,-3}{\bf (\bar
3,1)}_{1,-4,-3}{\bf (1,2)}_{0,3,6}$ & $\Sigma_{E_6}\Sigma_{E_6}S$\\
& ${\bf (3,2)}_{-1,1,-3}{\bf (\bar
3,1)}_{1,-4,-3}{\bf (1,2)}_{0,3,6}$ & $\Sigma_{E_6}\Sigma_{E_6}S$\\
\hline

 & ${\bf (3,2)}_{2,1,2}{\bf (\bar
3,1)}_{7,2,-1}{\bf (1,\bar 2)}_{7,-3,-1}$ & $\Sigma_{SO(12)}\Sigma_{SU(7)}\Sigma_{SU(7)}$ \\
 & ${\bf (3,2)}_{2,1,2}{\bf (\bar
3,1)}_{0,2,-6}{\bf (1,\bar 2)}_{-2,-3,4}$ & $\Sigma_{SO(12)}S\Sigma_{SO(12)}$ \\
 & ${\bf (3,2)}_{1,1,-3}{\bf (\bar
3,1)}_{-2,2,4}{\bf (1,\bar 2)}_{7,-3,-1}$ & $\Sigma_{E_6}\Sigma_{SO(12)}\Sigma_{SU(7)}$\\
$Q_Ld_RH_d$ & ${\bf (3,2)}_{-1,1,-3}{\bf (\bar 3,1)}_{-2,2,4}{\bf
(1,\bar
2)}_{7,-3,-1}$ & $\Sigma_{E_6}\Sigma_{SO(12)}\Sigma_{SU(7)}$\\
& ${\bf (3,2)}_{1,1,-3}{\bf (\bar
3,1)}_{7,2,-1}{\bf (1,\bar 2)}_{-2,-3,4}$ & $\Sigma_{E_6}\Sigma_{SU(7)}\Sigma_{SO(12)}$\\
& ${\bf (3,2)}_{-1,1,-3}{\bf (\bar
3,1)}_{7,2,-1}{\bf (1,\bar 2)}_{-2,-3,4}$ & $\Sigma_{E_6}\Sigma_{SU(7)}\Sigma_{SO(12)}$\\
 & ${\bf (3,2)}_{2,1,2}{\bf (\bar
3,1)}_{-2,2,4}{\bf (1,\bar 2)}_{0,-3,-6}$ & $\Sigma_{SO(12)}\Sigma_{SO(12)}S$ \\

\hline

 & ${\bf (1,\bar 2)}_{7,-3,-1}{\bf (1,1)}_{2,6,2}{\bf (1,\bar 2)}_{7,-3,-1}$ & $\Sigma_{SU(7)}\Sigma_{SO(12)}\Sigma_{SU(7)}$  \\
 & ${\bf (1,\bar 2)}_{-2,-3,4}{\bf (1,1)}_{1,6,-3}{\bf (1,\bar
2)}_{7,-3,-1}$ & $\Sigma_{SO(12)}\Sigma_{E_6}\Sigma_{SU(7)}$\\
$L_Le_RH_d$ & ${\bf (1,\bar 2)}_{-2,-3,4}{\bf (1,1)}_{-1,6,-3}{\bf
(1,\bar
2)}_{7,-3,-1}$ & $\Sigma_{SO(12)}\Sigma_{E_6}\Sigma_{SU(7)}$\\
&  ${\bf (1,\bar 2)}_{7,-3,-1}{\bf (1,1)}_{1,6,-3}{\bf (1,\bar
2)}_{-2,-3,4}$ &
$\Sigma_{SU(7)}\Sigma_{E_6}\Sigma_{SO(12)}$\\
& ${\bf (1,\bar 2)}_{7,-3,-1}{\bf (1,1)}_{-1,6,-3}{\bf (1,\bar
2)}_{-2,-3,4}$ &
$\Sigma_{SU(7)}\Sigma_{E_6}\Sigma_{SO(12)}$\\
 & ${\bf (1,\bar 2)}_{-2,-3,4}{\bf (1,1)}_{2,6,2}{\bf (1,\bar
2)}_{0,-3,-6}$ & $\Sigma_{SO(12)}\Sigma_{SO(12)}S$\\
\hline

 & ${\bf (1,\bar
2)}_{7,-3,-1}{\bf (1,1)}_{0,0,0}{\bf (1,2)}_{-7,3,1}$ & $\Sigma_{SU(7)}S\Sigma_{SU(7)}$ \\
 $L_LN_RH_u$ & ${\bf (1,\bar 2)}_{-2,-3,4}{\bf (1,1)}_{-7,0,-5}{\bf (1,2)}_{-7,3,1}$ & $\Sigma_{SO(12)}\Sigma_{SU(7)}\Sigma_{SU(7)}$\\
& ${\bf (1,\bar 2)}_{7,-3,-1}{\bf (1,1)}_{7,0,5}{\bf (1,2)}_{2,3,-4}$ & $\Sigma_{SU(7)}\Sigma_{SU(7)}\Sigma_{SO(12)}$\\
& ${\bf (1,\bar 2)}_{-2,-3,4}{\bf (1,1)}_{0,0,0}{\bf (1,
2)}_{2,3,-4}$ &
$\Sigma_{SU(7)}S\Sigma_{SU(7)}$\\
& ${\bf (1,\bar
2)}_{7,-3,-1}{\bf (1,1)}_{-7,0,-5}{\bf (1,2)}_{0,3,6}$ & $\Sigma_{SU(7)}\Sigma_{SU(7)}S$ \\

\hline
\end{tabular}
\caption{The Yukawa couplings of the MSSM model from $G_S=SU(6)$.}
\label{MSSMYukawaSU(6)}
\end{center}
\end{table}

\begin{table}[h]
\begin{center}
\renewcommand{\arraystretch}{1.10}
\begin{tabular}{|c|@{\,}l@{}|@{\,}l@{}|@{\,}l@{}|@{\,}l@{}|@{\,}l@{}|@{\,}l@{}|@{\,}l@{}|} \hline

& ~~~~~\,$Q_L$ & ~~~~~~$u_R$ & ~~~~~$d_R$ & ~~~~~~$e_R$ & ~~~~~\,$L_L$ & ~~~~~\,$H_u$ & ~~~~~\,$H_d$ \\
\hline

$M_0$ & ${\bf (3,2)}_{2,1,2}$ & ${ \bf (\bar 3,1)}_{2,-4,2}$ &
${\bf (\bar 3,1)}_{0,2,-6}$ & ${\bf (1,1)}_{1,6,-3}$ & ${\bf
(1,\bar 2)}_{7,-3,-1}$ & ${\bf (1,2)}_{-2,3,-4}$ & ${\bf (1,\bar
2)}_{-2,-3,4}$ \\
\hline

$M_1$ & ${\bf (3,2)}_{2,1,2}$ & ${ \bf (\bar 3,1)}_{1,-4,-3}$ &
${\bf (\bar 3,1)}_{7,2,-1}$ & ${\bf (1,1)}_{2,6,2}$ & ${\bf
(1,\bar 2)}_{7,-3,-1}$ & ${\bf (1,2)}_{-7,3,1}$ & ${\bf (1,\bar
2)}_{7,-3,-1}$ \\
\hline

$M_2$ & ${\bf (3,2)}_{1,1,-3}$ & ${\bf (\bar 3,1)}_{2,-4,2}$ &
${\bf (\bar 3,1)}_{-2,2,4}$ & ${\bf (1,1)}_{2,6,2}$ & ${\bf
(1,\bar 2)}_{7,-3,-1}$ & ${\bf (1,2)}_{-7,3,1}$ & ${\bf (1,\bar
2)}_{7,-3,-1}$ \\
\hline

$M_3$ & ${\bf (3,2)}_{-1,1,-3}$ & ${\bf (\bar 3,1)}_{2,-4,2}$ &
${\bf (\bar 3,1)}_{-2,2,4}$ & ${\bf (1,1)}_{2,6,2}$ & ${\bf
(1,\bar 2)}_{7,-3,-1}$ & ${\bf (1,2)}_{-7,3,1}$ & ${\bf (1,\bar
2)}_{7,-3,-1}$\\
\hline

$M_4$ & ${\bf (3,2)}_{2,1,2}$ & ${\bf (\bar 3,1)}_{1,-4,-3}$ &
${\bf (\bar 3,1)}_{7,2,-1}$ & ${\bf (1,1)}_{-1,6,-3}$ & ${\bf
(1,\bar 2)}_{-2,-3,4}$ & ${\bf (1,2)}_{-7,3,1}$ & ${\bf (1,\bar
2)}_{7,-3,-1}$ \\
\hline

$M_5$ & ${\bf (3,2)}_{1,1,-3}$ & ${\bf (\bar 3,1)}_{2,-4,2}$ &
${\bf (\bar 3,1)}_{-2,2,4}$ & ${\bf (1,1)}_{-1,6,-3}$ & ${\bf
(1,\bar 2)}_{-2,-3,4}$ & ${\bf (1,2)}_{-7,3,1}$ & ${\bf (1,\bar
2)}_{7,-3,-1}$ \\
\hline

$M_6$ & ${\bf (3,2)}_{-1,1,-3}$ & ${\bf (\bar 3,1)}_{2,-4,2}$ &
${\bf (\bar 3,1)}_{-2,2,4}$ & ${\bf (1,1)}_{-1,6,-3}$ & ${\bf
(1,\bar 2)}_{-2,-3,4}$ & ${\bf (1,2)}_{-7,3,1}$ & ${\bf (1,\bar
2)}_{7,-3,-1}$ \\
\hline

\end{tabular}
\caption{Field content in the MSSM from $G_S=SU(6)$} \label{MSSM
Field content SU(6)}
\end{center}
\end{table}
Let us start with the case of $\Sigma_{SU(7)}$ and consider
$(\alpha_1,\alpha_2,\alpha_3)=(k,k,0)$ with $k=0,1,2,3$. When
$(\alpha_1,\alpha_2,\alpha_3)=(0,0,0)$, which is the second case
in Table {\ref{SU(6) bulk01}}, it is clear that we have
$L_2=\mathcal{O}_{S}( 5E_l-2E_i-3E_j)^{1/30}$ or
$L_2=\mathcal{O}_{S}( -5E_l+3E_i+2E_j)^{1/30}$. We define
$(n_1,n_2,n_3)=(N_{{\bf (\bar 3,1)}_{7,2,-1}},N_{{\bf (1,\bar
2)}_{7,-3,-1}},N_{{\bf (1,1)}_{7,0,5}})$. To avoid exotic fields,
we require that $n_1\in \mathbb{Z}_{\geqslant 0}$. Given field
configurations $(n_1,n_2,n_3)$ on the curve $\Sigma_{SU(7)}$, the
necessary conditions\footnote{$L_{1\Sigma_{SU(7)}}=
\mathcal{O}_{\Sigma_{SU(7)}}(\frac{1}{5}(n_1-n_2))$,\;$L_{2\Sigma_{SU(7)}}=\mathcal{O}_{\Sigma_{SU(7)}}(\frac{1}{30}(-3n_1-2n_2+5n_3))$,
and
$L'_{\Sigma_{SU(7)}}=\mathcal{O}_{\Sigma_{SU(7)}}(\frac{1}{42}(3n_1+2n_2+n_3))$.}
for the homological class of the curve $\Sigma_{SU(7)}$ are
\begin{equation}
\left\{\begin{array}{l} (E_i-E_j)\cdot\Sigma_{SU(7)}=n_2-n_1\\
(E_i-E_l)\cdot\Sigma_{SU(7)}=n_2-n_3.\label {sigma SU(7)
condition01}
\end{array}   \right.
\end{equation}
if $L_2=\mathcal{O}_{S}( 5E_l-2E_i-3E_j)^{1/30}$. For the case of
$L_2=\mathcal{O}_{S}( -5E_l+3E_i+2E_j)^{1/30}$, the conditions are
as follows:
\begin{equation}
\left\{\begin{array}{l} (E_i-E_j)\cdot\Sigma_{SU(7)}=n_2-n_1\\
(E_i-E_l)\cdot\Sigma_{SU(7)}=n_3-n_1.\label {sigma SU(7)
condition02}
\end{array}   \right.
\end{equation}
Note that the first condition of Eq. ({\ref{sigma SU(7)
condition01}}) and Eq. (\ref{sigma SU(7) condition02}) is
universal since it comes from the restriction of the universal
supersymmetric line bundle $L_1=\mathcal{O}_{S}(E_j-E_i)^{1/5}$ to
the curve $\Sigma_{SU(7)}$. Note that there are no further
constraints for $n_i,\;i=1,2,3$ except $n_1\in
\mathbb{Z}_{\geqslant 0}$, $n_1\neq n_2$, $3n_1+2n_2\neq 5n_3$ and
$3n_1+2n_2+n_3\neq 0$. The last three constraints follow from the
conditions $L_{1\Sigma}\neq \mathcal{O}_{\Sigma}$,
$L_{2\Sigma}\neq \mathcal{O}_{\Sigma}$, and $L'_{\Sigma}\neq
\mathcal{O}_{\Sigma}$. Let us look at an example. Consider the
case of $(n_1,n_2,n_3)=(0,1,0)$, Eq. (\ref{sigma SU(7)
condition01}) and Eq. (\ref{sigma SU(7) condition02}) can be
easily solved by $\Sigma=H-E_{i}-E_m$ and $\Sigma=H-E_{i}-E_l$,
respectively. In this case, double and triplet states in the Higgs
field ${\bf {\bar 5}}_{7,-1}$ can be split without producing
exotic fields. Let us look at one more case,
$(\alpha_1,\alpha_2,\alpha_3)=(3,3,0)$. It follows from Table
{\ref{SU(6) bulk01}} that $L_2=\mathcal{O}_{S}(E_j-E_i)^{7/30}$.
The conditions for the homological class of the curve
$\Sigma_{SU(7)}$ to support the field configurations
$(n_1,n_2,n_3)$ are
\begin{equation}
\left\{\begin{array}{l} (E_i-E_j)\cdot\Sigma_{SU(7)}=n_2-n_1\\
2n_1=n_2+n_3.\label {sigma SU(7) condition03}
\end{array}   \right.
\end{equation}
This time we get one more constraint, $2n_1=n_2+n_3$. It follows
that when ${\bf (\bar 3,1)}_{7,2,-1}$ vanishes, the doublets
always show up together with singlets. For the cases of
$(\alpha_1,\alpha_2,\alpha_3)=(k,k,0)$ with $k=1,2$, we summarize
the results\footnote{For simplicity, we are not going to show the
universal conditions $(E_i-E_j)\cdot\Sigma=w_2-w_1,\;w\in \{n,s\}$
for $\Sigma_{SU(7)}$ and $\Sigma_{SO(12)}$, respectively and
$(E_i-E_j)\cdot\Sigma=p_3-p_1$ for $\Sigma_{E_6}$ in Table
\ref{MSSM cond. SU(7)}, \ref{MSSM cond. SO(12)}, and \ref{MSSM
cond. E_6}.}in Table \ref{MSSM cond. SU(7)}

\begin{table}[h]
\begin{center}
\renewcommand{\arraystretch}{1.25}
\begin{tabular}{|c|c|c|c|c|c|c|c|} \hline

$(\alpha_1,\alpha_2,\alpha_3)$ & Conditions & $L_2$\\
\hline
$(0,0,0)$& $(E_i-E_l)\cdot\Sigma_{SU(7)}=n_2-n_3$ & $\mathcal{O}_{S}(5E_l-2E_i-3E_j)^{1/30}$\\
& $(E_i-E_l)\cdot\Sigma_{SU(7)}=n_3-n_1$ & $\mathcal{O}_{S}(-5E_l+3E_i+2E_j)^{1/30}$\\
\hline
$~(1,1,0)^{\ast}$ & $n_2=n_3$ & $\mathcal{O}_{S}(E_i-E_j)^{1/10}$\\
\hline
$(1,1,0)$ & $([E_l,E_m]'')\cdot\Sigma_{SU(7)}=n_3-n_1$ & $\mathcal{O}_{S}(5[E_l,E_m]''-2E_i+2E_j)^{1/30}$\\
\hline
$(2,2,0)$ & $(-E_l+E_j)\cdot\Sigma_{SU(7)}=n_3-n_1$ & $\mathcal{O}_{S}(-5E_l-2E_i+7E_j)^{1/30}$\\
& $(E_l-E_i)\cdot\Sigma_{SU(7)}=n_3-n_1$ & $\mathcal{O}_{S}(5E_l-7E_i+2E_j)^{1/30}$\\
\hline
$(3,3,0)$ & $2n_1=n_2+n_3$ & $\mathcal{O}_{S}(E_j-E_i)^{7/30}$\\
\hline
\end{tabular}
\caption{The conditions for  $\Sigma_{SU(7)}$ supporting the field
configurations $(n_1,n_2,n_3)$ with
$L_1=\mathcal{O}_{S}(E_j-E_i)^{1/5}$.} \label{MSSM cond. SU(7)}
\end{center}
\end{table}
Similarly, we can extend the calculation to the curve
$\Sigma_{SO(12)}$. Let us define $(s_1,s_2,s_3,s_4,s_5)=(N_{{\bf
(3,2)}_{2,1,2}},N_{{\bf (\bar 3,1)}_{2,-4,2}},N_{{\bf
(3,1)}_{2,-2,-4}},N_{{\bf (1,2)}_{2,3,-4}},N_{{\bf
(1,1)}_{2,6,2}}) $ and consider the case of
$(\alpha_1,\alpha_2,\alpha_3)=(1,1,0)$, which is the third case in
Table {\ref{SU(6) bulk01}}. It is clear that we have
$L_2=\mathcal{O}_{S}( 5[E_l,E_m]''-2E_i+2E_j)^{1/30}$. The
necessary
conditions\footnote{$L_{1\Sigma_{SO(12)}}=\mathcal{O}_{\Sigma_{SO(12)}}(\frac{1}{5}(s_1-s_2))$,\;$L_{2\Sigma_{SO(12)}}=
\mathcal{O}_{\Sigma_{SO(12)}}(\frac{1}{30}(2s_1+3s_2-5s_3))$, and
$L'_{\Sigma_{SO(12)}}=\mathcal{O}_{\Sigma_{SO(12)}}(\frac{1}{6}(2s_1+s_3))$.}
for the homological class of the curve $\Sigma_{SO(12)}$ with
field configurations $(s_1,s_2,s_3,s_4,s_5)$ are
\begin{equation}
\left\{\begin{array}{l} (E_i-E_j)\cdot\Sigma_{SO(12)}=s_2-s_1\\
([E_l,E_m]'')\cdot\Sigma_{SO(12)}=s_2-s_3,\label {sigma SO(12)
condition01}
\end{array}   \right.
\end{equation}
and
\begin{equation}
\left\{\begin{array}{l} s_4=s_3+s_1-s_2\\s_5=2s_1-s_2.\label
{sigma SO(12) condition01_01}
\end{array}   \right.
\end{equation}
Note that Eq. (\ref{sigma SO(12) condition01_01}) impose severe
restrictions on the configurations $(s_1,s_2,s_3,s_4,s_5)$. For
example, one cannot simply set $(s_1,s_2,s_3,s_4,s_5)=(0,0,0,m,0)$
to achieve the doublet-triplet splitting of Higgs ${\bf
5}_{2,-4}$; it is easy to see that $m$ is forced to be zero by the
constraints in Eq. (\ref{sigma SO(12) condition01_01}). This will
cause trouble when we attempt to engineer the Higgs on the curve
$\Sigma_{SO(12)}$ with doublet-triplet splitting. Consider the
case of $s_4> 0$ and set $s_1=0$. From the constraints in Eq.
(\ref{sigma SO(12) condition01_01}), we obtain $s_2+(-s_3)<0$.
Note that to avoid exotic fields from $\Sigma_{SO(12)}$, it is
required that $s_1,\; s_2\in \mathbb{Z}_{\geqslant 0}$ and $s_3\in
\mathbb{Z}_{\leqslant 0}$. It follows that $0\leqslant
s_2+(-s_3)<0$, which leads to a contradiction. As a result, the
appearance of ${\bf (3,2)}_{2,1,2}$ cannot be avoided on the curve
$\Sigma_{SO(12)}$ as $N_{{\bf (1,2)}_{2,3,-4}}=s_4>0$. If $s_4>0$,
actually the most general non-trivial configurations are
$(s_1,s_2,s_3,s_4,s_5)=(l,l+n-m,n,m,l+m-n)$, where $m,\; l\in
\mathbb{Z}_{>0}$ and $m-l\leqslant n\leqslant 0$. Note that ${\bf
(3,2)}_{2,1,2}$ is treated as matter in the MSSM, which requires
that\footnote{We allow the cases in which three copies of matter
fields can be distributed over different matter curves.}
$l\leqslant 3$. It follows that $1\leqslant m \leqslant 3$ and
$m\leqslant l \leqslant 3$. It turns out that there are finitely
many non-trivial configurations. More precisely, the field
configurations are as follows:
\begin{equation}
(s_1,s_2,s_3,s_4,s_5)= \left\{\begin{array}{l} (1,0,0,1,2),
(2,1,0,1,3),(2,0,-1,1,4),\\(3,2,0,1,4),
(3,1,-1,1,5),(3,0,-2,1,6),\\(2,0,0,2,4),(3,1,0,2,5),
(3,0,-1,2,6),\\(3,0,0,3,6)\label {Field content SO(12)01}
\end{array}   \right\}.
\end{equation}
If $-3\leqslant s_4\leqslant 0$, with $0\leqslant s_1,\;
s_2\leqslant 3$ and $-3\leqslant s_3\leqslant 0$, we have another
branch of the configurations as follows:
\begin{equation}
(s_1,s_2,s_3,s_4,s_5)= \left\{\begin{array}{l} (0,1,-1,-2,-1),
(0,1,-2,-3,-1),(0,2,-1,-3,-2),\\(1,0,-1,0,2),(1,0,-3,-2,2),
(1,2,0,-1,0),\\(1,2,-1,-2,0),(1,3,0,-2,-1),
(1,3,-1,-3,-1),\\(2,0,-2,0,4),(2,0,-3,-1,4),(2,1,-2,-1,3),\\
(2,3,0,-1,1),(2,1,-3,-2,3),(2,3,-1,-2,1),\\
(2,1,-1,0,3),(2,3,-2,-3,1),(3,0,-3,0,6),\\
(3,1,-2,0,5),(3,1,-3,-1,5),(3,2,-1,0,4),\\
(3,2,-2,-1,4),(3,2,-3,-2,4)\label
{Field content SO(12)02}
\end{array}   \right\},
\end{equation}
where all configurations\footnote{$s_3<0$ represents $N_{{\bf
(3,1)}_{2,-2,-4}}=0$ and $N_{{\bf (\bar 3,1)}_{-2,2,4}}=-s_3$. The
same rule can be applied to other $s_i$. } in (\ref{Field content
SO(12)01}) and (\ref{Field content SO(12)02}) satisfy the
conditions $L_{1\Sigma}\neq \mathcal{O}_{\Sigma}$,
$L_{2\Sigma}\neq \mathcal{O}_{\Sigma}$, and $L'_{\Sigma}\neq
\mathcal{O}_{\Sigma}$. With these configurations, one can solve
the conditions for the intersection numbers, namely, the
conditions in Eq. (\ref{sigma SO(12) condition01}). Let us
consider the case of $(s_1,s_2,s_3,s_4,s_5)=(1,0,0,1,2)$, it is
clear that $\Sigma=2H-E_l-E_m-E_j$ is a solution. For a more
complicated case, for example
$(s_1,s_2,s_3,s_4,s_5)=(3,1,-1,1,5)$, the conditions can be solved
by $\Sigma=4H+E_p-2E_j-2E_l$ if $[E_l,E_m]''=E_l-E_m$ and by
$\Sigma=4H+E_p-2E_j-2E_m$ if $[E_l,E_m]''=E_m-E_l$.

Let us turn to another case. Consider the first case in Table
{\ref{SU(6) bulk01}}, namely
$(\alpha_1,\alpha_2,\alpha_3)=(1,1,0)^{\ast}$. The supersymmetric
fractional line bundle $L_2$ is $\mathcal{O}_{S}(E_i-E_j)^{1/10}$.
The necessary conditions are
\begin{equation}
\left\{\begin{array}{l} (E_i-E_j)\cdot\Sigma_{SO(12)}=s_2-s_1\\
s_1=s_3,\label {sigma SO(12) condition02}
\end{array}   \right.
\end{equation}
and Eq. (\ref{sigma SO(12) condition01_01}). Note that ${\bf (\bar
3,2)}_{-2,-1,-2}$ and ${\bf (3,1)}_{2,-2,-4}$ are exotic fields in
the MSSM. The constraint, $s_1=s_3$ in Eq. (\ref{sigma SO(12)
condition02}) and Eq. (\ref{sigma SO(12) condition01_01}) imply
that $s_1=s_3=0$. If $s_4\geqslant 0$, by the constraints in Eq.
(\ref{sigma SO(12) condition01_01}), we obtain
$(s_1,s_2,s_3,s_4,s_5)=(0,0,0,0,0)$. If $s_4<0$, we have general
configurations $(s_1,s_2,s_3,s_4,s_5)=(0,n,0,-n,-n)$, where
$1\leqslant n\leqslant 3$. However, these configurations violate
the condition $L'_{\Sigma}\neq \mathcal{O}_{\Sigma}$. As a check,
using the configurations in (\ref{Field content SO(12)01}),
(\ref{Field content SO(12)02}), and taking the condition $s_1=s_3$
into account, one can see that there are no solutions in this
case.

Next we consider the fifth case in Table {\ref{SU(6) bulk01}},
namely $(\alpha_1,\alpha_2,\alpha_3)=(3,3,0)$. In this case, $L_2$
is $\mathcal{O}_{S}(E_j-E_i)^{7/30}$. The necessary conditions are
\begin{equation}
\left\{\begin{array}{l} (E_i-E_j)\cdot\Sigma_{SO(12)}=s_2-s_1\\
2s_2=s_1+s_3,\label {sigma SO(12) condition03}
\end{array}   \right.
\end{equation}
and Eq. (\ref{sigma SO(12) condition01_01}). It is easy to see
that $s_2=s_4$. If $s_2=0$, we obtain the non-trivial
configurations $(s_1,s_2,s_3,s_4,s_5)=(k,0,-k,0,2k)$, where
$1\leqslant k\leqslant 3$. Note that these configurations satisfy
the conditions, $L_{1\Sigma}\neq \mathcal{O}_{\Sigma}$,
$L_{2\Sigma}\neq \mathcal{O}_{\Sigma}$, and $L'_{\Sigma}\neq
\mathcal{O}_{\Sigma}$. Let us turn to the case of $s_2=m\in
\mathbb{Z}_{>0}$. The general configurations are
$(s_1,s_2,s_3,s_4,s_5)=(l,m,2m-l,m,2l-m)$ with $l\geqslant 2m>0$.
Note that ${\bf (3,2)}_{2,1,2}$ is treated as matter in the MSSM.
As a result, we focus on the case of $l\leqslant 3$, which implies
that $m=1$ and $l=2,3$. It turns out that the allowed
configurations are
$(s_1,s_2,s_3,s_4,s_5)=\{(2,1,0,1,3),(3,1,-1,1,5)\}$, where the
configurations satisfy the conditions $L_{1\Sigma}\neq
\mathcal{O}_{\Sigma}$, $L_{2\Sigma}\neq \mathcal{O}_{\Sigma}$, and
$L'_{\Sigma}\neq \mathcal{O}_{\Sigma}$. Putting these two branches
together, we obtain
\begin{equation}
(s_1,s_2,s_3,s_4,s_5)=\left\{\begin{array}{l}
(1,0,-1,0,2),(2,0-2,0,4),(3,0,-3,0,6),\\(2,1,0,1,3),(3,1,-1,1,5)\label{Field
content SO(12)04}
\end{array} \right\}.
\end{equation}
As a check, from the field configurations in (\ref{Field content
SO(12)01}), (\ref{Field content SO(12)02}) and the constraint
$2s_2=s_1+s_3$, one can find that there are exactly five solutions
as shown in (\ref{Field content SO(12)04}).

Let us take a look at some solutions for the curve satisfying Eq.
(\ref{sigma SO(12) condition03}).  For the the case of
$(s_1,s_2,s_3,s_4,s_5)=(2,1,0,1,3)$ , it is easy to see that
$\Sigma=H-E_j-E_s$ solves the first equation in Eq. (\ref{sigma
SO(12) condition03}). For the case of
$(s_1,s_2,s_3,s_4,s_5)=(2,0,-2,0,4)$, $\Sigma=3H-2E_j-E_p$ can be
a solution. From these examples, we expect that if we choose
$\Sigma_{SO(12)}$ to house Higgs fields, it will be difficult to
achieve doublet-triple splitting without introducing extra chiral
fields. For other $U(1)^2$ flux configurations corresponding to
the case of $(\alpha_1,\alpha_2,\alpha_3)=(k,k,0)$ with $k=0,2$,
the analysis is similar to the case of $k=1$. We summarize the
results in Table {\ref{MSSM cond. SO(12)}}.

\begin{table}[h]
\begin{center}
\renewcommand{\arraystretch}{1.25}
\begin{tabular}{|c|c|c|c|c|c|c|c|} \hline

$(\alpha_1,\alpha_2,\alpha_3)$ & Conditions & $L_2$\\
\hline
$(0,0,0)$& $(E_i-E_l)\cdot\Sigma_{SO(12)}=s_3-s_1$ & $\mathcal{O}_{S}(5E_l-2E_i-3E_j)^{1/30}$\\
& $(E_i-E_l)\cdot\Sigma_{SO(12)}=s_2-s_3$ & $\mathcal{O}_{S}(-5E_l+3E_i+2E_j)^{1/30}$\\
\hline
$~(1,1,0)^{\ast}$ & $s_1=s_3$ & $\mathcal{O}_{S}(E_i-E_j)^{1/10}$\\
\hline
$(1,1,0)$ & $([E_l,E_m]'')\cdot\Sigma_{SO(12)}=s_2-s_3$ & $\mathcal{O}_{S}(5[E_l,E_m]''-2E_i+2E_j)^{1/30}$\\
\hline
$(2,2,0)$ & $(-E_l+E_j)\cdot\Sigma_{SO(12)}=s_2-s_3$ & $\mathcal{O}_{S}(-5E_l-2E_i+7E_j)^{1/30}$\\
& $(E_l-E_i)\cdot\Sigma_{SO(12)}=s_2-s_3$ & $\mathcal{O}_{S}(5E_l-7E_i+2E_j)^{1/30}$\\
\hline
$(3,3,0)$ & $2s_2=s_1+s_3$ & $\mathcal{O}_{S}(E_j-E_i)^{7/30}$\\
\hline

\end{tabular}
\caption{The conditions for $\Sigma_{SO(12)}$ supporting the field
configurations $(s_1,s_2,s_3,s_4,s_5)$ with
$L_1=\mathcal{O}_{S}(E_j-E_i)^{1/5}$ and constraints
$2s_1=s_2+s_5$, $s_4=s_3+s_1-s_2$.} \label{MSSM cond. SO(12)}
\end{center}
\end{table}
In addition to doublet-triplet splitting problem, we also would
like to study the matter spectrum. According to Table \ref{MSSM
Field content SU(6)}, the matter fields can come from the curves
$\Sigma_{SU(7)}$, $\Sigma_{SO(12)}$, and $\Sigma_{E_6}$. The
configurations of the fields and the conditions of the
intersection numbers on the curves $\Sigma_{SU(7)}$ and
$\Sigma_{SO(12)}$ have been studied earlier in this section. Next
we are going to analyze the case of $\Sigma_{E_6}$. Note that for
the case of $M_0$ in Table \ref{MSSM Field content SU(6)}, to
engineer $3\times d_R$ on the bulk, it is required to set
$\alpha_3=3$. However, it gives rise to exotic fields ${\bf
(1,2)}_{3,6}$ and ${\bf (1,\bar 2)}_{-3,-6}$ on the bulk. In what
follows, we are going to focus on the case of
$(\alpha_1,\alpha_2,\alpha_3)=(k,k,0)$ on the bulk.

Let us start with the case of
$(\alpha_1,\alpha_2,\alpha_3)=(0,0,0)$. It is clear that
$L_2=\mathcal{O}_{S}( 5E_l-2E_i-3E_j)^{1/30}$ or
$L_2=\mathcal{O}_{S}( -5E_l+3E_i+2E_j)^{1/30}$. We define
$(p_1,p_2,p_3,p_4,p_5,p_6)=(N_{{\bf (3,2)}_{1,1,-3}}, N_{{\bf
(3,2)}_{-1,1,-3}},N_{{\bf (\bar 3,1)}_{1,-4,-3}},N_{{\bf (\bar
3,1)}_{-1,-4,-3}},N_{{\bf (1,1)}_{1,6,-3}},N_{{\bf
(1,1)}_{-1,6,-3}})$. The necessary
conditions\footnote{$L_{1\Sigma_{E_6}}=\mathcal{O}_{\Sigma_{E_6}}(\frac{1}{5}(p_1-p_3))$,
\;$L_{2\Sigma_{E_6}}=\mathcal{O}_{\Sigma_{E_6}}(-\frac{1}{30}(3p_1+5p_2+2p_3))$,
and
$L'_{\Sigma_{E_6}}=\mathcal{O}_{\Sigma_{E_6}}(\frac{1}{2}(p_1-p_2))$.}
for the curve $\Sigma_{E_6}$ are as follows:
\begin{equation}
\left\{\begin{array}{l} (E_i-E_j)\cdot\Sigma_{E_6}=p_3-p_1\\
(E_i-E_l)\cdot\Sigma_{E_6}=p_2+p_3
 ,\label {sigma E_6 condition01}
\end{array}   \right.
\end{equation}
and
\begin{equation}
\left\{\begin{array}{l}
p_4=p_2+p_3-p_1\\p_5=2p_1-p_3\\p_6=p_1+p_2-p_3
 ,\label {sigma E_6 condition01_01}
\end{array}   \right.
\end{equation}
if $L_2=\mathcal{O}_{S}( 5E_l-2E_i-3E_j)^{1/30}$. For the case of
$L_2=\mathcal{O}_{S}( -5E_l+3E_i+2E_j)^{1/30}$, the conditions are
\begin{equation}
\left\{\begin{array}{l} (E_i-E_j)\cdot\Sigma_{E_6}=p_3-p_1\\
(E_i-E_l)\cdot\Sigma_{E_6}=-p_1-p_2,\label {sigma E_6 condition02}
\end{array}   \right.
\end{equation}
and Eq. (\ref{sigma E_6 condition01_01}), where
$L_1=\mathcal{O}_{S}(E_j-E_i)^{1/5}$ has been used. Note that the
first condition in Eq. ({\ref{sigma E_6 condition01}}) and Eq.
(\ref{sigma E_6 condition01_01}) are universal since they come
from the restriction of the universal supersymmetric line bundle
$L_1=\mathcal{O}_{S}(E_j-E_i)^{1/5}$ to the curve $\Sigma_{E_6}$
and from the consistency of the definition of
$(p_1,p_2,p_3,p_4,p_5,p_6)$, respectively and that Eq. (\ref{sigma
E_6 condition01_01}) impose severe restrictions on the
configurations $(p_1,p_2,p_3,p_4,p_5,p_6)$. For example, one can
simply set $(p_1,p_2,p_3,p_4,p_5,p_6)=(n,0,0,0,0,0)$ to engineer
$n$ copies of ${\bf (3,2)}_{1,1,-3}$ on the curve $\Sigma_{E_6}$.
Then by constraints in Eq. (\ref{sigma E_6 condition01_01}), $n$
is forced to be vanishing in order to avoid the exotic fields. Let
us look at some examples of the non-trivial configurations. It is
easy to see that if $p_1=p_3=0$, we obtain non-trivial
configurations $(p_1,p_2,p_3,p_4,p_5,p_6)=(0,l,0,l,0,l)$, where
$l\in \mathbb{Z}_{> 0}$. When $p_2=p_4=0$, the non-trivial
configurations are $(p_1,p_2,p_3,p_4,p_5,p_6)=(m,0,m,0,m,0)$ with
$m\in \mathbb{Z}_{> 0}$. If $p_3=p_4=0$, it follows that
$(p_1,p_2,p_3,p_4,p_5,p_6)=(n,n,0,0,2n,2n)$, where $n\in
\mathbb{Z}_{>0}$. However, these configurations violate the
conditions $L_{1\Sigma}\neq \mathcal{O}_{\Sigma}$,
$L_{2\Sigma}\neq \mathcal{O}_{\Sigma}$ and $L'_{\Sigma}\neq
\mathcal{O}_{\Sigma}$. Therefore, we need to find more general
non-trivial configurations. For the matter fields in the MSSM, we
require that the number of the matter field is equal to or less
than three. As a result, we impose the conditions $1\leqslant
p_i\leqslant 3,\;i=1,2,3,4$ in this case. By the constraints in
Eq. ({\ref{sigma E_6 condition01_01}}), we obtain the following
configurations
\begin{equation}
(p_1,p_2,p_3,p_4,p_5,p_6)= \left\{\begin{array}{l}
(0,r,1-r,1,r-1,2r-1),(1,r,1-r,0,r+1,2r),\\
(0,q,2-q,2,q-2,2q-2),(1,q,2-q,1,q,2q-1),\\(2,q,2-q,0,q+2,2q),
(0,v,3-v,3,v-3,2v-3),\\(1,v,3-v,2,v-1,2v-2),(2,v,3-v,1,v+1,2v-1),\\
(3,v,3-v,0,v+3,2v),(1,t,4-t,3,t-2,2t-3),\\(2,t,4-t,2,t,2t-2),
(3,t,4-t,1,t+2,2t-1),\\(2,u,5-u,3,u-1,2u-3),(3,u,5-u,2,u+1,2u-2),\\
(3,3,3,3,3,3)\label {Field content E_6 01}
\end{array}   \right\},
\end{equation}
where $r=0,1$, $q=0,1,2$, $v=0,1,2,3$, $t=1,2,3$, and $u=2,3$.
Taking the conditions of $L_{1\Sigma}\neq \mathcal{O}_{\Sigma}$,
$L_{2\Sigma}\neq \mathcal{O}_{\Sigma}$ and $L'_{\Sigma}\neq
\mathcal{O}_{\Sigma}$ into account, the resulting configurations
are as follows:
\begin{equation}
(p_1,p_2,p_3,p_4,p_5,p_6)= \left\{\begin{array}{l}
(0,1,1,2,-1,0),(1,0,2,1,0,-1),(1,2,0,1,2,3),\\
(2,1,1,0,3,2),(0,1,2,3,-2,-1), (0,2,1,3,-1,1),\\
(1,3,0,2,2,4),(1,0,3,2,-1,-2),(2,0,3,1,1,-1),\\
(2,3,0,1,4,5),(3,1,2,0,4,2),(3,2,1,0,5,4),\\
(1,2,2,3,0,1),(2,1,3,2,1,0),(2,3,1,2,3,4),\\
(3,2,2,1,4,3)\label {Field content E_6 02}
\end{array}   \right\}.
\end{equation}
Once we get allowed configurations, it is not difficult to
calculate the homological classes of the curves, which satisfy Eq.
(\ref{sigma E_6 condition01}) or Eq. (\ref{sigma E_6
condition02}). For example, consider the case of
$(p_1,p_2,p_3,p_4,p_5,p_6)=(0,1,1,2,-1,0)$,  one can check that
$\Sigma=3H-E_i+E_l$ solves Eq. (\ref{sigma E_6 condition01}). Let
us look at one more complicated example,
$(p_1,p_2,p_3,p_4,p_5,p_6)=(3,2,2,1,4,3)$. In this case,
$\Sigma=6H+3E_i+2E_j-2E_l$ is a solution of Eq. (\ref{sigma E_6
condition02}). Next we consider the case of
$(\alpha_1,\alpha_2,\alpha_3)=(1,1,0)$. It is clear that we have
$L_2=\mathcal{O}_{S}( 5[E_l,E_m]''-2E_i+2E_j)^{1/30}$. The
necessary conditions are
\begin{equation}
\left\{\begin{array}{l} (E_i-E_j)\cdot\Sigma_{E_6}=p_3-p_1\\
([E_l,E_m]'')\cdot\Sigma_{E_6}=-p_1-p_2,\label {sigma E_6
condition03}
\end{array}   \right.
\end{equation}
and Eq. (\ref{sigma E_6 condition01_01}). Note that the
constraints are the same as the previous case,
$(\alpha_1,\alpha_2,\alpha_3)=(0,0,0)$. As a result, the allowed
configurations are the same as (\ref{Field content E_6 02}). Let
us take a look at the classes of the curves, which solve Eq.
(\ref{sigma E_6 condition03}). For simplicity, we focus on the
case of $[E_l,E_m]''=E_l-E_m$ and consider
$(p_1,p_2,p_3,p_4,p_5,p_6)=(1,0,2,1,0,-1)$, it is not difficult to
see that $\Sigma=H-E_i-E_m$ is a solution. For the case of
$(p_1,p_2,p_3,p_4,p_5,p_6)=(2,1,1,0,3,2)$,
$\Sigma=4H+2E_l-E_j-E_m$ can solve Eq. (\ref{sigma E_6
condition03}).

Let us turn to the first case in Table {\ref{SU(6) bulk01}},
namely $(\alpha_1,\alpha_2,\alpha_3)=(1,1,0)^{\ast}$. In this
case, $L_2$ is $\mathcal{O}_{S}(E_i-E_j)^{1/10}$ and the necessary
conditions for the homological class of $\Sigma_{E_6}$ with given
configurations $(p_1,p_2,p_3,p_4,p_5,p_6)$ are
\begin{equation}
\left\{\begin{array}{l} (E_i-E_j)\cdot\Sigma_{E_6}=p_3-p_1\\
p_2+p_3=0, \label {sigma E_6 condition04}
\end{array}   \right.
\end{equation}
and Eq. (\ref{sigma E_6 condition01_01}). Note that to avoid
exotic fields, we require that $p_1,\; p_2,\; p_3,\; p_4\in
\mathbb{Z}_{\geqslant 0}$. The constraint, $p_2+p_3=0$ in Eq.
(\ref{sigma E_6 condition04}) implies that $p_2=p_3=0$. By the
constraints in Eq. (\ref{sigma E_6 condition01_01}), we obtain
$(p_1,p_2,p_3,p_4,p_5,p_6)=(0,0,0,0,0,0)$, which means that there
are no non-trivial configurations in this case. As a check, by the
configurations in (\ref{Field content E_6 02}) and the constraint
$p_2+p_3=0$, it is easy to see that there is indeed no solution,
namely all configurations in (\ref{Field content E_6 02}) are
completely ruled out by the constraint $p_2+p_3=0$.

For the case of $(\alpha_1,\alpha_2,\alpha_3)=(3,3,0)$, we have
$L_2=\mathcal{O}_{S}(E_j-E_i)^{7/30}$. Given the configuration
$(p_1,p_2,p_3,p_4,p_5,p_6)$, the necessary conditions are
\begin{equation}
\left\{\begin{array}{l} (E_i-E_j)\cdot\Sigma_{E_6}=p_3-p_1\\
p_3=2p_1+p_2 ,\label {sigma E_6 condition05}
\end{array}   \right.
\end{equation}
and Eq. (\ref{sigma E_6 condition01_01}). Since ${\bf
(3,2)}_{1,1,-3}$, ${\bf (3,2)}_{-1,1,-3}$, ${\bf (\bar
3,1)}_{1,-4,-3}$, and ${\bf (\bar 3,1)}_{1,-4,-3}$ are all matter
in the MSSM, we require that $p_i\leqslant 3,\; i=1,2,3,4$. By the
second condition in Eq. (\ref{sigma E_6 condition05}), we have
$(p_1,p_2)=(1,0),\; (0,1),\;(0,2),\;(0,3)$, or $(1,1)$. Since
$p_4\leqslant 3$, it follows that the allowed configurations are
$(p_1,p_2,p_3,p_4,p_5,p_6)=(0,1,1,2,-1,0)$, $(1,0,2,1,0,-1)$, and
$(1,1,3,3,-1,-1)$. Recall that in order to obtain matter in the
MSSM, it is required that $L_{1\Sigma}\neq \mathcal{O}_{\Sigma}$,
$L_{2\Sigma}\neq \mathcal{O}_{\Sigma}$ and $L'_{\Sigma}\neq
\mathcal{O}_{\Sigma}$. As a result, the resulting configurations
are
\begin{equation}
(p_1,p_2,p_3,p_4,p_5,p_6)=\left\{\begin{array}{l}
(0,1,1,2,-1,0),\;(1,0,2,1,0,-1) \label {Field content E_6 04}
\end{array}   \right\}.
\end{equation}
As a check, using the configurations in (\ref{Field content E_6
02}) and the constraint $p_3=2p_1+p_2$, one can see that the
resulting configurations are the same as that in (\ref{Field
content E_6 04}). Now let us solve the classes of the curves
satisfying Eq. (\ref{sigma E_6 condition05}). For these two
configurations, the first condition in Eq. (\ref{sigma E_6
condition05}) can be solved by $\Sigma=H-E_i-E_l$. For other
$U(1)^2$ flux configurations corresponding to the case of
$(\alpha_1,\alpha_2,\alpha_3)=(k,k,0)$ with $k=2$, the analysis is
similar to the case of $k=0,1$. We summarize the results in Table
{\ref{MSSM cond. E_6}}.

\begin{table}[h]
\begin{center}
\renewcommand{\arraystretch}{1.25}
\begin{tabular}{|c|c|c|c|c|c|c|c|} \hline

$(\alpha_1,\alpha_2,\alpha_3)$ & Conditions & $L_2$\\
\hline
$(0,0,0)$& $(E_i-E_l)\cdot\Sigma=p_2+p_3$ & $\mathcal{O}_{S}(5E_l-2E_i-3E_j)^{1/30}$\\
& $(E_i-E_l)\cdot\Sigma=-p_1-p_2$ & $\mathcal{O}_{S}(-5E_l+3E_i+2E_j)^{1/30}$\\
\hline
$~(1,1,0)^{\ast}$ & $p_2+p_3=0$ & $\mathcal{O}_{S}(E_i-E_j)^{1/10}$\\
\hline
$(1,1,0)$ & $([E_l,E_m]'')\cdot\Sigma=-p_1-p_2$ & $\mathcal{O}_{S}(5[E_l,E_m]''-2E_i+2E_j)^{1/30}$\\
\hline
$(2,2,0)$ & $(-E_l+E_j)\cdot\Sigma=-p_1-p_2$ & $\mathcal{O}_{S}(-5E_l-2E_i+7E_j)^{1/30}$\\
& $(E_l-E_i)\cdot\Sigma=-p_1-p_2$ & $\mathcal{O}_{S}(5E_l-7E_i+2E_j)^{1/30}$\\
\hline
$(3,3,0)$ & $p_3=2p_1+p_2$ & $\mathcal{O}_{S}(E_j-E_i)^{7/30}$\\
\hline
\end{tabular}
\caption{The conditions for $\Sigma_{E_6}$ supporting the field
configurations $(p_1,p_2,p_3,p_4,p_5,p_6)$ with
$L_1=\mathcal{O}_{S}(E_j-E_i)^{1/5}$ and constraints
$p_4=p_2+p_3-p_1$, $p_5=2p_1-p_3$, and $p_6=p_1+p_2-p_3$.}
\label{MSSM cond. E_6}
\end{center}
\end{table}
After analyzing the spectrum from the curves, it is clear that we
are unable to obtain a minimal spectrum of the MSSM, but
non-minimal spectra with doublet-triplet splitting can be
obtained. In the next section we will give examples of non-minimal
spectra for the MSSM.

\subsection{Non-minimal Spectrum for the MSSM: Examples}

In the previous section we already analyzed the spectrum from the
curves $\Sigma_{SU(7)}$, $\Sigma_{SO(12)}$, and $\Sigma_{E_6}$.
With some physical requirements, we obtain all field
configurations supported by the curves. In what follows, we shall
give examples of the non-minimal MSSM spectra using the results
shown in section $5.2.2$.

In what follows, we shall focus on the case $M_1$ in Table $13$.
In this case, $Q_L$ and $e_R$ are localized on the curves with
$G_{\Sigma}=SO(12)$. $u_R$ comes from $\Sigma_{E_6}$ and
$d_R,L_L,H_u$ and $H_d$ live on $\Sigma_{SU(7)}$. It is not
difficult to see that in the examples considered, we are unable to
get a minimal spectrum of the MSSM without exotic fields. However,
it is possible to construct non-minimal spectra of the MSSM. One
possible way is that we can make the exotic fields form trilinear
couplings with conserved $U(1)$ charges so that they can decouple
from the low-energy spectrum. According to the results in Table
\ref{SU(6) bulk01}, let us consider the $U(1)^2$ flux
configuration $L_1=\mathcal{O}_{S}(E_1-E_2)^{1/5}$ and
$L_2=\mathcal{O}_{S}(5E_3-2E_2-3E_1)^{1/30}$, which corresponds to
the case of $(\alpha_1,\alpha_2,\alpha_3)=(0,0,0)$ on the bulk. To
obtain three copies of $Q_L$ and $e_R$, we engineer two curves
$\Sigma^1_{SO(12)}$ and $\Sigma^2_{SO(12)}$ with field content
$(2,0,-2,0,4)$ and $(1,0,-1,0,2)$, respectively. The exotic fields
are $2\times {\bf(\bar 3,1)}_{-2,2,4}$ and one singlet on
$\Sigma^1_{SO(12)}$. For the curve $\Sigma^2_{SO(12)}$, we get
exotic fields $1\times {\bf(\bar 3,1)}_{-2,2,4}$ and two singlets.
To get three copies of $u_R$, we arrange two curves,
$\Sigma^1_{E_6}$ and $\Sigma^2_{E_6}$ with field content
$(3,1,2,0,4,2)$ and $(2,1,1,0,3,2)$, respectively. We have exotic
fields $3\times {\bf(3,2)}_{1,1,-3}$, $1\times {\bf
(3,2)}_{-1,1,-3}$ and six singlets on $\Sigma^1_{E_6}$. On
$\Sigma^2_{E_6}$, the exotic fields are $2\times
{\bf(3,2)}_{1,1,-3}$, $1\times {\bf (3,2)}_{-1,1,-3}$ and five
singlets. Since the rest of the fields in the case of $M_1$ come
from the curves with $G_{\Sigma}=SU(7)$, we can easily engineer
$3\times d_R$, $3\times L_L$, $1\times H_u$ and $1\times H_d$ on
individual curves, denoted respectively by $\Sigma^1_{SU(7)}$,
$\Sigma^2_{SU(7)}$, $\Sigma^u_{SU(7)}$, and $\Sigma^d_{SU(7)}$.
Note that ${\bf(3,2)}_{\pm 1,1,-3}$, ${\bf(\bar 3,1)}_{-2,2,4}$,
and ${\bf(1,\bar 2)}_{7,-3,-1}$ can form trilinear couplings. To
make the exotic fields form the couplings, we introduce one extra
curve $\Sigma^{\Phi}_{SU(7)}$ with $\Phi={\bf(1,\bar
2)}_{7,-3,-1}$. Now we arrange $\Sigma^1_{SO(12)}$ intersects
$\Sigma^1_{E_6}$ and $\Sigma^2_{E_6}$, so does
$\Sigma^2_{SO(12)}$. The curve $\Sigma^u_{SU(7)}$ passes through
the intersection point of $\Sigma^1_{SO(12)}$ and $\Sigma^1_{E_6}$
and that of $\Sigma^2_{SO(12)}$ and $\Sigma^2_{E_6}$. The vertices
of the triple intersections
$(\Sigma^1_{SO(12)},\Sigma^1_{E_6},\Sigma^u_{SU(7)})$ and
$(\Sigma^2_{SO(12)},\Sigma^2_{E_6},\Sigma^u_{SU(7)})$ represent
the coupling $Q_Lu_RH_u$. Another two vertices are formed by
triple intersections
$(\Sigma^1_{SO(12)},\Sigma^2_{E_6},\Sigma^{\Phi}_{SU(7)})$ and
$(\Sigma^2_{SO(12)},\Sigma^1_{E_6},\Sigma^{\Phi}_{SU(7)})$, which
represent the coupling $\Theta\Psi\Phi$ and
$\widetilde{\Theta}\Psi\Phi$, where $\Theta={\bf(3,2)}_{1,1,-3}$,
$\widetilde{\Theta}={\bf(3,2)}_{-1,1,-3}$, and $\Psi={\bf(\bar
3,1)}_{-2,2,4}$. When $\Phi$ gets a vev, the exotic fields are
decoupled through the coupling, which means that at low energy,
those fields will not show up in the spectrum. To obtain the
coupling $Q_Ld_RH_d$, one can arrange two curves
$\Sigma^1_{SU(7)}$, and $\Sigma^d_{SU(7)}$ intersect
$\Sigma^1_{SO(12)}$ at one point. For the coupling $L_Le_RH_d$,
one can let the curve $\Sigma^2_{SU(7)}$ intersect
$\Sigma^d_{SU(7)}$ at another point on $\Sigma^1_{SO(12)}$. The
intersection point of $\Sigma^u_{SU(7)}$ and $\Sigma^2_{SU(7)}$
represents the coupling $L_LN_RH_u$. To sum up, the superpotential
is as follows:
\begin{eqnarray}
\mathcal{W}&\supset & \mathcal{W}_{{\rm MSSM}}+
\Theta\Psi\Phi+\widetilde{\Theta}\Psi\Phi+\cdots.\label{MSSM
couplings}
\end{eqnarray}

\renewcommand{\thefootnote}{\fnsymbol{footnote}}

\begin{table}[h]
\begin{center}
\renewcommand{\arraystretch}{1.2}
\begin{tabular}{|@{}c@{}|c|l@{\,}|c|@{\,}l@{}|@{\,}l@{}|@{\,}l@{}|} \hline

Multiplet & Curve & ~~~~~~~~~$\Sigma$ & $g_{\Sigma}$ & ~~~~~~$L_{1\Sigma}$ & ~~~~~~~$L_{2\Sigma}$ & ~~~~~~~~$L'_{\Sigma}$\\
\hline \hline

$2\times Q_L$ & \multirow{2}{*}{$\Sigma_{SO(12)}^{1}$} &
$5H+2E_2-2E_3$ & \multirow{2}{*}{0} &
\multirow{2}{*}{$\mathcal{O}_{\Sigma_{SO(12)}^{1}}(1)^{2/5}$} &
\multirow{2}{*}{$\mathcal{O}_{\Sigma_{SO(12)}^{1}}(1)^{7/15}$} &
\multirow{2}{*}{$\mathcal{O}_{\Sigma_{SO(12)}^{1}}(1)^{1/3}$} \\
$+3\times e_R$\footnotemark[3] & & $-2E_4-2E_5$ &&&&\\
\hline

\multirow{2}{*}{$1\times Q_L$\footnotemark[4]} &
\multirow{2}{*}{$\Sigma_{SO(12)}^{2}$} & $4H+E_2-E_3$ &
\multirow{2}{*}{0} &
\multirow{2}{*}{$\mathcal{O}_{\Sigma_{SO(12)}^{2}}(1)^{1/5}$} &
\multirow{2}{*}{$\mathcal{O}_{\Sigma_{SO(12)}^{2}}(1)^{7/30}$} &
\multirow{2}{*}{$\mathcal{O}_{\Sigma_{SO(12)}^{2}}(1)^{1/6}$} \\
 & & $-2E_4-2E_5$ &&&&\\
\hline

$2\times u_R\footnotemark[5]$ & $\Sigma_{E_6}^{1}$ & $5H+3E_3-E_1$
& 0 & $\mathcal{O}_{\Sigma_{E_6}^{1}}(1)^{1/5}$ &
$\mathcal{O}_{\Sigma_{E_6}^{1}}(-1)^{3/5}$ &
$\mathcal{O}_{\Sigma_{E_6}^{1}}(1)$\\
\hline

$1\times u_R\footnotemark[6]$ & $\Sigma_{E_6}^{2}$ & $4H+2E_3-E_1$
& 0 & $\mathcal{O}_{\Sigma_{E_6}^{2}}(1)^{1/5}$ &
$\mathcal{O}_{\Sigma_{E_6}^{2}}(-1)^{13/30}$ &
$\mathcal{O}_{\Sigma_{E_6}^{2}}(1)^{1/2}$\\
\hline

\multirow{2}{*}{$3\times d_R$} &
\multirow{2}{*}{$\Sigma_{SU(7)}^{1}$} & $4H+E_2+E_3$ &
\multirow{2}{*}{0} &
\multirow{2}{*}{$\mathcal{O}_{\Sigma_{SU(7)}^{1}}(1)^{3/5}$} &
\multirow{2}{*}{$\mathcal{O}_{\Sigma_{SU(7)}^{1}}(-1)^{3/10}$} &
\multirow{2}{*}{$\mathcal{O}_{\Sigma_{SU(7)}^{1}}(1)^{3/14}$} \\
&& $-2E_1$ &&&&\\
\hline

\multirow{2}{*}{$3\times L_L$} &
\multirow{2}{*}{$\Sigma_{SU(7)}^{2}$} & $4H+E_3+E_1$ &
\multirow{2}{*}{0} &
\multirow{2}{*}{$\mathcal{O}_{\Sigma_{SU(7)}^{2}}(-1)^{3/5}$} &
\multirow{2}{*}{$\mathcal{O}_{\Sigma_{SU(7)}^{2}}(-1)^{1/5}$} &
\multirow{2}{*}{$\mathcal{O}_{\Sigma_{SU(7)}^{2}}(1)^{1/7}$} \\
&& $-2E_2$ &&&&\\
\hline

$1\times H_u$ & $\Sigma_{SU(7)}^{u}$ & $~\,H-E_1-E_3$ & 0 &
$\mathcal{O}_{\Sigma_{SU(7)}^{u}}(1)^{1/5}$ &
$\mathcal{O}_{\Sigma_{SU(7)}^{u}}(1)^{1/15}$ &
$\mathcal{O}_{\Sigma_{SU(7)}^{u}}(-1)^{1/21}$\\
\hline

$1\times H_d$ & $\Sigma_{SU(7)}^{d}$ & $~\,H-E_2-E_4$ & 0 &
$\mathcal{O}_{\Sigma_{SU(7)}^{d}}(-1)^{1/5}$ &
$\mathcal{O}_{\Sigma_{SU(7)}^{d}}(-1)^{1/15}$ &
$\mathcal{O}_{\Sigma_{SU(7)}^{d}}(1)^{1/21}$\\
\hline

\multirow{2}{*}{$1\times\Phi$} &
\multirow{2}{*}{$\Sigma_{SU(7)}^{\Phi}$} & $3H-E_1-E_3$ &
\multirow{2}{*}{0} &
\multirow{2}{*}{$\mathcal{O}_{\Sigma_{SU(7)}^{\Phi}}(-1)^{1/5}$} &
\multirow{2}{*}{$\mathcal{O}_{\Sigma_{SU(7)}^{\Phi}}(-1)^{1/15}$}&
\multirow{2}{*}{$\mathcal{O}_{\Sigma_{SU(7)}^{\Phi}}(1)^{1/21}$} \\
&& $-2E_2$ &&&&\\
\hline
\end{tabular}
\caption{An example for a non-minimal MSSM spectrum from
$G_S=SU(6)$ with the $U(1)^2$ gauge flux configuration
$L_1=\mathcal{O}_{S}(E_{1}-E_{2})^{1/5}$ and
$L_2=\mathcal{O}_{S}(5E_3-2E_2-3E_1)^{1/30}$.} \label{non-min
MSSM01}
\end{center}
\end{table}
\footnotetext[3]{With one additional singlet.}
\footnotetext[4]{With two additional singlets.}
\footnotetext[5]{With six additional singlets.}
\footnotetext[6]{With five additional singlets.} As mentioned
earlier, through the last two couplings in (\ref{MSSM couplings}),
we obtain a non-minimal MSSM spectrum at low energy. Note that in
this case, $H_u$ and $H_d$ come from the curves $\Sigma^u_{SU(7)}$
and $\Sigma^d_{SU(7)}$, respectively. As shown in section $5.2.2$,
doublet-triplet splitting can be achieved by $U(1)^2$ gauge
fluxes. Therefore, a non-minimal spectrum of the MSSM with
doublet-triple splitting can be achieved in a local F-theory model
where $G_S=SU(6)$ and with $U(1)^2$ gauge fluxes. As shown in
section $5.2.2$, given the field configurations, one can calculate
the homological classes of the curves supporting the
configurations. For the present example, we simply summarize the
field content and the classes of the curves in Table \ref{non-min
MSSM01}. Note that in the previous example there are some exotic
singlets. Following similar procedure, these singlets can be
lifted via trilinear couplings. Let us consider the following
example. To obtain three copies of $Q_L$ and $e_R$, we engineer
two curves $\widetilde{\Sigma}^1_{SO(12)}$ and
$\widetilde{\Sigma}^2_{SO(12)}$ with field content $(2,1,-2,-1,3)$
and $(1,2,-1,-2,0)$, respectively. Clearly the exotic fields are
$1\times {\bf(\bar 3,1)}_{2,-4,2}$, $2\times {\bf(\bar
3,1)}_{-2,2,4}$, and $1\times {\bf(1,\bar 2)}_{-2,-3,4}$ on
$\widetilde{\Sigma}^1_{SO(12)}$. For the curve
$\widetilde{\Sigma}^2_{SO(12)}$, we get exotic fields $2\times
{\bf(\bar 3,1)}_{2,-4,2}$, $1\times {\bf(\bar 3,1)}_{-2,2,4}$, and
$2\times {\bf(1,\bar 2)}_{-2,-3,4}$. To get three copies of $u_R$,
we arrange two curves, $\widetilde{\Sigma}^1_{E_6}$ and
$\widetilde{\Sigma}^2_{E_6}$ with field content $(2,1,1,0,3,2)$
and $(3,1,2,0,4,2)$, respectively. We have exotic fields $2\times
{\bf(3,2)}_{1,1,-3}$, $1\times {\bf (3,2)}_{-1,1,-3}$, and five
singlets on $\widetilde{\Sigma}^1_{E_6}$. On
$\widetilde{\Sigma}^2_{E_6}$, the exotic fields are $3\times
{\bf(3,2)}_{1,1,-3}$, $1\times {\bf (3,2)}_{-1,1,-3}$, and six
singlets. Since the rest of the fields in the case of $M_1$ come
from the curves with $G_{\Sigma}=SU(7)$, we can easily engineer
$3\times d_R$, $3\times L_L$, $1\times H_u$ and $1\times H_d$ on
individual curves, denoted respectively by
$\widetilde{\Sigma}^1_{SU(7)}$, $\widetilde{\Sigma}^2_{SU(7)}$,
$\widetilde{\Sigma}^u_{SU(7)}$, and
$\widetilde{\Sigma}^d_{SU(7)}$. Note that these exotic fields can
form trilinear couplings with triplets on $\Sigma_{SU(7)}$. To
make the exotic fields form the couplings, we introduce three
extra curves $\Sigma^{\Upsilon_1}_{SU(7)}$,
$\Sigma^{{\bar\Upsilon}_2}_{SU(7)}$, and
$\Sigma^{\Upsilon'_3}_{SU(7)}$  with
${\Upsilon}_1={\bf(3,1)}_{-7,-2,1}$, $\bar\Upsilon_2={\bf(\bar
3,1)}_{7,2,-1}$, and $\Upsilon_3+\Lambda$, respectively, where
$\Upsilon_3={\bf(3,1)}_{-7,-2,1}$ and
$\Lambda={\bf(1,1)}_{-7,0,-5}$. The superpotential is as follows:
\begin{eqnarray}
\mathcal{W}&\supset & \mathcal{W}_{{\rm MSSM}}+
\Xi\Delta\Upsilon_1+\Xi\widetilde{\Delta}\Upsilon_1+\Theta\Pi\bar\Upsilon_2+\widetilde{\Theta}\Pi\bar\Upsilon_2
+\Psi\Lambda\Upsilon_3+\cdots,\label{MSSM couplings02}
\end{eqnarray}
where $\Xi={\bf(\bar 3,1)}_{2,-4,2}$,
$\Delta={\bf(1,1)}_{1,6,-3}$,
$\widetilde{\Delta}={\bf(1,1)}_{-1,6,-3}$, and $\Pi={\bf(1,\bar
2)}_{-2,-3,4}$. When ${\Upsilon}_1$, $\bar\Upsilon_2$, and
$\Upsilon_3$ get vevs, the exotic fields are decoupled via the
couplings, which means that at low energy, those fields will not
show up in the spectrum. For the couplings in $\mathcal{W}_{{\rm
MSSM}}$, the arrangement of the curves is similar to the previous
example. We are not going to repeat that. In this example, we
obtain a non-minimal MSSM spectrum at low energy. The field
content and the classes of the curves are summarized in Table
\ref{non-min MSSM02}.\footnotemark[7]

\footnotetext[7]{In this example $Q_L$ and $u_R$ are localized on
different curves. The Yukawa coupling $Q_Lu_RH_u$ descended from
${\bf 10}{\bf 10}{\bf 5}$ can be expressed as
$[\widetilde{\Sigma}^1_{SO(12)}(1,2)+\widetilde{\Sigma}^2_{SO(12)}(3)][\widetilde{\Sigma}^1_{E_6}(1)+
\widetilde{\Sigma}^2_{E_6}(2,3)][\widetilde{\Sigma}^u_{SU(7)}]$
generating nonvanishing diagonal elements in the Yukawa mass
matrix, where the indices in the parenthesis represent the
generations.}

\renewcommand{\thefootnote}{\fnsymbol{footnote}}

\begin{table}[h]
\begin{center}
\renewcommand{\arraystretch}{1.2}
\begin{tabular}{|@{}c@{}|c|l@{\,}|c|@{\,}l@{}|@{\,}l@{}|@{\,}l@{}|} \hline

Multiplet & Curve & ~~~~~~~~~$\Sigma$ & $g_{\Sigma}$ & ~~~~~~$L_{1\Sigma}$ & ~~~~~~~$L_{2\Sigma}$ & ~~~~~~~~$L'_{\Sigma}$\\
\hline \hline

$2\times Q_L$ & \multirow{2}{*}{$\widetilde{\Sigma}_{SO(12)}^{1}$}
& $5H-E_1-4E_3$ & \multirow{2}{*}{0} &
\multirow{2}{*}{$\mathcal{O}_{\widetilde{\Sigma}_{SO(12)}^{1}}(1)^{1/5}$}
&
\multirow{2}{*}{$\mathcal{O}_{\widetilde{\Sigma}_{SO(12)}^{1}}(1)^{17/30}$}
&
\multirow{2}{*}{$\mathcal{O}_{\widetilde{\Sigma}_{SO(12)}^{1}}(1)^{1/3}$} \\
$+3\times e_R$ & & $-E_5$ &&&&\\
\hline

\multirow{2}{*}{$1\times Q_L$} &
\multirow{2}{*}{$\widetilde{\Sigma}_{SO(12)}^{2}$} &
\multirow{2}{*}{$4H+E_1-2E_3$} &  \multirow{2}{*}{0} &
\multirow{2}{*}{$\mathcal{O}_{\widetilde{\Sigma}_{SO(12)}^{2}}(-1)^{1/5}$}
&
\multirow{2}{*}{$\mathcal{O}_{\widetilde{\Sigma}_{SO(12)}^{2}}(1)^{13/30}$}
&
\multirow{2}{*}{$\mathcal{O}_{\widetilde{\Sigma}_{SO(12)}^{2}}(1)^{1/6}$} \\
& & $+E_6$ &&&&\\
\hline

$1\times u_R$ & $\widetilde{\Sigma}_{E_6}^{1}$ & $4H+2E_3-E_1$ & 0
& $\mathcal{O}_{\widetilde{\Sigma}_{E_6}^{1}}(1)^{1/5}$ &
$\mathcal{O}_{\widetilde{\Sigma}_{E_6}^{1}}(-1)^{13/30}$ &
$\mathcal{O}_{\widetilde{\Sigma}_{E_6}^{1}}(1)^{1/2}$\\
\hline

$2\times u_R$ & $\widetilde{\Sigma}_{E_6}^{2}$ & $5H+3E_3-E_1$ & 0
& $\mathcal{O}_{\widetilde{\Sigma}_{E_6}^{2}}(1)^{1/5}$ &
$\mathcal{O}_{\widetilde{\Sigma}_{E_6}^{2}}(-1)^{3/5}$ &
$\mathcal{O}_{\widetilde{\Sigma}_{E_6}^{2}}(1)$\\
\hline

\multirow{2}{*}{$3\times d_R$} &
\multirow{2}{*}{$\widetilde{\Sigma}_{SU(7)}^{1}$} & $4H+E_2+E_3$ &
\multirow{2}{*}{0} &
\multirow{2}{*}{$\mathcal{O}_{\widetilde{\Sigma}_{SU(7)}^{1}}(1)^{3/5}$}
&
\multirow{2}{*}{$\mathcal{O}_{\widetilde{\Sigma}_{SU(7)}^{1}}(-1)^{3/10}$}
&
\multirow{2}{*}{$\mathcal{O}_{\widetilde{\Sigma}_{SU(7)}^{1}}(1)^{3/14}$} \\
&& $-2E_1$ &&&&\\
\hline

\multirow{2}{*}{$3\times L_L$} &
\multirow{2}{*}{$\widetilde{\Sigma}_{SU(7)}^{2}$} & $4H+E_3+E_1$ &
\multirow{2}{*}{0} &
\multirow{2}{*}{$\mathcal{O}_{\widetilde{\Sigma}_{SU(7)}^{2}}(-1)^{3/5}$}
&
\multirow{2}{*}{$\mathcal{O}_{\widetilde{\Sigma}_{SU(7)}^{2}}(-1)^{1/5}$}
&
\multirow{2}{*}{$\mathcal{O}_{\widetilde{\Sigma}_{SU(7)}^{2}}(1)^{1/7}$} \\
&& $-2E_2$ &&&&\\
\hline

$1\times H_u$ & $\widetilde{\Sigma}_{SU(7)}^{u}$ & $~\,3H+E_2-E_4$
& 0 & $\mathcal{O}_{\widetilde{\Sigma}_{SU(7)}^{u}}(1)^{1/5}$ &
$\mathcal{O}_{\widetilde{\Sigma}_{SU(7)}^{u}}(1)^{1/15}$ &
$\mathcal{O}_{\widetilde{\Sigma}_{SU(7)}^{u}}(-1)^{1/21}$\\
\hline

$1\times H_d$ & $\widetilde{\Sigma}_{SU(7)}^{d}$ & $~\,H-E_2-E_4$
& 0 & $\mathcal{O}_{\widetilde{\Sigma}_{SU(7)}^{d}}(-1)^{1/5}$ &
$\mathcal{O}_{\widetilde{\Sigma}_{SU(7)}^{d}}(-1)^{1/15}$ &
$\mathcal{O}_{\widetilde{\Sigma}_{SU(7)}^{d}}(1)^{1/21}$\\
\hline

$1\times\Upsilon_1$ & $\widetilde{\Sigma}_{SU(7)}^{\Upsilon_1}$ &
$H-E_2-E_3$ & 0 &
$\mathcal{O}_{\widetilde{\Sigma}_{SU(7)}^{\Upsilon_1}}(-1)^{1/5}$
&
$\mathcal{O}_{\widetilde{\Sigma}_{SU(7)}^{\Upsilon_1}}(1)^{1/10}$
&
$\mathcal{O}_{\widetilde{\Sigma}_{SU(7)}^{\Upsilon_1}}(-1)^{1/14}$ \\
\hline

\multirow{2}{*}{$1\times{\bar\Upsilon}_2$} &
\multirow{2}{*}{$\widetilde{\Sigma}_{SU(7)}^{{\bar\Upsilon}_2}$} &
$2H-E_1-E_4$ &\multirow{2}{*}{0} &
\multirow{2}{*}{$\mathcal{O}_{\widetilde{\Sigma}_{SU(7)}^{{\bar\Upsilon}_2}}(1)^{1/5}$}
&
\multirow{2}{*}{$\mathcal{O}_{\widetilde{\Sigma}_{SU(7)}^{{\bar\Upsilon}_2}}(-1)^{1/10}$}&
\multirow{2}{*}{$\mathcal{O}_{\widetilde{\Sigma}_{SU(7)}^{{\bar\Upsilon}_2}}(1)^{1/14}$} \\
&& $-E_5$ &&&&\\
\hline

\multirow{2}{*}{$1\times\Upsilon_3$} &
\multirow{2}{*}{$\widetilde{\Sigma}_{SU(7)}^{\Upsilon'_3}$} &
\multirow{2}{*}{$H-E_2-E_4$} & \multirow{2}{*}{0} &
\multirow{2}{*}{$\mathcal{O}_{\widetilde{\Sigma}_{SU(7)}^{\Upsilon'_3}}(-1)^{1/5}$}
&
\multirow{2}{*}{$\mathcal{O}_{\widetilde{\Sigma}_{SU(7)}^{\Upsilon'_3}}(-1)^{1/15}$}
& \multirow{2}{*}{$\mathcal{O}_{\widetilde{\Sigma}_{SU(7)}^{\Upsilon'_3}}(-1)^{2/21}$} \\
$+1\times\Lambda$ &&&&&&\\
\hline
\end{tabular}
\caption{An example for a non-minimal MSSM spectrum from
$G_S=SU(6)$ with the $U(1)^2$ gauge flux configuration
$L_1=\mathcal{O}_{S}(E_{1}-E_{2})^{1/5}$ and
$L_2=\mathcal{O}_{S}(5E_3-2E_2-3E_1)^{1/30}$.} \label{non-min
MSSM02}
\end{center}
\end{table}

\section{Conclusions}

In this paper we demonstrate how to obtain $U(1)^2$ gauge flux
configurations $(L_1,L_2)$ with an exotic-free bulk spectrum of
the local F-theory model with $G_S=SU(6)$. In this case each
configuration is constructed by two fractional line bundles, which
are well-defined in the sense that up to a linear transformation
of the $U(1)$ charges, an $U(1)^2$ flux configuration can be
associated with a polystable bundle of rank two with structure
group $U(1)^2$. Under physical assumptions, we obtain all flux
configurations as shown in Table \ref{SU(6) bulk01} and Table
\ref{SU(6) bulk02}. For the case of $G_S=SO(10)$, as shown in
\cite{BHV: 2008 II} there is a no-go theorem which states that for
an exotic-free spectrum, there are no solutions for $U(1)^2$ gauge
fluxes.

To build a model of the MSSM, we study the field configurations
localized on the curves with non-trivial gauge fluxes induced from
the restriction of the flux configurations on the bulk $S$. With
the non-trivial induced fluxes, the enhanced gauge group
$G_{\Sigma}$ will be broken into $G_{\rm std}\times U(1)$. Under
physical assumptions, we obtain all field configurations localized
on the curves with $G_{\Sigma}=SU(7)$, $G_{\Sigma}=SO(12)$ and
$G_{\Sigma}=E_6$. Form the breaking patterns, we know that Higgs
fields are localized on the curves $\Sigma_{SU(7)}$ and
$\Sigma_{SO(12)}$. On the curve $\Sigma_{SU(7)}$, we found that
doublet-triplet splitting can be achieved. However, it is
impossible to get the splitting on the curve $\Sigma_{SO(12)}$
without raising exotic fields, which means that when building
models, we should engineer the Higgs fields on the curve
$\Sigma_{SU(7)}$ instead of $\Sigma_{SO(12)}$. Unlike Higgs
fields, matter fields in the MSSM are distributed over the curves
$G_{\Sigma}=SU(7)$, $G_{\Sigma}=SO(12)$ and $G_{\Sigma}=E_6$. With
the solved field configurations, it is clear that it is extremely
difficult to get the minimal spectrum of the MSSM without exotic
fields. However, if those exotic fields can form trilinear
couplings with the doublets or triplets on the curves with
$G_{\Sigma}={SU(7)}$, the exotic fields can be lifted from the
massless spectrum when these doublets or triplets get vevs. In
order to achieve this, we introduce extra curves to support these
doublets or triplets coupled to exotic fields. With this
procedure, we can construct a non-minimal spectrum of the MSSM
with doublet-triple splitting. It would be interesting to study
mechanisms breaking non-minimal gauge group $G_S$ down to $G_{\rm
std}$ other than $U(1)^2$ gauge fluxes.

\renewcommand{\thesection}{}
\section{\hspace{-1cm} Acknowledgments}

I would like to thank K. Becker, C. Bertinato, C.-M. Chen, J.
Heckman, and E. Sharpe for valuable communications and
discussions. I especially thank D. Robbins for his careful reading
of the manuscript of this paper and useful comments. This work is
supported in part by the NSF grant PHY-0555575 and Texas A\&M
University.
\newpage


\renewcommand{\theequation}{\thesection.\arabic{equation}}
\setcounter{equation}{0}



\newpage
\begin{thebibliography}{}

\bibitem{GSW}
 M.~B.~Green, J.~H.~Schwarz and E.~Witten,
 ``Superstring Theory''
 Cambridge University Press 1987.

\bibitem{Aldazabal:2000sa}
  G.~Aldazabal, L.~E.~Ibanez, F.~Quevedo and A.~M.~Uranga,
  JHEP {\bf 0008}, 002 (2000)
  [arXiv:hep-th/0005067].

\bibitem{Verlinde:2005jr}
  H.~Verlinde and M.~Wijnholt,
  JHEP {\bf 0701}, 106 (2007)
  [arXiv:hep-th/0508089];
  D.~Malyshev and H.~Verlinde,
  Nucl.\ Phys.\ Proc.\ Suppl.\  {\bf 171}, 139 (2007)
  [arXiv:0711.2451 [hep-th]], and references therein.

\bibitem{Blumenhagen:2005mu}
  R.~Blumenhagen, M.~Cvetic, P.~Langacker and G.~Shiu,
  Ann.\ Rev.\ Nucl.\ Part.\ Sci.\  {\bf 55}, 71 (2005)
  [arXiv:hep-th/0502005], and references therein.


\bibitem{BlumenhagenIbanezKachru}
  R.~Blumenhagen, M.~Cvetic and T.~Weigand,
  Nucl.\ Phys.\  B {\bf 771}, 113 (2007)
  [arXiv:hep-th/0609191];
  L.~E.~Ibanez and A.~M.~Uranga,
  JHEP {\bf 0703}, 052 (2007)
  [arXiv:hep-th/0609213];
  B.~Florea, S.~Kachru, J.~McGreevy and N.~Saulina,
  JHEP {\bf 0705}, 024 (2007)
  [arXiv:hep-th/0610003];
  R.~Blumenhagen, M.~Cvetic, D.~Lust, R.~Richter and T.~Weigand,
  Phys.\ Rev.\ Lett.\  {\bf 100}, 061602 (2008)
  [arXiv:0707.1871 [hep-th]].

\bibitem{Blumenhagen:2008zz}
  R.~Blumenhagen, V.~Braun, T.~W.~Grimm and T.~Weigand,
  arXiv:0811.2936 [hep-th].

\bibitem{Vafa:1996xn}
  C.~Vafa,
  Nucl.\ Phys.\  B {\bf 469}, 403 (1996)
  [arXiv:hep-th/9602022];
  D.~R.~Morrison and C.~Vafa,
  Nucl.\ Phys.\  B {\bf 473}, 74 (1996)
  [arXiv:hep-th/9602114];
  D.~R.~Morrison and C.~Vafa,
  Nucl.\ Phys.\  B {\bf 476}, 437 (1996)
  [arXiv:hep-th/9603161].

\bibitem{Denef:2008wq}
  F.~Denef,
  arXiv:0803.1194 [hep-th].

\bibitem{Bershadsky:1996nh}
  M.~Bershadsky, K.~A.~Intriligator, S.~Kachru, D.~R.~Morrison, V.~Sadov and C.~Vafa,
  Nucl.\ Phys.\  B {\bf 481}, 215 (1996)
  [arXiv:hep-th/9605200].

\bibitem{Sen:1996vd}
  A.~Sen,
  Nucl.\ Phys.\  B {\bf 475}, 562 (1996)
  [arXiv:hep-th/9605150].

\bibitem{BHV:2008I}
 C.~Beasley, J.~J.~Heckman and C.~Vafa,
  JHEP {\bf 0901}, 058 (2009)
  [arXiv:0802.3391 [hep-th]].

\bibitem{BHV:2008II}
 C.~Beasley, J.~J.~Heckman and C.~Vafa,
 arXiv:0806.0102 [hep-th].

\bibitem{Donagi:local}
  R.~Donagi and M.~Wijnholt,
  arXiv:0802.2969 [hep-th];
  R.~Donagi and M.~Wijnholt,
  arXiv:0808.2223 [hep-th];
  M.~Wijnholt,
  arXiv:0809.3878 [hep-th].

\bibitem{Heckman:local}
 J.~J.~Heckman and C.~Vafa,
 arXiv:0809.1098 [hep-th];
 J.~J.~Heckman and C.~Vafa,
 arXiv:0809.3452 [hep-ph];
J.~J.~Heckman, G.~L.~Kane, J.~Shao and C.~Vafa,
  arXiv:0903.3609 [hep-ph];
J.~J.~Heckman, A.~Tavanfar and C.~Vafa,
  arXiv:0906.0581 [hep-th].

\bibitem{Heckman:localFlavor01}
 J.~J.~Heckman and C.~Vafa,
 arXiv:0811.2417 [hep-th].

\bibitem{Heckman:localFlavor02}
V.~Bouchard, J.~J.~Heckman, J.~Seo and C.~Vafa,
  arXiv:0904.1419 [hep-ph].

\bibitem{Heckman:localFlavor03}
J.~J.~Heckman and C.~Vafa,
  arXiv:0904.3101 [hep-th].

\bibitem{Ibanez:locaFlavorl01}
 A.~Font and L.~E.~Ibanez,
 arXiv:0811.2157 [hep-th].

\bibitem{Ibanez:localFlavor02}
 A.~Font and L.~E.~Ibanez,
  JHEP {\bf 0909}, 036 (2009)
  [arXiv:0907.4895 [hep-th]].

\bibitem{Randall:localFlavor}
  L.~Randall and D.~Simmons-Duffin,
  arXiv:0904.1584 [hep-ph].

\bibitem{Nanopoulos:local}
 J.~Jiang, T.~Li, D.~V.~Nanopoulos and D.~Xie,
 arXiv:0811.2807 [hep-th];
 J.~Jiang, T.~Li, D.~V.~Nanopoulos and D.~Xie,
  arXiv:0905.3394 [hep-th];
  T.~Li,
  arXiv:0905.4563 [hep-th].

\bibitem{Blumenhagen:local}
 R.~Blumenhagen,
 arXiv:0812.0248 [hep-th].

\bibitem{Bourjaily:local}
  J.~L.~Bourjaily,
  arXiv:0901.3785 [hep-th];
J.~L.~Bourjaily,
  arXiv:0905.0142 [hep-th].

\bibitem{Chen:2009me}
  C.-M.~Chen and Y.-C.~Chung,
  Nucl.\ Phys.\  B {\bf 824}, 273 (2010)
  [arXiv:0903.3009 [hep-th]].


\bibitem{Conlon:local01}
  J.~P.~Conlon and E.~Palti,
  arXiv:0907.1362 [hep-th].

\bibitem{Conlon:local02}
 J.~P.~Conlon and E.~Palti,
  arXiv:0910.2413 [hep-th].

\bibitem{Vafa:NoncommutativeandYukawa}
  S.~Cecotti, M.~C.~N.~Cheng, J.~J.~Heckman and C.~Vafa,
  arXiv:0910.0477 [hep-th].


\bibitem{Caltech:global}
  J.~Marsano, N.~Saulina and S.~Schafer-Nameki,
  JHEP {\bf 0908}, 030 (2009)
  [arXiv:0904.3932 [hep-th]];
J.~Marsano, N.~Saulina and S.~Schafer-Nameki,
  JHEP {\bf 0908}, 046 (2009)
  [arXiv:0906.4672 [hep-th]].

\bibitem{Donagi:global}
  R.~Donagi and M.~Wijnholt,
  arXiv:0904.1218 [hep-th].

\bibitem{Cordova:Decouplinggeom}
  C.~Cordova,
  arXiv:0910.2955 [hep-th].

\bibitem{Other:global}
  H.~Hayashi, R.~Tatar, Y.~Toda, T.~Watari and M.~Yamazaki,
  Nucl.\ Phys.\  B {\bf 806}, 224 (2009)
  [arXiv:0805.1057 [hep-th]];
  A.~Collinucci, F.~Denef and M.~Esole,
  JHEP {\bf 0902}, 005 (2009)
  [arXiv:0805.1573 [hep-th]];
  A.~P.~Braun, A.~Hebecker, C.~Ludeling and R.~Valandro,
  arXiv:0811.2416 [hep-th];
  G.~Aldazabal, P.~G.~Camara and J.~A.~Rosabal,
  arXiv:0811.2900 [hep-th];
  A.~Collinucci,
  arXiv:0812.0175 [hep-th];
 B.~Andreas and G.~Curio,
  arXiv:0902.4143 [hep-th];
  R.~Tatar, Y.~Tsuchiya and T.~Watari,
  Nucl.\ Phys.\  B {\bf 823}, 1 (2009)
  [arXiv:0905.2289 [hep-th]];
  A.~Collinucci,
  arXiv:0906.0003 [hep-th];
  R.~Blumenhagen, T.~W.~Grimm, B.~Jurke and T.~Weigand,
  JHEP {\bf 0909}, 053 (2009)
  [arXiv:0906.0013 [hep-th]];
 P.~Aluffi and M.~Esole,
  arXiv:0908.1572 [hep-th];
 R.~Blumenhagen, T.~W.~Grimm, B.~Jurke and T.~Weigand,
  arXiv:0908.1784 [hep-th];
 T.~W.~Grimm, T.~W.~Ha, A.~Klemm and D.~Klevers,
  arXiv:0909.2025 [hep-th];
  K.~S.~Choi,
  arXiv:0910.2571 [hep-th].


\bibitem{Tartar:globalFlavor01}
H.~Hayashi, T.~Kawano, R.~Tatar and T.~Watari,
  arXiv:0901.4941 [hep-th].

\bibitem{Tartar:globalFlavor02}
H.~Hayashi, T.~Kawano, Y.~Tsuchiya and T.~Watari,
  arXiv:0910.2762 [hep-th].

\bibitem{Buchbinder:SUSYbreaking}
  E.~I.~Buchbinder,
  JHEP {\bf 0809}, 134 (2008)
  [arXiv:0805.3157 [hep-th]].

\bibitem{Caltech:SUSYbreaking}
  J.~J.~Heckman, J.~Marsano, N.~Saulina, S.~Schafer-Nameki and C.~Vafa,
  arXiv:0808.1286 [hep-th];
  J.~Marsano, N.~Saulina and S.~Schafer-Nameki,
  arXiv:0808.1571 [hep-th];
  J.~Marsano, N.~Saulina and S.~Schafer-Nameki,
  arXiv:0808.2450 [hep-th].


\bibitem{Blumenhagen:SUSYbreaking}
  R.~Blumenhagen, J.~P.~Conlon, S.~Krippendorf, S.~Moster and F.~Quevedo,
  JHEP {\bf 0909}, 007 (2009)
  [arXiv:0906.3297 [hep-th]].


\bibitem{Heckman:cosmology}
  J.~J.~Heckman, A.~Tavanfar and C.~Vafa,
  arXiv:0812.3155 [hep-th].

\bibitem{Katz:1996xe}
  S.~H.~Katz and C.~Vafa,
  Nucl.\ Phys.\ B {\bf 497}, 146 (1997)
  [arXiv:hep-th/9606086].


\bibitem{Donalson}
S. Donaldson,``Anti Self-Dual Yang-Mills Connections over Complex
Algebraic Surfaces and Stable Vector Bundles,'' Proc. London Math.
Soc. 50 1 (1985).

\bibitem{UY}
K. Uhlenbeck and S.-T. Yau,``On the existence of Hermitian
Yang-Mills connections in stable bundles,'' Comm. Pure App. Math.
39 257 (1986), 42 703 (1986).

\bibitem{Slansky:1981yr}
  R.~Slansky,
  Phys.\ Rept.\  {\bf 79}, 1 (1981).


\bibitem{delPezzo:01}
Demazure, Pinkham et Teissier, ``Seminaire sur les Singularities
des Surfaces,'' Ecole Polytechnique, 1976-1977.

\bibitem{delPezzo:02}
Y. I. Manin, Cubic forms: Algebra, geometry, arithmetic.
North-Holland Publishing Co., Amsterdam, second ed., 1986.
Translated from the Russian by M. Hazewinkel.


\bibitem{Hartshone:01}
R. Hartshorne, ``Algebraic geometry,'' New York : Springer-Verlag,
1977.

\bibitem{Griffith:01}
P. Griffith and J. Harris, ``Principles of Algebraic Geometry,''
Wiley NY 1994.

\bibitem{YM:01}
M. F. Atiyah, N. J. Hitchin, and I. M. Singer, ``Self-Duality in
Four-Dimensional Riemannian Geometry,'' Proc. R. Soc. Lond. A 362
(1978) 425-461.




\end{thebibliography}
\end{document}